\documentclass[11pt]{article}
\pdfoutput=1
\usepackage{myoptions}
\usepackage{jheppub}
\usepackage{float, extarrows, tikz-cd}
\usepackage{graphicx}
\usepackage{slashed}
\usepackage{tabularx,ragged2e}
\usepackage{amssymb}
\usepackage{amsmath,amssymb}
\usepackage{slashed}
\usepackage{caption}
\usepackage{xcolor}
\usepackage{dsfont}
\usepackage{verbatim}
\usepackage{subfig}
\usepackage{mathtools, xcolor,ytableau, amsfonts,tikz}
\usepackage{graphicx}
\usepackage{physics}
\usepackage{simpler-wick}
\newcolumntype{C}{>{\Centering\arraybackslash}X}

\ytableausetup{centertableaux,boxsize=.5em}
\title{Spontaneously Broken Boosts in CFTs} 

					\author[a]{Zohar Komargodski,}   
                                          \author[a]{M\'ark Mezei,}                                       
                                           \author[b]{Sridip Pal,}
                                            \author[a]{Avia Raviv-Moshe}   
                                           \affiliation[a]{Simons Center for Geometry and Physics, SUNY, Stony Brook, NY 11794, USA}                                                                    
                                            \affiliation[b]{School of Natural Sciences, Institute for Advanced Study, Princeton, NJ  08540, USA}
                                            
					 \emailAdd{zkomargodski@scgp.stonybrook.edu}  
                                           \emailAdd{mmezei@scgp.stonybrook.edu}   
                                           \emailAdd{sridip@ias.edu}                                        
                                           \emailAdd{araviv-moshe@scgp.stonybrook.edu}
                                           
\abstract{Conformal Field Theories (CFTs) have rich dynamics in heavy states.
We describe the constraints due to spontaneously broken boost and dilatation symmetries in such states. The spontaneously broken boost symmetries require the existence of new low-lying  {\it primaries} whose scaling dimension gap, we argue, scales as $O(1)$. We demonstrate these ideas  in various states, including fluid, superfluid, mean field theory, and Fermi surface states. We end with some remarks about the large charge limit in 2d and discuss a theory of a single compact boson with an arbitrary conformal anomaly. }
\begin{document} 
\maketitle
\renewcommand{\arraystretch}{1.5}

\section{Introduction and Summary}

Nambu-Goldstone theorems are some of our strongest non-perturbative constraints on the dynamics of Quantum Field Theories (QFTs). 
Let us review the setup for continuous internal (non space-time) symmetries in a QFT in $d$ space-time dimensions.
By Noether's theorem, there exists a charge $Q$ which is an extended operator depending topologically on a co-dimension 1 surface $\Sigma$. If in some state $|\Omega\rangle$ we have for some local operator $O(0)$ 
$$\langle\Omega|[Q,O(0)]|\Omega\rangle\neq0~,$$
then it can be shown by a deformation of $\Sigma$ that, roughly speaking, $\langle\Omega|j_0 (x) O(0)|\Omega\rangle$
cannot decay faster than $1/|x|^{d-1}$ as we take $|x|\to \infty$. This algebraic decay of a correlation function implies gapless excitations of $\ket{\Omega}$ in infinite volume. Furthermore, under various additional assumptions about the nature of the state $\ket{\Omega}$, the existence of an ordinary massless boson (or a superfluid mode) can be established. Incidentally, if $\langle\Omega|j_0 (x) O(0)|\Omega\rangle\neq0$ holds true, this means that also $\langle\Omega|O^\dagger (x) O(0)|\Omega\rangle$ must decay algebraically at most. And under some assumptions it cannot decay faster than $1/|x|^{d-2}$.\footnote{The argument for this invokes inserting a complete set of states and recalling that the matrix elements of $j_\mu$ are suppressed by a factor of momentum at small momentum. It would be nice to understand to what extent this argument about the matrix elements is general. \label{fn1}} This is unacceptable in $d=2$ since it leads to a violation of clustering, due to the connected correlator not decaying. Hence no such states $|\Omega\rangle$ can exist in $d=2$, which is the familiar statement of the Coleman-Mermin-Wagner theorem~\cite{Mermin:1966fe,Coleman:1973ci}.

In finite volume (in the absence of boundaries) no state $ |\Omega\rangle$ can have the property that 
$\langle\Omega|[Q,O(0)]|\Omega\rangle\neq0$. This is simply because we can diagonalize $Q$ in finite volume. Thus, the phenomenon of symmetry breaking is really due to the infinite volume limit. When symmetry breaking occurs, the Hilbert space of the finite volume theory becomes closer and closer to a direct sum of sub-Hilbert spaces which do not communicate via the action of {\it local} operators. When we take the infinite volume limit we only keep one of these sub-Hilbert spaces. But these sub-Hilbert spaces can still communicate by the action of extended operators such as $Q$. This is why symmetry breaking may occur in infinite volume and this is why the notion of superselection sectors exists. These comments will be important below.

The situation for space-time symmetries is, in principle, similar. There are however some interesting differences. Given the energy-momentum (EM) tensor of the theory, $T_{\mu\nu}$, we can construct the space-time symmetries of the infinite volume theory living in $\mathbb{R}^{d-1,1}$ from the conserved currents 
$${\mathcal{J}}^{(\xi)}_\mu=\xi^\nu T_{\mu\nu}~,$$
where $\xi$ is a Killing vector satisfying as usual $\partial^{(\nu}\xi^{\mu)}=0$. The currents ${\mathcal{J}}^{(\xi)}_\mu$ of course lead to the usual translations, rotations, and boosts. We denote these charges by $Q^{(\xi)}$. 
There can exist a state $|\Omega\rangle$ and a local operators $O(0)$ (which may or may not have spin indices, which we suppress for now) such that 
\begin{equation}\label{Comm}\langle\Omega|[Q^{(\xi)},O(0)]\Omega\rangle\neq0~.\end{equation}

For constant $\xi$, namely the translation symmetry, this can be easily achieved in any state which is not translationally invariant. Similarly, for rotations the commutator would be generally nonzero in a non-isotropic state for any operator $O$ with spin indices. 
A general discussion of various allowed symmetry breaking patterns can be found in~\cite{Nicolis:2015sra}. A rather common situation is the spontaneous breaking of boost symmetry in states $|\Omega\rangle$ which are homogeneous and isotropic. This is one of the focal points of our paper.\footnote{There are important consequences of the spontaneous breaking of boost symmetry also in states which break the spatial translation symmetry. See for example~\cite{Dubovsky:2012sh,Aharony:2013ipa}. Here our focus is on translationally invariant, isotropic states.}  

From dimensional analysis, a nonzero commutator for the boost Killing vector~\eqref{Comm} (which we can take to be $\xi^1=x^0$, $\xi^0=x^1$, with the rest of the components vanishing) means that correlators of some components of the EM tensor and $O$ decay not faster than $1/|x|^{d}$. This does not lead to any particular problems in $d=2$ and hence there is no obstruction for the spontaneous breaking of boost symmetry in two space-time dimensions. We will indeed see some examples later.

The boost symmetry differs conceptually from ordinary (non space-time) symmetries in that it cannot be preserved at finite volume. Indeed, if we compactify space while keeping time intact, the symmetry between space and time is manifestly destroyed. Therefore, the spontaneous breaking of boost symmetry does not necessarily mean that super-selection sectors must arise!  
Thus, the boost symmetry Nambu-Goldstone theorem does lead to an algebraic decay and hence a gapless spectrum of excitations of $|\Omega\rangle$, but it does not imply super-selection sectors since there is no sense in diagonalizing the boost symmetry in compact space.

As we have argued above, the boost symmetry Nambu-Goldstone theorem leads to a somewhat faster decay of correlation functions compared to the case of spontaneous breaking of ordinary global symmetries. Therefore, one-particle excitations are not necessary, but rather, composite massless excitations could play a role. This was realized in the beautiful recent paper \cite{Alberte:2020eil}.

While there is no boost symmetry in compact space, it is still true that if we focus our attention on a small enough patch of our compact space, there is an {\it approximate} boost symmetry. Depending on the state of the system, this approximate boost symmetry may appear to be spontaneously broken. The algebraic decay and gapless excitations in the flat space limit therefore entail some constraints on the spectrum of the finite volume theory. The nature of these constraints can be understood on dimensional grounds. Take space to be a hypercube $V=L^{d-1}$ where $L$ is the length and $V$ is the volume. Then to see a state with finite energy density $\epsilon$ in a small patch of the hypercube we have to start from a state with total energy $E= \epsilon V$. We assume that on distances $\epsilon^{-1/d}\ll \Delta x \ll L$ we see an algebraic decay. Indeed, $\epsilon^{-1/d}$ is a length scale in the infinite volume theory and since we see an algebraic decay in infinite volume it must be true that there is an algebraic decay at finite volume in the range 
 $\epsilon^{-1/d}\ll \Delta x \ll L$.
 An algebraic decay in the range $\epsilon^{-1/d}\ll \Delta x \ll L$ is not to be taken for granted. It means that the gap above our state, $E_\text{gap}$,  must be much smaller than $\epsilon^{1/d}$. So the energy of the excitations of $|\Omega\rangle$ should scale as $\epsilon V+ E_\text{gap}$ with $ E_\text{gap}\ll \epsilon^{1/d}$. Clearly, for the consistency  of an algebraic decay, the gap has to be arbitrarily smaller than $\epsilon^{1/d}$. This means that the gap has to go to zero as $L\to\infty$, in agreement with the gapless nature of the excitations in infinite volume. We can roughly speaking say that therefore 
 $$E_\text{gap}\sim L^{-\gamma}~,$$
 with $\gamma>0$. We can think of $\gamma$ as a certain critical exponent that measures how fast the gap closes in finite space as we increase the volume. 
  
Of course, the most natural choice is $\gamma=1$ but this does not follow from our general considerations. That $\gamma=1$ is natural can be motivated based on Wilsonian considerations. Let us take space to be infinite. The deep infrared theory should be a fixed point of the renormalization group (perhaps in the Lifshitz sense)  and hence there is a scale-free effective field theory description of the gapless degrees of freedom that give rise to the algebraic correlators at infinite volume. In that case, $\gamma=1$ follows from dimensional analysis. The above argument assumes that the long 
correlators are captured by some scale-free infrared theory where the energy density is a cutoff scale. In practice, it could be that the gap is even smaller than what is predicted by dimensional analysis if the low-energy modes have additional degeneracy (which in finite volume is only approximate). Indeed, consider states that are generic in the Eigenstate Thermalization Hypothesis (ETH) sense (see~\cite{deutsch2018eigenstate} for a recent review and references). Those states have a macroscopic entropy and hence an exponentially small gap. 
Our bound on the gap holds in all theories and appropriate states, regardless of whether the states are generic. 
 
 Furthermore, one can already at this stage say a few words about the density of these low-lying states. For instance, if there was just one such state with energy above $|\Omega\rangle$ scaling like  $L^{-\gamma}$ then the correlator in the range $\epsilon^{-1/d}\ll\Delta x\ll L$ would have behaved non-algebraically. 
 The same argument holds for any finite collection of states. We therefore need {\it infinitely} many states that become gapless as $L\to \infty$. 
 
 The breaking of the boost symmetry is not a rare phenomenon. As we will see, in a unitary theory, any state that in the infinite volume limit has a non-vanishing energy density will break the boost symmetry. Therefore, the above constraints on the spectrum hold for quite generic states in finite volume.
 
Our purpose here is to apply these ideas to Conformal Field Theories (CFTs). CFTs can be studied on the cylinder $$S^{d-1}\times \mathbb{R}~,$$ where the radius of the sphere is $R$. The energy spectrum is related to the spectrum of scaling dimensions 
\begin{equation}\label{cylinder} E={\Delta\over R}~.\end{equation}
 We consider states with a nontrivial macroscopic limit, i.e. states with nonzero energy density. This means that we must take $\Delta_\Omega= s_{d-1}\epsilon \,R^{d}$ with $\epsilon$ the energy density and $ s_{d-1}$ the volume of the unit $S^{d-1}$. So as we take the macroscopic limit, we are discussing states that correspond to operators with a large scaling dimension. 
From our general considerations above we found that there are infinitely many states with energy going to zero as  $R^{-\gamma}$. This means that there are infinitely many states with energy below
\begin{equation}\label{scalingIntro} 
E= s_{d-1}\epsilon \,R^{d-1}+c\, \epsilon^{(1-\gamma)/d}R^{-\gamma}~, 
\end{equation} 
where $c$ is some dimensionless constant and the power of $\epsilon$ in the second term is adjusted so that the result makes sense dimensionally.
This translates to having operators with scaling dimensions
\begin{equation}\label{scalingi} \Delta=\Delta_\Om +c\,\Delta_\Om^{1-\gamma\over d}~. \end{equation} 
This means that the gap around heavy operators with scaling dimension $\Delta_\Om$ is at most $\Delta_\Om^{1-\gamma\over d}$ with $\gamma>0$. 
As we have explained above, with some additional physical input it follows that $\gamma=1$ (or larger) and hence the gap around heavy operators scales like $O(1)$.
This bound on the scaling dimension gap around heavy operators should hold very generally and it does not require the genericity or typicality that is assumed in ETH. (Several of the applications we will study here are in fact concerned with large charge ground states, which are very atypical states.)

In CFTs on the cylinder~\eqref{cylinder} the gap constraint~\eqref{scalingi} may appear trivial given that there are descendant states. (For a review of CFTs on the cylinder see for instance~\cite{Simmons-Duffin:2016gjk}.) Indeed, for any primary state $|\Omega\rangle$ there is a family of states $|\partial^N \Omega\rangle$ (where the index contractions in the derivatives are not explicitly displayed) with scaling dimension $\Delta_\Om+N$ and hence energy $E=\Delta_\Om/R+N/R$ for any non-negative integer $N$.
There is therefore an interesting twist in the story: we can ask if the descendant states are those responsible for the low-energy theorems. The question of whether the descendant states are those responsible for the low-energy theorems is the essential new question addressed in this paper. 

We will see that, perhaps surprisingly, the answer is negative. One must have new {\it primary} states with dimension~\eqref{scalingi}. Furthermore, in all examples we study, $\gamma=1$ for these new primary states, in agreement with the general arguments. The boost symmetry Nambu-Goldstone theorem is not sufficient to completely determine the low lying spectrum of primary excitations of $|\Omega\rangle$ and indeed the structure of excitations differs in different examples we study. 

A situation where the existence of new low-lying primaries as above is a nontrivial constraint arises in the study of ground states at fixed, large charge $Q$ for some $U(1)$ symmetry. From general considerations (which however do not apply in mean field theory) one expects~\cite{Hellerman:2015nra} (for a review and more references see~\cite{Gaume:2020bmp})
$$\Delta\sim Q^{d\over d-1}~.$$
These ground states are very non-generic heavy states and the constraints arising from the spontaneous breaking of boost symmetry are satisfied quite differently in different examples we study. 
Sometimes the excitations needed for the boost Nambu-Goldstone theorem should be considered as a Regge trajectory of (primary) excitations of $|\Omega\rangle$ -- the Regge trajectories appearing in these cases are reminiscent of~\cite{Jafferis:2017zna}. This occurs in the superfluid case and 2d, where these are one-particle states, and in mean field theory, where the excitations are to be thought of as a particle and a (zero momentum) hole in the Bose-Einstein condensate.  Sometimes we find that the primary excitations are two-particle states, as in the free Fermi surface. 

The case of 2d CFTs  provides an interesting testing ground for our considerations. As briefly alluded to above, spontaneous breaking of boost symmetry is possible in 2d. The states responsible for the algebraic decay of correlators are the Virasoro (but not global conformal) descendants of the heavy state. Hence, we do not find any constraint on the gap above a heavy operator. Nevertheless, the relevant Virasoro descendants form a Regge trajectory.  We also examine the fate of the large charge effective field theory of~\cite{Hellerman:2015nra} in 2d. We find that for compact CFTs, it just becomes the free compact scalar representation of the $u(1)\times u(1)$ Ka\v{c}-Moody algebra. And the theory only makes sense for $c=1$. For situations where the $U(1)$ symmetry does not get enhanced to a current algebra, the effective theory is nontrivial and it describes a single compact boson with arbitrary conformal anomaly. It resembles the effective string theory of Polchinski-Strominger~\cite{Polchinski:1991ax}.

To put the present paper in context, let us point out that there has been a monumental effort in recent years to understand the spectrum of scaling dimensions in CFTs. The numerical bootstrap constraints on the low scaling dimension operators are beautifully reviewed in~\cite{Simmons-Duffin:2016gjk} and~\cite{Chester:2019wfx}, among others. There is then a complementary effort to understand the scaling dimensions of various special heavy operators, e.g. the ground states at large fixed charge, as reviewed in~\cite{Gaume:2020bmp}, large fixed spin~\cite{Komargodski:2012ek,Fitzpatrick:2012yx}, and various combinations of large charge and large spin~\cite{Cuomo:2017vzg,Cuomo:2019ejv}. The present paper is aimed at understanding the general constraints from the spontaneous breaking of boost symmetry. This is relevant for the study of  heavy operators quite generally. But except for some brief discussion of the hydrodynamic regime, our focus here will be exclusively on the large charge ground states.

The outline of the paper is as follows. The paper consists of two parts. In Part~\ref{part1} we present an abstract argument for the bound on the gap in the operator spectrum of CFTs around heavy operators. In section~\ref{sec:NGthm} we write the precise consequences of the spontaneous breaking of boost symmetry (reviewing and very slightly extending~\cite{Alberte:2020eil}) and dilatation symmetry. This results in some constraints on the low momentum, low frequency behavior of Energy-Momentum correlation functions. We show how these constraints are satisfied in states that obey the assumptions of hydrodynamics: the ballistic sound mode in hydrodynamics exists essentially because of the spontaneous breaking of boost symmetry.
In section~\ref{sec:CFT}, we discuss in more detail CFTs on the cylinder and review how the operator-state correspondence can be used to relate matrix elements on the cylinder and the CFT data. We explain some important properties of conformal blocks in heavy states and conclude that descendants cannot play a role in the Nambu-Goldstone theorem for boosts. 
 Making the additional assumption that heavy state $\ket{\Om}$ is the ground state of some sector of the theory, we present an argument that the Nambu-Goldstone theorem implies a bound on the gap in the CFT operator spectrum.
In Part~\ref{part2} we examine a variety of examples, show that our bound is obeyed and identify special sets of operators that saturate the Nambu-Goldstone theorem. In section~\ref{sec:superfluid} we discuss the case of the superfluid. We identify in detail the primary states that are responsible for the Nambu-Goldstone theorem. 
In section~\ref{sec:FreeScalar} we repeat the exercise for mean field theory. In section~\ref{sec:FreeFermions} we repeat it for a free Fermi surface. In section~\ref{sec:boostbreak} we show that in $d=2$ the Nambu-Goldstone sum rule is satisfied by a Regge trajectory of one-particle-like states in the Verma module. We also discuss the role of the large charge effective field theory in 2d.
Three appendices contain technical details about Green's functions in nontrivial states, an extension of the free fermion discussion to $d=4$, and, finally, a derivation of some contact terms in correlation functions of the EM tensor.
%

\part{Bounding the Gap in the Operator Spectrum }\label{part1}

\section{The Boost Nambu-Goldstone Theorem}\label{sec:NGthm}

While Nambu-Goldstone theorems for internal symmetries are thoroughly understood, the case of spacetime symmetries provides us with new lessons to this day. One difference explained in the introduction, is that breaking of the boost symmetry does not lead to superselection sectors. Another difference is that while it is impossible to break continuous internal symmetries in $d=2$, boosts can be spontaneously broken in $d=2$. Finally, as mentioned above, in \cite{Alberte:2020eil} it was shown that the boost Nambu-Goldstone theorem can be saturated by multiparticle states; this is a novel phenomenon.\footnote{ See \cite{Rothstein:2017twg,Rothstein:2017niq} for a different, but equivalent discussion of the role of multiparticle states in closing the symmetry algebra of the theory. Ultimately, Landau's derivation of the relation between couplings and the Fermi velocity in Fermi liquid theory contains the same insights.}${}^,$\footnote{It would be nice to understand whether this can happen for internal symmetries away from the vacuum state -- see footnote \ref{fn1}.}

In this section we review the formulation of the boost Nambu-Goldstone theorem in translationally invariant energy eigenstates   as a sum rule as discussed in \cite{Alberte:2020eil}, and extend it 
to dilatations in CFTs. We demonstrate how these sum rules are obeyed in generic ETH states, i.e.~by hydrodynamics. Throughout the paper we use mostly minus signature, $g^{\mu\nu}={\rm diag}(+1,-1,-1,\dots)$.

The logic of the Nambu-Goldstone theorem requires us finding an order parameter for which $\bra{\Om}\de_{K^i}{\cal O}(t,x)\ket{\Om}\neq 0$, where $K^i$ is a boost in the $i$th direction. Let us choose ${\cal O}=T^{0j}$, for which we have  
\es{SumRule}{
\bra{\Om}\de_{K^i}T^{0j}(t,x)\ket{\Om}&=i\bra{\Om}[K^i,T^{0j}(t,x)]\ket{\Om}\\
&=\bra{\Om}T^{00}(t,x)\de^{ij}+T^{ij}(t,x)\ket{\Om}\\
&=\le(\ep+P\ri)\de^{ij}\,.
}
$\epsilon$ is the energy density and $P$ is the pressure.

We can rewrite it as a sum rule obeyed by the following correlator
\es{SumRuleB}{
\le(\ep+P\ri)\de^{ij}&=i\bra{\Om}[K^i,T^{0j}(0)]\ket{\Om}\\
&=-i\int d^{d-1}x \ x^i\, \expval{[T^{00}(t,x),T^{0j}(0)]}\,,
}
which we conveniently rewrite in momentum space as (see appendix \ref{app:GreensFunctions} for our notation and definition of the Green's functions)\footnote{The Fourier transform is defined through $$f(\om,k)=\int d^dx \ e^{i(\om t-\vec{k}\cdot\vec{x})}f(t,x)\,.$$}
\es{SumRule2}{
2\pi \de(\om)\le(\ep+P\ri)\de^{j}_i&=\lim_{k\to 0}\,{\p\ov \p k^i} G^{(comm)}_{T^{00},T^{0j}}(\om,k)\,,
}
where $G^{(comm)}_{T^{00},T^{0j}}(\om,k)$ is the commutator Green's function, or by using \eqref{Grels},
\es{SumRule3}{
 \de(\om)\le(\ep+P\ri)\de^{j}_i&=\lim_{k\to 0}\,{\p\ov \p k^i} \rho_{T^{00},T^{0j}}(\om,k)\,.
}
($ \rho_{T^{00},T^{0j}}$ stands for the  spectral density.)
Our task is then to find states whose contribution to the spectral density saturates the above sum rule. As usual, the Nambu-Goldstone theorem is a constraint on the low-frequency behavior of Green's functions.

\subsection{The Example of Hydrodynamics}

In the standard treatment of hydrodynamics, we obtain the retarded correlators $G^{(R)}_{AB}(\om)$, see e.g.~\cite{Kovtun:2012rj}. We want to extract the spectral density from this data. Using time reflection symmetry to deduce $G^{(R)}_{T^{00},T^{0j}}(\om,k)=G^{(R)}_{T^{0j},T^{00}}(\om,k)$, using \eqref{Grels} we end up with the formula
\es{rhoRet}{
\rho_{T^{00},T^{0j}}(\om,k)={1\ov \pi}{\rm Im}\le[G^{(R)}_{T^{00},T^{0j}}(\om,k)\ri]\,.
}
We then calculate
\es{GRhydro}{
G^{(R)}_{T^{00},T^{0j}}(\om,k)&=\le(\ep+P\ri)\, {\om k^j\ov \om^2-(c_s k)^2+i \ga_s \om k^2}\\
\rho_{T^{00},T^{0j}}(\om,k)&={\le(\ep+P\ri)\ov \pi}\, {\ga_s\om^2 k^2 k^j\ov (\om^2-(c_s k)^2)^2+(\ga_s \om k^2)^2}\,.
}
If we plot this function for fixed $k$, we see two peaks at $\om=\pm c_s k$. As we decrease $k$ they get narrower, but also get closer. To understand this better, let us examine them in a new variable $\hat\om\equiv \om/(c_s k)$, in terms of which, we have 
\es{rhodetails}{
\rho_{T^{00},T^{0j}}(\om,k)&={\le(\ep+P\ri)\ov \pi c_s}\, {\tilde k^j\,  \hat\om^2   \ov  (\hat\om^2-1)^2+\tilde k^2\,  \hat\om^2 }\,,\\
\tilde k&\equiv {\ga_s   k\ov c_s}\,.
}
Now it is a standard result that for small $\tilde k$
\es{rhodetails2}{
{1\ov \pi}\,{\tilde k^j\,  \hat\om^2   \ov  (\hat\om^2-1)^2+\tilde k^2\,  \hat\om^2 }&\approx {\tilde k^j\hat\om^2\ov 2 \tilde k} \le[\de(\hat\om-1)+\de(\hat\om+1)\ri]\\
&={\tilde k^j\ov 2 \tilde k} \le[\de(\hat\om-1)+\de(\hat\om+1)\ri]\\
&={c_s k^j \ov 2} \le[\de(\om-c_s k)+\de(\om+c_s k)\ri]\\
&\approx {c_s k^j }\, \de(\om)\,.
}
Putting the factors together, we learn that for small $k$
\es{GRhydro3}{
\rho_{T^{00},T^{0j}}(\om,k)&\approx {\le(\ep+P\ri)k^j} \de(\om)\,.
}
Plugging this into the RHS of \eqref{SumRule3}, we verify that the equation indeed holds.

One wonders what states gave us the saturation of the sum rule. Hydrodynamics differs from the other EFTs we study in Part~\ref{part2} in that the sound mode is a collective excitation, and there is no preferred set of states $\ket{\Om'}$ that gives rise to the above spectral density. In harmony with ETH, all states $\ket{\Om'}$ close in energy to $\ket{\Om}$ contribute (see~\cite{Delacretaz:2020nit} for a detailed analysis). Yet, one can say that it is the  ballistic pole in hydrodynamics which is responsible for the spontaneously broken boost symmetry.

%

\subsection{The Energy Density Spectral Function}\label{sec:energySumrule}

One can argue from conservation that $\rho_{T^{00},T^{00}}(\om,k)$ should go to $- {\le(\ep+P\ri)}\, k^2 \de'(\om)$ at very small momentum. One argument goes as follows: we have determined in \eqref{SumRule3} that for $k\ll \om$:
\es{Smallk}{
\rho_{T^{00},T^{0j}}(\om,k)\approx \le(\ep+P\ri)k^j\de(\om)\,, 
}
hence the Green's function according to \eqref{GABSpect}
\es{SmallkGreen}{
G_{T^{00},T^{0j}}(\om,k)&=\int_{-\infty}^\infty d\om'\ {\rho_{T^{00},T^{0j}}(\om',k)\ov \om -\om'}\\
&\approx \le(\ep+P\ri){k^j\ov \om }
}
Conservation of the stress tensor gives  
\es{cons}{
G_{T^{00},T^{00}}(\om,k)=-{k_j\ov \om}\,G_{T^{00},T^{0j}}(\om,k)\approx \le(\ep+P\ri){k^2\ov \om^2 }\,,
}
which is reproduced by 
\es{t00t00}{\rho_{T^{00},T^{00}}(\om,k)\approx - {\le(\ep+P\ri)}\, k^2 \de'(\om)\,.}
Let us see how this comes about in hydrodynamics.
\es{GRhydro2}{
G^{(R)}_{T^{00},T^{00}}(\om,k)&=\le(\ep+P\ri)\, {k^2\ov \om^2-(c_s k)^2+i \ga_s \om k^2}\\
\rho_{T^{00},T^{00}}(\om,k)&={\le(\ep+P\ri)\ov \pi}\, {\ga_s\om k^4\ov (\om^2-(c_s k)^2)^2+(\ga_s \om k^2)^2}\\
&\approx {\le(\ep+P\ri)\ov  c_s^2}\, {c_s k \ov 2} \le[\de(\om-c_s k)-\de(\om+c_s k)\ri]\\
&\approx - {\le(\ep+P\ri)}\, k^2 \de'(\om)\,.
}
Note that $- \de'(\om)$ is positive for $\om>0$, as required for the spectral density.

\subsection{Additional Sum Rules in CFTs}

Let us ask what sum rules follow from the breaking of boosts $K^i$, dilatation $D$, and special conformal transformations $S_\mu$ in a homogeneous and isotropic finite energy density CFT state $\ket{\Om}$, which we imagine as the macroscopic limit of some CFT state on $S^{d-1}$ that corresponds to a heavy scalar operator. 
The idea is to find an operator for which $\bra{\Om}\de{\cal O}(t,x)\ket{\Om}\neq 0$, where $\de{\cal O}$ is the transformation of the operator under the generator $K^i,\, D,\, S^i$.

Let us first consider $K^i$. We have already worked out the case ${\cal O}=T^{0j}$, and obtained the sum rule \eqref{SumRule3}.
Similarly for ${\cal O}=J^{j}$, by using $\de_{K^i}J^{j}=J^0 \de^{ij}$, we get
\es{SumRuleJ}{
 \de(\om)\rho\,\de^{j}_i&=\lim_{k\to 0}\,{\p\ov \p k^i} \rho_{T^{00},J^{j}}(\om,k)\,,
}
where $\rho$ is the charge density. 

Next we ask about $D$, for which we have $\de_{D}{\cal O}=\De_{\cal O} {\cal O}$. We then write 
\es{SumRuleD}{
\De_{\cal O}\bra{\Om} \sO\ket{ \Om} &=i\expval{[D,{\cal O}(0)]}\\
&=i\int d^{d-1}x \ x^\mu\, \expval{[T_{\mu 0}(t,x),{\cal O}(0)]}\\
&=-i\int d^{d-1}x \ x^i\, \expval{[T^{i 0}(t,x),{\cal O}(0)]}\,.
}
This is then almost identical to \eqref{SumRuleB}, and we get 
\es{SumRuleD2}{
&\text{for ${\cal O}=T^{00}$:} \qquad \de(\om)d\ep=\lim_{k\to 0}\,{\p\ov \p k^i} \rho_{T^{i0},T^{00}}(\om,k)\\
&\text{for ${\cal O}=J^{0}$:} \qquad \de(\om)(d-1)\rho=\lim_{k\to 0}\,{\p\ov \p k^i} \rho_{T^{i0},J^0}(\om,k)\,.
}
The right hand side of the first equation here is known from \eqref{SumRule3} to be equal to\\ $ (d-1)(\ep+P)\de(\om)$, so we only learn that $\ep=(d-1)P$, which is true in CFT. The second equation is a new sum rule.  We will check it for the superfluid in section~\ref{sec:superfluid}.

Finally we discuss $S_\mu$. Since for a primary ${\cal O}$, $[S_\mu,{\cal O}]=0$ we have to work with descendants. Let us consider the variation 
\es{Svar}{
\de_{S_\mu}\le(i[P_\nu,{\cal O}]\ri)&=-[S_\mu,[P_\nu,{\cal O}]]\\
&=[[P_\nu,S_\mu],{\cal O}]\\
&=-2i[\eta_{\mu\nu} D-M_{\mu\nu},{\cal O}]\\
&=-2\eta_{\mu\nu}\de_D {\cal O}+2\de_{M_{\mu\nu}}{\cal O}\,,
}
where in the second line we used the Jacobi identity and $[S^\mu,{\cal O}]=0$, while in the third the conformal algebra. Since the last line is a linear combination of terms we have already considered, we conclude that we do not learn any new constraint from the sum rule associated to the breaking of $S_\mu$.

In appendix~\ref{app:contact} we derive identities for the contact terms in certain products of the energy-momentum tensor. The constraints above follow from these contact terms.

\section{Conformal Field Theory }\label{sec:CFT}

\subsection{General Considerations} 

The purpose of this section is to explore the implication of sum rules for the spectrum of CFTs. The sum rules are obeyed by a state with finite energy and/or charge density. The strategy is to start from a state of a CFT on $ S^{d-1}\times \R $. The state on the cylinder $ S^{d-1}\times \R $  corresponds to an operator of the CFT in the plane with scaling dimension $\Delta_\Omega$. The energy $E$ of the state is given by 
$$E={\Delta_\Omega\ov R}\,,\quad \epsilon=\frac{\Delta_\Omega}{s_{d-1}R^{d}}\,,$$ 
where $\epsilon$ is the energy density and $s_{d-1}$ is the volume of unit sphere $S^{d-1}$. Now we take the \textit{macroscopic limit} \cite{Lashkari:2016vgj,Jafferis:2017zna} by considering a family of operators with $\Delta_\Omega \to\infty$ and take $R\to\infty$ while keeping $\epsilon$ fixed.\footnote{The scale invariance of CFTs imply that $R$ is just a convenient auxiliary parameter. We could set $R=1$ and discuss the macroscopic limit equally well: it would involve zooming in onto a small patch of the cylinder in correlation functions.} Likewise the theory may have a conserved $U(1)$ symmetry and we can construct states with fixed charge density 
$\rho=Q/(s_{d-1}R^{d-1})$. 

%


In particular, we consider correlators of light operators of the form $\langle \Omega| \sO\sO|\Omega\rangle$ on the cylinder and take the required limit.  Since we are considering a family of operators in the macroscopic limit, we are actually looking at a family of correlators, as we take $R\to\infty$. One of the underlying assumptions is that the correlators lead to a nice function of $\Delta_\Omega$ and/or $Q$ so that it makes sense to take the limit. In this limit the positions of the light operators $ \sO$ remain fixed on the cylinder. 

The question that we are interested in is what states are responsible for saturating the sum rules discussed in section~\ref{sec:NGthm}. The sum rules are obtained as limits of two point functions on the cylinder, which we can rewrite by inserting   a complete set of states 
\es{CompSet}{
\bra{\Omega}\sO\sO\ket{\Omega}=\sum_{\#}\langle \Omega| \sO | \# \rangle \langle  \#|\sO|\Omega\rangle\,.
}
In the limits prescribed by the sum rule, only certain states $|\#\rangle$ give contributions. For example as we discuss in Part~\ref{part2}, in the infinite volume limit, for the superfluid the $|\#\rangle$'s contributing are one particle states, while for free fermions they are particle-hole states~\cite{Alberte:2020eil}. If we trace them back to the states of CFT on the cylinder at finite $R$ and then via radial quantization to the operators on the plane, they correspond to certain operators appearing in the $s$-channel conformal block decomposition of the four point correlator $\langle \Omega(0) \sO (z) \sO(1) \Omega(\infty) \rangle$ (by $s$-channel we always mean the channel where we consider the OPE of $\Omega(0) \sO (z)$). Every $s$-channel conformal block sums up the contribution of a primary and its descendants in \eqref{CompSet}. 

In our arguments below, we assume that the the states $\ket{\Om}$ are ground states in some sector of the theory. This translates to the fact that in the $S$ channel expansion of $\langle \Omega| \sO\sO|\Omega\rangle$, we only have operators with scaling dimension larger than $\Omega$.  In cases where it is possible to construct EFT to describe the correlator in these heavy states, the heavy state effectively acts like vacuum for the EFT modes. We note that in the intuitive discussion of the Introduction, we did not need to make this assumption, and we expect that our conclusions remain true for arbitrary heavy scalar operators; it would be interesting to generalize our arguments to such states. 


 At the end of the day, we will be interested in spinning conformal blocks where $\sO$ carries spin index. Nonetheless in the macroscopic limit, the difference between scalar blocks and spinning blocks is inconsequential. So, let us illustrate the basic concepts of conformal blocks using scalar operators $\sO$
\begin{equation}\label{eqs:4point}
\langle \Omega(0) \sO (z) \sO(1) \Omega(\infty) \rangle= (z\bar z)^{-\frac{1}{2} (\Delta_\Omega+\Delta_{\sO})} \sum |C_{\Omega\sO\Delta}|^2\mathcal{G}_{\Delta_\Omega+\Delta,\ell} (z,\bar z)\,,
\end{equation}
where $\mathcal{G}_{\Delta_\Omega+\Delta,\ell} (z,\bar z)$ is the conformal block. Here we have parametrized the scaling dimension of the operators appearing in the intermediate channel in a way such that $\Delta$ denotes the excitation over the state $\Omega$ i.e $C_{\Omega\sO\Delta}$ is the OPE coefficient involving the operator $\Omega$, $\sO$ and the operator appearing in the $s$-channel with scaling dimension $\Delta_\Omega+\Delta$. This is a convenient choice as one of the main results of our paper involves putting a bound on this gap compared to $\Delta_\Omega$, i.e putting a bound on $\Delta$. We will come back to the discussion of gap in due time. 

Starting from \eqref{eqs:4point}, one can transform to the cylinder and write down the correlator of $\sO$ in the heavy state $|\Omega\rangle$.
\begin{equation}\label{eqs:cyl}
\langle \Omega| \sO(\tau,\vec{n}_1) \sO(0,\vec{n}_2) | \Omega\rangle= \sum |C_{\Omega\sO\Delta}|^2g_{\Delta,\ell} (z,\bar z)\,,
\end{equation}
where we 
defined the $\sO$ operator on the cylinder by conformally transforming it from the plane. Furthermore, we have defined $g_{\Delta,\ell}$ as
\es{gdef}{
g_{\Delta,\ell} (z,\bar z)\equiv (z\bar z)^{- \Delta_{\Omega}/2} \mathcal{G}_{\Delta_\Omega+\Delta,\ell} (z,\bar z)\,.
} 
Note, the LHS of \eqref{eqs:cyl} is defined on the cylinder and thus the cross ratio $z,\bar z$ on the RHS should be understood as a function of cylinder coordinates. In particular, the conformally transformed operators $\sO$ are inserted at $(\tau,\vec{n}_1)$ and $(0,\vec{n}_2)$ with $\vec{n}_1\cdot\vec{n}_2=\cos\theta$. The cross ratio $z,\bar z$ is related to $\tau$ and $\theta$ in following way
\begin{equation}
\sqrt{z\bar z}=e^{\tau/R}\,,\quad \ \frac{z+\bar z}{2\sqrt{z\bar z}}=\cos\theta\,.
\end{equation}
The limit $R\to\infty$ is taken in a way, so that $\theta R\equiv x$ and $\tau$ are kept fixed and identified with the coordinate in the macroscopic limit. The macroscopic limit reads in terms of $u=\tau+ i x$ and $\bar u=\tau-ix$
\begin{equation}\label{Mlim}
z=1+\frac{u}{R}\,,\quad  \bar{z}=1+\frac{\bar u}{R}\,, \quad R\to\infty\,, \quad u,\bar u\ \text{fixed}\,.
\end{equation} 

In what follows, we will be establishing that in the macroscopic limit, the descendants are suppressed i.e the conformal blocks take a very simple form. This will be followed by the discussion on the implication of this suppression for the spectrum of primaries appearing in the $s$-channel. Along the way, we will explore how and which of these primaries survive the macroscopic limit and eventually saturate the sum rule.

\subsection{Supression of Descendants}
In this subsection we show that the primaries dominate the $s$-channel expansion in the macroscopic limit. We will first consider the heavy operator limit which amounts to $\Delta_\Omega\to\infty$. (This limit is not the macroscopic limit since we do not yet scale the coordinates as required in the macroscopic limit. The suppression of descendants in this limit was studied in~\cite{Jafferis:2017zna}.) Afterwards, we will keep the energy/charge density fixed and take $R\to\infty$ limit, i.e. the macroscopic limit.

Let us start with the blocks showing up in $\expval{\Omega \sO \sO \Omega}$, where $\Omega$ is the heavy primary with $\Delta_\Omega\to\infty$ and $\sO$ has $O(1)$ scaling dimension. We write the block as an expansion in Gegenbauer polynomials $C_\ell\equiv C_\ell^{(d/2-1)}\le(z+\bar{z}\ov 2 \sqrt{z\bar{z}}\ri)$:
\es{blocks}{
g_{\De,\ell}(z,\bar{z})&=\sum_{m,n} r_{m,n} \,(z\bar{z})^{(\Delta+m+n)/2} C_{\ell+m-n}\,,\\
r_{00}&=1\,,\\
r_{10}&={1\ov 2(2\ell+d-2)}\,{(\ell+1)(\De+\De_\sO+\ell)^2\ov \De_\Omega+\De+\ell}\,,\\
r_{01}&={1\ov 2(2\ell+d-2)}\,{(\ell+d-3)(\De+\De_\sO-\ell-d+2)^2\ov \De_\Omega+\De-\ell-d+2}\,\\
\vdots
}
(The sum is restricted to $\ell+m-n\geq 0$.) From the recursion relations of \cite{Dolan:2003hv}, one can show that 
\es{desccoeff}{
r_{m,n}&={\widetilde r_{m,n}\ov \Delta_\Omega^{m+n}}+\dots\,,\\
\widetilde r_{m,n}&\approx m! \,n! \,, \quad\text{for $1\ll m,n\ll \Delta_\Omega$,}
}
where there are some powers of $m,\,n$ that we suppressed. We conclude that the contribution of the  $k=m+n$ level descendants is suppressed by $1/\Delta_\Omega^k$. This suggests that we should be able to approximate the conformal block by the first term ($m=n=0$) in~\eqref{blocks}. Since we will be eventually interested in setting $z,\bar z$ close to $1$ as in~\eqref{Mlim} we need to make sure that the descendants corresponding to  $m,n \gtrsim \Delta_\Omega$ continue to be suppressed. It can be shown again from the recursion relations that the contribution from $m,n \gg \Delta_\Omega$ is suppressed compared to that of the primary. Instead of going though a meticulous analysis of this kind, we use the knowledge of the $4d$ blocks to demonstrate the absence of anomalously large resummation effects.

 The $4d$ conformal blocks are known in closed form~\cite{Dolan:2000ut}:
\es{4dblock}{
\mathcal{G}_{\de,\ell}(z,\bar z)&=(-1)^{\ell}\frac{z\bar z}{z-\bar z}\left[k_{\de+\ell}(z)k_{\de-\ell-2}(\bar z)-k_{\de-\ell-2}(z)k_{\de+\ell}(\bar z)\right]\\
k_{\beta}(z)&=z^{\beta/2} {}_2F_{1}\left(\frac{\beta-\Delta_{12}^{-}}{2},\frac{\beta-\Delta_{12}^{-}}{2},\beta;z \right)
}
As a consistency check, we recover \eqref{blocks}. Let us now investigate the  macroscopic limit.
Let us first set  $\Delta=O(1),\,\ell=O(1)$, and use $z= 1+{u\over R}$. We obtain using the definition \eqref{gdef}
\begin{equation}\label{nearvacBlock}
g_{\Delta,\ell} (z,\bar z)= (z\bar z)^{-\Delta_{\Omega}/2}\mathcal{G}_{\Delta_\Omega+\De,\ell}(z,\bar z)=(\ell+1)+{(\ell+1)(2\De-\De_\sO)(u+\bar{u})\over 2 R}+\dots\,.
\end{equation}

In order for our analysis to apply for the macroscopic limit we need to simultaneously take $z= 1+{u\over R}$ along with keeping the energy density fixed, in other words, we must study 
the double scaling limit of the conformal blocks:
\begin{equation}\label{scaling}
z\equiv 1+{u\over R}\,,\quad \De={\cal E} R\,,\quad \ell\equiv p\, R \,,\quad \Delta_\Omega\equiv  \ep\, R^d~.
\end{equation}
Then $w$ becomes a coordinate in the macroscopic limit, ${\cal E}$ is a fixed $O(1)$ coefficient (whose interpretation is the energy of the intermediate state above the energy of $|\Omega\rangle$), $p$ is the modulus of the momentum of the intermediate state and  $\ep$ is the energy density in $|\Omega\rangle$. The reason that we scale $\Delta$ and $\ell$ as above is not immediately obvious but it will soon become clear that this leads to an interesting macroscopic limit. 
We get
\es{DoubleScaledBlock}{
g_{\Delta,\ell} (z,\bar z)=R\,\exp\le({{\cal E}\ov 2}(u+\bar{u})\ri)\,{2\sinh\left({ p\ov 2} (u- \bar{u})\right)\over u-\bar{u}}+\dots\,.
}

Noting that in 4d,
\es{Gegenbauer}{
C_\ell^{(1)}\le(\cos(x)\ri)={\sin\le((\ell+1)x\ri)\ov \sin(x)}
}
and that
\es{Ratio}{
{z+\bar{z}\ov 2 \sqrt{z\bar{z}}}=\cos\le({u-\bar{u}\ov 2iR}\ri)+\dots\,,
}
we realize that in the two limits discussed above
\es{GegenLimit}{
C_\ell^{(1)}\le({z+\bar{z}\ov 2 \sqrt{z\bar{z}}}\ri)=\begin{cases}
(\ell+1)+\dots\qquad\quad &\text{for $\ell=O(1)$\,,}\\
R\,{2\sinh\left({ p\ov 2} (u- \bar{u})\right)\over u-\bar{u}}\qquad\quad &\text{for $\ell=O(R)$\,.}
\end{cases}
} 
These are exactly the leading pieces of \eqref{nearvacBlock} and \eqref{DoubleScaledBlock}. Hence we find that the conclusion that the primaries dominate the blocks is indeed correct, i.e. it is not spoiled by  resummation effect.
 These conclusions remain true for spinning blocks, which are obtained by differential operators acting on scalar blocks \cite{Costa:2011mg,Costa:2011dw}. 

Before we proceed it is worth noting that from~\eqref{GegenLimit} we readily see that the interesting intermediate blocks have $\ell=O(R)$ and $\Delta=O(R)$, which makes sense, since these correspond to intermediate states with finite momentum and frequency in infinite volume. 

Thus far we have found that the contributions in the macroscopic limit are solely due to the primary operators in the intermediate channel and hence the macroscopic conformal blocks are exceedingly simple~\eqref{DoubleScaledBlock}. The factor $\exp\le({{\cal E}\ov 2}(u+\bar{u})\ri)$ is simply the time dependence that follows from translating the operators in energy eigenstates. 
The spatial dependence $\sim{\sinh\left({ p\ov 2} (u - \bar{u})\right)\over u -\bar{u}}$ contains the absolute value of the momentum $p=|\vec p|$. It reflects the sum over all momentum eigenstate with fixed $|\vec p|$. Using the fact that $\sO$ is a scalar operator, the matrix elements have a simple dependence on $\vec p$ and the Gegenbauer polynomial arises through an integral of the form $C_{\ell=pR}^{(d/2-1)}(\cos(x/R))\sim \int d^{d-1} k \ \delta(|\vec k|-p)\,e^{i \vec k\cdot \vec x}$ (for details see \eqref{GegenLimiti}).

\subsection{An alternative argument}

 For completeness let us present an intuitive argument for the suppression $\Delta_\Omega^{-k}$ found in \eqref{desccoeff} that also applies to spinning blocks.
 One can argue for this suppression by computing matrix elements on the cylinder directly in the $\Delta_\Omega\to\infty$ limit. This computation can be done in a straightforward way even for  spinning operators. For example, let us derive that the first descendant is indeed suppressed. To proceed, we recall that 
\begin{equation}\label{eqs:three}
\begin{aligned}
\langle \Omega'(x) T_{\alpha\beta} (z) \Omega(y) \rangle= \frac{(y-z)^{\Delta-d}}{(y-x)^{2\Delta_\Omega+\Delta-d}}  f(x,z)\,,
\end{aligned}
\end{equation}
where $f(x,z)$ is  the OPE coefficient $C_{\Omega T_{00}\Omega'}$ times a kinematical function that carries the spin indices and is independent of the operator dimensions.  $\Omega'$ is the primary with scaling dimension $\Delta_\Omega+\Delta$ appearing in the internal channel. We assume $\Delta$ scales slower than $\Delta_\Omega$, hence $\Omega'$ and $\Omega$ has the same scaling dimension to leading order in $\Delta_\Omega$. From \eqref{eqs:three}  it follows that 
\begin{equation}\label{eqs:threeder}
\begin{aligned}
\langle \partial_\mu\Omega'(x) T_{\alpha\beta} (z) \Omega(y) \rangle= \frac{2\Delta_\Omega (x-y)_\mu (y-z)^{\Delta-d}}{(y-x)^{2\Delta_\Omega+\Delta-d+2}} f(x,z) + \frac{(y-z)^{\Delta-d}}{(y-x)^{2\Delta_\Omega+\Delta-d}} \partial_\mu f(x,z) \,.
\end{aligned}
\end{equation}

Now we map \eqref{eqs:three} and \eqref{eqs:threeder} from the plane onto the cylinder ($T_{00}(z(\tau, \theta))$ should be understood as the operator on the cylinder and $T_{rr} (z)$ should be understood as an operator on the plane) via
\begin{equation}
\begin{aligned}
\langle \Omega' | T_{00}(z(\tau, \theta)) | \Omega\rangle &= \left(\frac{R}{r}\right)^{-d} \lim_{y\to\infty} y^{2\Delta_\Omega}\langle \Omega'(0) T_{rr} (z) \Omega(y) \rangle\sim \left(\frac{R}{r}\right)^{-d} f(0,z)\,,\\
\langle \partial_\mu\Omega' | T_{00}(z(\tau, \theta)) | \Omega\rangle &= \left(\frac{R}{r}\right)^{-d} \lim_{y\to\infty} y^{2\Delta_\Omega} \langle \partial_\mu\Omega'(0) T_{rr} (z) \Omega(y) \rangle\sim \left(\frac{R}{r}\right)^{-d} \partial_\mu f(0,z)\,,
\end{aligned}
\end{equation}
and it follows from the above 
\begin{equation}
\begin{aligned}
& \frac{\langle \partial_\mu \Omega' | T_{00} | \Omega\rangle}{\langle\Omega' | T_{00} | \Omega\rangle}  \simeq O(1)\,,
\end{aligned}
\end{equation}
as the ratio of function $f(x,z)$ and its derivative with respect to $x$ at $x=0$ is order one. Note that we are not making any assumption on the OPE coefficient $C_{\Omega T_{00}\Omega'}$, since it cancels out in the ratio. 
Now the contribution to the correlator coming from an s-channel conformal block corresponding to a primary $\Omega'$ is given by 
\begin{equation}
\langle \Omega| T^{00} T^{00} |\Omega\rangle \ni \sum_{\alpha,\beta=\Omega',P\Omega',PP\Omega'\cdots} \langle \Omega |T^{00}|\alpha\rangle \mathcal{N}_{\alpha\beta}^{-1} \langle\beta|T^{00}|\Omega\rangle\,,
\end{equation}
where $N_{\alpha\beta}=\langle\alpha | \beta\rangle$. We follow the notation of \cite{Simmons-Duffin:2016gjk}  and by $P\Omega'$ we mean the operator $\partial  \Omega'$.  In particular, for the first descendant $P\Omega'$, we have 
\begin{equation}
N_{P_\mu \Omega', P_\nu \Omega'}=\langle \Omega' | K_\mu P_\nu | \Omega' \rangle \simeq (\Delta_\Omega+\De)\delta_{\mu\nu} \,,
\end{equation}
where we  used the commutation relation of $K$ and $P$. Altogether, we find that the contribution coming from the first descendant is suppressed,  i.e. 
\begin{equation}
\langle \Omega |T^{00} |\partial \Omega'\rangle \,\mathcal{N}_{P \Omega', P \Omega'}^{-1} \langle \partial \Omega'|T^{00}|\Omega \rangle \simeq\langle \partial_\mu \Omega' | T^{00}| \Omega\rangle|^2 \Delta_\Omega^{-1} \simeq  \langle\Omega' | T_{00} | \Omega\rangle\Delta_\Omega^{-1} \,.
\end{equation}
A similar argument implies that the level $k$ descendant is suppresed by $\Delta_\Omega^{-k}$. The factors $m!\, n!$ in the conformal block in \eqref{desccoeff} can be accounted for from the number of distinct states at level $k=m+n$ with spin $\ell+m-n$.

\subsection{Implication for the Gap}
\label{subsec:gap}

In this subsection, we combine the Nambu-Goldstone boost sum rules, the macroscopic limit of CFTs and the form of conformal blocks in this limit to derive constraints on the gap in the CFT operator spectrum. We will be studying the correlator
\es{FocusCorr}{
\langle \Omega| T_{00}T_{00}|\Omega\rangle\,,
}
use the $s$-channel decomposition, and aim to put a bound on the gap $\Delta=\Delta_{\Omega'}-\Delta_\Omega$, where $\ket{\Omega'}$ is an excited state (above the ``vacuum'' state $\ket{\Omega}$) exchanged in the correlator. The sum rule corresponding to the correlator in \eqref{FocusCorr} was given in \eqref{t00t00} and in CFTs takes the form
\begin{equation}\label{eqs:sumrulet00}
{\rho_{T^{00}T^{00}} (\om,p)\ov p^2}= -\frac{d\epsilon}{d-1}\delta'(\omega)+O(p)\,.
\end{equation}
We chose to focus on a spectral density of two identical operators, since every contribution to such a spectral density is positive definite for $\om>0$. The obvious next step is to express the LHS of \eqref{eqs:sumrulet00} with CFT data from the cylinder.

To this end, we write the Euclidean correlator (for $\tau>0$) in the large charge limit at finite $R$ using the above results for conformal blocks:\footnote{We note that by mapping from the  $\abs{z}<1$ region of plane to the cylinder, we naturally get $\tau<0$ and the radial (or cylinder time) ordering $T_{00}(0,\vec{n}_2)T_{00}(\tau,\vec{n}_1)$. Here we instead wrote the answer for $\tau>0$.}
\es{eqs:T00T00}{
\langle \Omega| T_{00}(\tau,\vec{n}_1) T_{00}(0,\vec{n}_2)  |\Omega \rangle& =\epsilon^2\left[1+ \sum_\ell  {\#\, \ell!\ov \De_\Omega^\ell} \, e^{-\ell |\tau|/R}\,C^{(d/2-1)}_{\ell}(\cos\theta)\right]\\
&\quad+\sum_{\Delta}{C_{\Omega T_{00}\De}^2\ov R^{2d}} \,\le[e^{-\Delta |\tau|/R}\, C^{(d/2-1)}_{\ell(\De)}(\cos\theta)+O\le(1/ \De_\Omega\ri)\ri]\,,
}
where in the first line we wrote the contribution from the conformal block of $\Omega$ itself (schematically including the descendants and using that its OPE coefficient $C_{\Omega T_{00}0}/R^{d}=\bra{\Omega}T_{00}\ket{\Omega}=\ep$), while in the second we included the blocks corresponding to other exchanged operators (and only wrote explicitly the contributions of the primaries of dimension $\De_\Omega+\De$ and spin $\ell(\De)$). 

While the expression~\eqref{eqs:T00T00} is correct in the macroscopic limit, for the macroscopic limit to actually exist, one must make some assumptions about the OPE coefficients $C_{\Omega T_{00}\De}^2$. (A similar logic holds in the discussion of ETH in CFT~\cite{Lashkari:2016vgj}.)
Indeed, the most important exchanged operators which will contribute in the macroscopic limit are clearly such that $\Delta\sim R$ and $\ell\sim R$. In addition, there is a factor of $R^{d-3}$ from the Gegenbauer polynomials as in~\eqref{GegenLimit} (where $d=4$). Let us introduce the density of states by per unit energy and per unit momentum
\es{rhoDef}{
\rho(\omega,p; R)&=\sum_{\sO}\de\le(\om-{\De_\sO\ov R}\ri)\,\de\le(p-{\ell_\sO\ov R}\ri)\,,
}
This is a finite volume, un-smeared object. In terms of  $\rho(\omega,p; R)$ we obtain an expression for the macroscopic limit of the the correlator~\eqref{eqs:T00T00}:
\es{eqs:T00T00M}{
\langle \Omega| T_{00}(\tau,x) T_{00}(0)  |\Omega \rangle& = \epsilon^2+\int d\omega\, dp\ \rho(\omega,p) {C_{\Omega T_{00}\De}^2\ov R^{d+3}} e^{-\omega |\tau|}\, F_p(x)\,,
}
where $F_p(x)$ is the Gegenbauer polynomial with the $R$-dependence stripped out (we will soon write a concrete expression for it in any dimension; in $d=4$ this can be read out from~\eqref{GegenLimit} and we find $F_p(x) = {\sin(px)/ x}$).  

For the macroscopic limit to exist the combination $K(\omega,p)\equiv\rho(\omega,p) \,{C_{\Omega T_{00}\De}^2/ R^{d+3}}$ must become $R$-independent in the appropriate sense.\footnote{To make the required averaging procedure precise we introduce the smearing:
\es{rhoCSmeared}{
K(\omega,p)&={1\ov 4 \delta^2}\,\sum_{\sO\in \sI(\om,p;R,\de) } {C_{\Omega T_{00}\De}^2\ov R^{d+3}}\,,\\
\sI(\om,p;R,\de)&\equiv\le\{\sO\,\Big\vert \,{\De_\sO\ov R}\in(\om-\de,\om+\de)\,,\, {\ell_\sO\ov R}\in(p-\de,p+\de)\ri\}\,,
}
where the set $\sI$ contains operators whose energy and momentum in the macroscopic limit agrees with $\om,\, p$ respectively. Here we have introduced an infinitesimal window width $\de$; $K(\omega,p)$ should be independent of $\de$ to leading order in the macroscopic limit, and hence we do not include $\de$ as its argument.} 


Finally, to give an expression for $F_p(x)$ that is valid in any dimension we use the integral representation of the Gegenbauer polynomials:
\es{GegenLimiti}{
\lim_{R\to\infty}C^{(d/2-1)}_{pR}\left(\cos (x/R)\right)&=\lim_{R\to\infty} {\Ga\le(pR+d-2\ri)\ov 2^{d-3}\Ga\le(d-2\ov 2\ri)^2\Ga(pR+1)}\\
&\quad\times \int_0^\pi  d\vartheta\ \sin^{d-3}(\vartheta)\, \le(\cos (x/R)+i\sin (x/R)\cos(\vartheta) \ri)^{pR}\\
&={(pR)^{d-3}\ov 2^{d-3}\Ga\le(d-2\ov 2\ri)^2}\int_0^\pi d\vartheta\ \sin^{d-3}(\vartheta)\, e^{i p x \cos(\vartheta)}\\
&={1\ov 2^{d-2}\pi^{(d-2)/2}\Ga\le(d-2\ov 2\ri)}\,{R^{d-3}\ov p}\int d^{d-1}k \ \de\le(\abs{\vec{k}}-p\ri) \,e^{i \vec{k}\cdot \vec{x}}\,.
}
The last relation is very intuitive: if we zoom in onto a small patch of the sphere, we get plane waves with fixed $|\vec p|$, averaged over all directions. In $d=4$ this precisely agrees with~\eqref{DoubleScaledBlock}.
Therefore, 
\begin{equation}\label{Fform}F_p(\vec x)={1\ov 2^{d-2}\pi^{(d-2)/2}\Ga\le(d-2\ov 2\ri)p}\,\int d^{d-1}k \ \de\le(\abs{\vec{k}}-p\ri) \,e^{i \vec{k}\cdot \vec{x}}\end{equation}

%

The expression~\eqref{Fform} allows us to rewrite~\eqref{eqs:T00T00M} in a simpler form, after doing the $p$ integral 
\es{eqs:T00T00Mi}{
\langle \Omega| T_{00}(\tau,x) T_{00}(0)  |\Omega \rangle& = \epsilon^2+{1\ov 2^{d-2}\pi^{(d-2)/2}\Ga\le(d-2\ov 2\ri)}\int d\omega \,d^{d-1}k\ K(\omega,k)  e^{-\omega |\tau|}  \,{e^{i \vec{k}\cdot \vec{x}}\over k}\,.
}
It is now straightforward to obtain an expression for the spectral density in the macroscopic limit
\es{KeyForm}{
\rho_{T^{00}T^{00}} (\om,p)\sim {1\over p}\left(K(\omega,p)-K(-\omega,p)\right)~.
}

We must now make contact with \eqref{eqs:sumrulet00}. We learn that, at the very least, $K(\omega,p)$ must have support at $\omega=0$ and $p=0$. This means that there must be operators with $\Delta_\text{gap}/R\to 0$ as $R\to\infty$. In more conventional CFT terms, if we denote the dimension of the ground state by $\Delta_\Omega$, we conclude that the gap $\Delta_\text{gap}$ must be smaller than $\Delta_\Omega^{1/d}$ in the sense that there must be operators with \begin{equation}\label{boundg}\Delta_\text{gap}/\Delta_\Omega^{1/d}\to 0\end{equation}
as $\Delta_\Omega\to \infty$. If we in addition recall the physical reasoning advocated in the introduction, namely, that there is a scale invariant low energy theory describing the massless excitations in the small $\omega$ regime, we would conclude that $\Delta_\text{gap}=O( 1)$. 
In addition, we can constrain the angular momentum of these operators in a similar fashion since the sum rule \eqref{eqs:sumrulet00}
implies that we need to take infinitesimal $p$. This means that the angular momentum must satisfy $$\ell /\Delta_\Omega^{1/d} \to 0$$ and by a similar effective theory reasoning it follows that it is in fact $O(1)$.

 This implies a surprisingly small primary gap of $O(1)$ (for dimension and angular momentum) around any heavy state. One can say more about the density of operators or OPE coefficients using \eqref{eqs:sumrulet00} if one makes additional assumptions about the density or OPE coefficients separately. Incidentally, since the density $\rho(\omega,p; R)$ cannot decay as $R\to\infty$ (because the number of participating operators cannot go to zero) we get a rather general bound on the OPE coefficient  ${C_{\Omega T_{00}\De}^2 }= O(R^{d+3})$. (This bound can be strengthened by requiring a more realistic density of states.)

Recalling that the sum rule \eqref{eqs:sumrulet00} is linear in $\ep$, we can refine this bound slightly. Since the macroscopic limit is a double scaling limit, we can choose $\ep\sim \Delta_\Omega/R^d$ at will, while $R,\,\Delta_\Omega\to \infty$.  We can then resolve $R^{d+3}$ as $\Delta_\Omega R^{3} \le(\Delta_\Omega/R^d\ri)^\al$, and show that $\al=0$ by contradiction: were $\al>0$, we take $\ep$ large, while for $\al<0$, we take $\ep$ small to derive a violation of the sum rule. Hence we conclude
\es{BoundOnC}{
{C_{\Omega T_{00}\De}^2 }= O(\Delta_\Omega R^{3})~.}

\part{Examples}\label{part2}

\section{Superfluid Phase}\label{sec:superfluid}
In the superfluid phase the ground state is homogenous, isotropic, has a finite charge density, and breaks $U(1)$ spontaneously. As in the previous section, we imagine this state to be the
macroscopic limit of a family of large charge states on the cylinder $S^{d-1}\times \R $.  These states correspond to scalar operators of the underlying CFT.
 In the $Q\to\infty$ limit, there is a separation of energy scales; $\rho^{-1/(d-1)}$ (with $\rho\sim {Q/R^{d-1}}$ the charge density) is a UV scale while the IR scale is given by $R$. For distances much larger than the UV scale but much less than $R$, the system is described by an effective field theory with $1/Q$ being the expansion parameter \cite{Hellerman:2015nra,Monin:2016jmo}. As we will see the UV scale is precisely related to $\epsilon^{-1/d}$ (where $\epsilon$ is the energy density) mentioned in the introduction while $R$ plays the role of $L$. In the infinite volume limit, the state with finite energy density breaks $SO(d+1,1)\times U(1)$ down to a $SO(d)$ and a linear combination of $U(1)$ and time translation. The action of the effective field theory can be constructed in the CCWZ way \cite{PhysRev.177.2239,PhysRev.177.2247} in terms of a field $\chi$ and its fluctuation $\pi$ around the symmetry breaking saddle. The field $\pi$ is identified with the Goldstone mode, corresponding to the aforementioned spontaneous breaking.

In this section, we will elaborate on the general ideas described in the previous section using the explicit example of superfluid EFT, which captures the large charge sector of an underlying CFT. The aim is to identify the primary states that saturate the sum rule for broken boost symmetries and discuss the connection to what we have found based on general arguments in the previous section.

\subsection{General Consideration}
The effective field theory is described by the Euclidean action \cite{Hellerman:2015nra,Monin:2016jmo}
\begin{equation}\label{eqs:schi}
S=-c_1\int\ d^dx\ \sqrt{g}\ \abs{\partial\chi}^d+ \cdots+ i \int\ d^dx\ \sqrt{g}\ \rho\dot{\chi}  
\end{equation}
(Here we assume that the underlying CFT preserves parity, hence the parity violating terms of~\cite{Cuomo:2021qws} are forbidden.)
The saddle point that describes a ground state with finite homogenous charge density $\rho$ is given by 
\begin{equation}
\begin{aligned}
\chi &= -i \mu \tau +\pi\\
c_1 d\mu^{d-1}&= \rho
\end{aligned} 
\end{equation}
Expanding around this saddle, we obtain the effective action for the Goldstone field $\pi$ 
\begin{equation}\label{spi}
S_\pi = \frac{d(d-1)}{2}c_1\mu^{d-2}\int d^dx \sqrt{g}\left(\dot{\pi}^2+\frac{1}{d-1}\partial_i\pi\partial^i\pi \right)+\cdots\,.
\end{equation}

For the purpose of examining the sum rules, we will eventually be interested in the correlator involving the stress-energy tensor (in particular the components $T_{00}$, $T_{0i}$) and current $J_i$. From \eqref{eqs:schi}, we obtain the stress-energy tensor (below we consider the Minkowski signature metric  and we use $\tau=i t$)\footnote{Note, in Euclidean signature, we have Euclidean stress energy tensor defined as $T_{\mu\nu}=\frac{2}{\sqrt{g}}\frac{\delta S}{\delta g^{\mu\nu}}$. The Minkowski $T$ is related via $T^{\tau\tau}=-T^{00}$ and $T^{\tau i}=-iT^{0i}$. In the text, we use the index $0$ and work with the Minkowski $T$ operator. The same applies for $J$. \label{fn:Wick} }
\begin{equation}\label{eqs:T}
\begin{aligned}
T_{00}&= \epsilon + i \frac{d\epsilon}{\mu} \frac{d\pi}{d\tau}+ \cdots\,, \qquad T_{0i}= \frac{d\epsilon}{(d-1)\mu} \partial_i\pi + \cdots\,,\qquad T_{ij}=-\eta_{ij}\frac{\epsilon}{d-1}+\cdots\,,
\end{aligned} 
\end{equation}
where we used that the (leading order in $\mu$) energy density is
\begin{equation}\label{eqs:densitymu}
\epsilon= c_1(d-1)\mu^d\,.
\end{equation} 
The two-point correlator of $\pi$ on the cylinder in the large $Q$ limit is given by
\begin{equation}\label{prop}
D(\tau,x)=\frac{\frac{\mu^2}{s_{d-1}R^{d-1}}}{d\epsilon}\left[-|\tau|+\sum_{\ell=1}\frac{2\ell+d-2}{d-2}  \frac{e^{-\omega_\ell |\tau|/R}}{2\omega_\ell/R}C^{d/2-1}_{\ell}(\cos\theta)\right]\,,
\end{equation}
where $\theta=\arccos(\vec{n}_1\cdot\vec{n}_2)$ and $\tau$ are the angle and time separation between insertion of two $\pi$ fields respectively. Here $\omega_\ell=\sqrt{\ell(\ell+d-2)/(d-1)}$~.

Another relevant quantity for our purpose is the current $J_\mu$ corresponding to $U(1)$ that acts as a shift symmetry on the field $\chi$:
\begin{equation}\label{eqs:J}
\begin{aligned}
& J_0 = \rho+i\frac{(d-1)\rho}{\mu}\frac{d\pi}{d\tau}\,,  \qquad J_i = \frac{\rho}{\mu}\partial_i\pi\,.
\end{aligned}
\end{equation}
In rest of this section, we will be heavily using \eqref{eqs:T}, \eqref{prop}, \eqref{eqs:J}.

\subsection{Sum Rules}
The three major sum rules that we have been discussing in this paper involves looking at $\langle T_{00}T_{00}\rangle$, $\langle T_{00}T_{0i}\rangle$ and $ \langle T_{00}J_k\rangle$ in the large charge state.  For now, we will focus on the two point correlator of $T_{00}$ in the large charge state and will perform the analysis in detail by working on the cylinder and then taking the macroscopic limit while tracking the set of states, that are eventually going to saturate the corresponding sum rule. We will come back to the other sum rules involving $\langle T_{00}T_{0i}\rangle$ and $\langle T_{00}J_k\rangle$ later and verify them by working directly in the macroscopic limit without performing the computation on the cylinder.

\subsection*{$\langle T_{00}T_{00}\rangle$}
The two point correlator in $T_{00}$ in the large charge state is given by for $\tau>0$
\begin{equation}\label{eqs:T00T00s}
\begin{aligned}
\langle  T_{00}(\tau) T_{00}(0) \rangle& =\epsilon^2\left[1+ \frac{d^2}{2(d-2)}\frac{1}{\Delta_\Omega} e^{-|\tau|/R}C^{d/2-1}_{1}(\cos\theta)+\cdots\right]\\
&\quad+ \frac{d\epsilon}{s_{d-1}R^{d-1}}\left[\sum_{\ell=2}\left(\frac{2\ell+d-2}{2(d-2)} \right)\left(\frac{\omega_\ell}{R}\right) e^{-\omega_\ell |\tau|/R}C^{d/2-1}_{\ell}(\cos\theta)+\cdots\right]
\end{aligned}
\end{equation}
For $\tau<0$, the right hand side provides us with $\langle T_{00}(0) T_{00}(\tau) \rangle$. We will keep this mind and for brevity use $\langle T_{00} T_{00} \rangle$. We can compare the above with \eqref{eqs:T00T00}. In the first line on the RHS, we have two terms corresponding to the contribution coming from exchange of $\Omega$ and its descendant. We have omitted the contributions coming from higher descendants. They are suppressed in $\Delta_\Omega\to \infty $ limit. In the previous section, we have argued that there is no cumulative effect coming from considering all the descendants together. In the second line, we have a contribution from a single Regge trajectory, one primary for each given integer $\ell \geq 2$ with the scaling dimension $\Delta_\Omega+\Delta$, where we have $\Delta=\sqrt{\ell(\ell+d-2)/(d-1)}$~. Here we denote $\Delta$ as $\omega_\ell$ since we have single Regge trajectory and sum over spin $\ell$ suffices. The OPE coefficients can be read off as
\begin{equation}\label{OPE}
C^{2}_{\Omega T_{00}\Delta}=\frac{d\Delta_\Omega}{s_{d-1}^2}\left(\frac{2\ell+d-2}{2(d-2)} \right)\sqrt{\ell(\ell+d-2)/(d-1)}~.
\end{equation}

In what follows we will show that the states with $\ell\sim pR$ contribute to the macroscopic limit and then in the $p\to0$ limit saturate the sum rule. To obtain the macroscopic limit we therefore only need to take the macroscopic limit of the OPE coefficients \eqref{OPE} and we obtain for the spectral density
%
\begin{equation}\label{eq:spec}
\rho_{T^{00}T^{00}} (\om,p)=\frac{d  \epsilon }{2 \sqrt{d-1}} \, p\left[\delta\left(\omega-\frac{p}{\sqrt{d-1}}\right)-\delta\left(\omega+\frac{p}{\sqrt{d-1}}\right)\right]\,.
\end{equation}

In the $p\to 0$ limit \eqref{eq:spec} yields
\begin{equation}
\rho_{T^{00}T^{00}} (\om,p) \underset{p\to 0}{\simeq} -\frac{d  \epsilon }{ d-1}\, p^2\delta'(\om)=-(\epsilon+P)\delta'(\om)\,,
\end{equation}
and the sum rule \eqref{t00t00} is satisfied. We note that $p\to 0$ limit means that $\ell=o(R)$, thus it is clear that the states with $\Delta=\ell=o(R)$ from the Regge trajectory saturate the sum rule. The OPE coefficients for these states 
\es{OPEsuperfluid}{
C^{2}_{\Omega T_{00}\Delta}=O\left(\Delta_\Omega \ell^2\right)=o\left(\Delta_\Omega R^2\right)\,.
}
Comparing with the bound from \eqref{BoundOnC} in sec.~\ref{subsec:gap}, we see that this is down by a factor of $1/R$. The reason for this is that that there are $o(R)$ states in the superfluid in this kinematical regime, whereas the bound \eqref{BoundOnC} allows only one state (as a worst case scenario). \\

\subsection*{The rest of the sum rules}
The two-point correlators of $T_{0\mu}$ and $J_\mu$ in the macroscopic limit can be found from the macroscopic limit of the two-point correlator of the $\pi$ field:
\begin{equation}\label{eqs:propmac}
\langle \pi(\tau,\vec{x}) \pi(0) \rangle_{\text{macro}}= {\mu^2\ov d(d-2)s_{d-1}\ep}\, \frac{\sqrt{d-1}}{\le(\frac{1}{d-1}\tau^2+\vec{x}\cdot\vec{x}\ri)^{d/2-1}}\,.
\end{equation}
From \eqref{eqs:T} and \eqref{eqs:J} we immediately realize that all we need to know for these computations is that for $\tau>0$
\es{eqs:propmac2}{
\p_\tau\p_i\langle \pi(\tau,\vec{x}) \pi(0) \rangle_{\text{macro}}&= {\mu^2\ov s_{d-1}\ep}\,\frac{\frac{\tau}{\sqrt{d-1}} x_i}{\le(\frac{1}{d-1}\tau^2+\vec{x}\cdot\vec{x}\ri)^{d/2+1}}\\
&\equiv h_i(x)\,.
}
In terms of this correlator:
\es{AllOthers}{
\langle T_{0i}(x) T_{00} (0)\rangle_{\text{macro}}&=-i{d^2\ep^2\ov(d-1)\mu^2}\,h_i(x)\,,\\
\langle T_{0i}(x) J_{0} (0)\rangle_{\text{macro}}&=\langle T_{00}(x) J_{i} (0)\rangle_{\text{macro}}=-i{d\ep\rho\ov \mu^2}\,h_i(x)\,.
}

The Fourier transformed function $\widetilde{h}(\omega,\vec{p})$ is defined through:
\begin{equation}
\label{eq:FourierWightman}
\widetilde{h}_i(\omega,\vec{p})=\int d^d x \ e^{i\omega \tau -i\vec{p}\cdot\vec{x}}\,h_i(x)\,.
\end{equation}
By rescaling $\tau=\sqrt{d-1}\,\tau',\, \om=\om'/\sqrt{d-1}$, we get a standard Lorentz invariant integral 
\es{eq:FourierWightman2}{
\widetilde{h}_i\le({\omega'\ov \sqrt{d-1}},\vec{p}\ri)&={\sqrt{d-1} \,\mu^2\ov s_{d-1}\ep} \int d\tau'\, d\vec{ x} \ e^{i\omega' \tau' -i\vec{p}\cdot\vec{x}}\,\frac{{\tau'} x_i}{\le({\tau'}^2+\vec{x}\cdot\vec{x}\ri)^{d/2+1}}\\
&={\sqrt{d-1} \,\pi \mu^2\ov 2d\ep}\frac{\partial}{\partial\omega'}\frac{\partial}{\partial p^i}\left[ \Theta(\omega')\Theta({\om'}^2-p^2)\le({\om'}^2-p^2\ri)\right]\,.
}
We use \eqref{commWightman} and \eqref{Grels} to write
\es{rhoFromWightman}{
\rho_{A,B}(\omega,\vec{p})=\frac{1}{2\pi}\left(\tilde{G}_{A,B}(\omega,\vec{p})-\tilde{G}^*_{A,B}(-\omega,-\vec{p})\right)\,,
}
and get the following expressions for the spectral densities from \eqref{eq:FourierWightman2}:
\es{MacroSumRuleSuperfluid}{
\rho_{T_{0i}T_{00}}(\om,p)&=\frac{1}{2}(\epsilon+P)p_i \left[\Theta(\omega)\delta\left(\omega-\frac{p}{\sqrt{d-1}}\right)+\Theta(-\omega)\delta\left(\omega+\frac{p}{\sqrt{d-1}}\right)\right]\\
&\underset{p\to 0}{\simeq}  (\epsilon+P)p_i \delta(\omega)\,,\\
\rho_{T_{00}J_{i}}(\om,p)&=\rho_{T_{0i}J_{0}}(\om,p)\underset{p\to 0}{\simeq}  \rho p_i \delta(\omega)\,.
}
Hence the sum rules \eqref{SumRule3}, \eqref{SumRuleJ}, and \eqref{SumRuleD2}  are all satisfied.

We end this section with a remark about three-dimensional parity violating superfluids~\cite{Cuomo:2021qws}. The ground state contains vortices, but it is homogenous and isotropic, hence our sum rules apply. The low lying excitations are phonons (the $\pi$ excitations that we have been studying) and other {\it softer} vortex excitations with spin $\ell$ and a gap of $\Delta\sim \ell(\ell+1)Q^{-3/2}$ above the ground state. (For parity preserving superfluids, the excitations with spin $\ell$ above the ground state have a gap $\Delta\sim \ell(\ell+1)$. Vortex excitations in parity preserving fluids only appear for $\ell > \sqrt{Q}$ \cite{Cuomo:2017vzg, Cuomo:2019ejv}.
A natural question is what saturates the sum rules in parity violating superfluids. The answer is that it is still the phonons, because  \cite{Cuomo:2021qws} found that the OPE coefficients involving the phonon modes, the ground state $\Omega$ and $T_{00}$ (or other relevant components of $T$)  stay unchanged compared to the parity preserving one. Thus the softer vortex modes should not contribute in the macroscopic limit and should not play any role in the sum rule. This is corroborated by the fact that there are only finitely many vortices on the cylinder, hence they disappear in the macroscopic limit. It would be nice to check these claims explicitly.

\section{Free Scalar}
\label{sec:FreeScalar}
In this section, we study the saturation of the sum rules associated with  broken boosts and scale invariance described in section \ref{sec:NGthm} for the case of a free relativistic complex scalar field in a finite charge density state in dimension $d > 2$. 

The free scalar field does not lead to a state with finite energy density in the macroscopic limit due to the flat directions arising from the shift symmetry of the free scalar action. Indeed, due to these flat directions, $\Delta\sim Q$ rather than $\Delta\sim Q^{d/(d-1)}$. This means that the effective field theory description of~\eqref{eqs:schi} is inappropriate, and should be replaced by the approach~\cite{Hellerman:2017veg,Hellerman:2017sur,Hellerman:2018xpi,Bourget:2018obm,Grassi:2019txd,Beccaria:2020azj}. In fact sometimes these two types of effective theories are connected~\cite{Sharon:2020mjs}. In our analysis of the boost symmetry realization on the large charge states in free field theory, we will not use an effective theory approach, rather, we will pursue a more straightforward analysis of the correlation functions. 

\subsection{General Consideration}
The Euclidean two-point correlation function in a theory of a free complex scalar field is given by (we use the normalization of \cite{Osborn:1993cr}):
\begin{equation}
\label{eq:PropScalarOsborn}
G(x-y) \equiv \langle \bar{\phi}(x)\phi(y)\rangle = \frac{1}{(d-2)s_{d-1}}\,\frac{1}{|x-y|^{2\Delta_\phi}},
\end{equation}
where $ s_{d-1} = {2\pi^{\frac{d}{2}}}/{\Gamma(\frac{d}{2})}$ and $\Delta_\phi = \frac{d-2}{2}$. We define the lightest operator of charge $Q$  by the following:
\begin{equation}
\label{eq:DefOQForBosons}
\mathcal{O}_{Q} \equiv \frac{\left((d-2)s_{d-1}\right)^{Q/2}}{\sqrt{Q!}}\bar{\phi}^Q.
\end{equation}
Its scaling dimension scales like $\Delta_Q\sim Q$. Using radial quantization, the ground state of charge $Q$ on the cylinder is:
$$\vert\Omega\rangle \equiv \mathcal{O}_{Q}(0) |0\rangle .$$ 
For the purpose of examining the sum rules, we will eventually be interested in the correlators involving the $U(1)$ current $J^\mu$ and the stress-tensor $T^{\mu\nu}$. 
On the plane, they are given by the following expressions:
\begin{align}
&T^{\mu\nu} = \partial^\mu\bar{\phi}\partial^\nu\phi +\partial^\nu\bar{\phi}\partial^\mu\phi-g^{\mu\nu}g^{\alpha\beta}\partial_\alpha\bar{\phi}\partial_\beta\phi + T^{\mu\nu}_{\text{imp}}, \qquad T^{\mu\nu}_{\text{imp}} = \chi\left[g^{\mu\nu}\partial^2-\partial^\mu\partial^\nu\right]\bar{\phi}\phi,\label{eq:BosonicStressTensor}\\
&J^\mu = i \left(\partial^\mu\bar{\phi}\phi-\bar{\phi}\partial^\mu\phi\right),
\end{align}
where  $\chi = \frac{d-2}{2(d-1)}$ is the coefficient of the curvature coupling term ${\cal R}\bar{\phi}\phi$ (with ${\cal R}$ the Ricci scalar). In the next subsection, we will argue that the sum rules associated with the broken boosts and dilatations are not affected by the improvement term, and therefore we can ignore it when calculating  correlation functions for the purpose of verifying the sum rules (see subsection \ref{subsubsec:FlatSpaceCorr} for further details).  For details about Wick rotating tensors to Minkowski signature see footnote~\ref{fn:Wick}.



\subsection{ Sum Rules}
\label{subsec:SumRules}
We would like to check how the sum rules described in section \ref{sec:NGthm} are satisfied for the case of the free bosonic field theory. For this purpose, one can  take the following strategy: first, calculate the correlators which are of interest for the saturation of the sum rules on the plane. Then, map it onto the cylinder $ S^{d-1}\times \R $ and take the radius of the $S^{d-1}$ sphere to infinity. This last step amounts to taking the infinite volume limit.
In practice, taking the macroscopic limit (as described in subsection \ref{subsubsec:MacroLimitScalar}) directly on the flat space correlators is equivalent to performing the procedure described above.

In what follows, we will be using $x^\mu,y^\mu$ as the coordinates on the plane. 
On the cylinder, we want to evaluate 
$$\langle \Omega| J^{i}(0)T^{\tau\tau}(\tau,\theta_j)  | \Omega\rangle\,,$$
where $(\tau, \theta_j)$ refer to the coordinate on the cylinder. We can choose a coordinate system where we have only one angle $\theta$. This correlator is related to a four-point correlator on the plane
$$\langle \Omega(0) T^{rr} (z,\bar z) J^{i}(1) \Omega(\infty)\rangle\,$$
by conformal transformation where we have 
\begin{equation}
r=e^{\tau/R}\,,\quad z\bar z=r^2 \,,\quad \frac{z+\bar z}{\sqrt{z\bar z}}=2\cos\theta \,.
\end{equation}
We follow the usual convention (where $x^{d}$ denotes the Euclidean time on the plane)
$$z=x^d+ i x^{1}\,, \ \bar z=x^d- i x^{1}\,.$$
In the macroscopic limit, as we let $z\to 1$, $T^{rr}$ on the plane essentially becomes $T^{dd}$. 
Therefore, we will leverage that and on the plane we will calculate 
$$\langle \Omega(0) T^{dd} (z,\bar z) J^{i}(1) \Omega(\infty)\rangle.$$ 
We deal with the right hand side of the sum rule in a similar way, going from the plane to the cylinder. 

The organization of this subsection is as follows: in subsection \ref{subsubsec:FlatSpaceCorr}, we calculate the correlation functions on the plane. In subsection \ref{subsubsec:MacroLimitScalar}, we define the macroscopic limit associated with the free bosonic theory. In subsection \ref{subsubsec:SaturationOfSumRulesScalar}, we calculate the spectral density and show that the sum rules are satisfied. 

\subsubsection{Flat Space Correlators}
\label{subsubsec:FlatSpaceCorr}
The free field correlators are obtained using Wick contractions. 
We concentrate on 
$$\langle \mathcal{O}_Q(0) T^{\mu\nu}(x)J^\rho(y)\mathcal{O}_{-Q}(\infty)\rangle\,.$$
Eventually, when taking the macroscopic limit, we will choose to work a particular configuration where $T,J$ lie in the $(x^{d}, x^1)$ plane as mentioned previously and get the macroscopic correlator and then covariantize to obtain the correlator for arbitrary insertion points of $T$ and $J$.

We first evaluate the contribution to $$\langle \mathcal{O}_Q(0) T^{\mu\nu}(x)J^\rho(y)\mathcal{O}_{-Q}(\infty)\rangle $$
from the non-improved stress energy tensor. We find that this results in: 
\begin{equation}
\label{eq:CorrelationFunction1Scalars}
\begin{aligned}
&\langle \mathcal{O}_Q(0) T^{\mu\nu}(x)J^\rho(y)\mathcal{O}_{-Q}(\infty)\rangle \\
& \ni iQ\left[ H^{\mu\rho}(x-y)F^\nu(x)+H^{\nu\rho}(x-y)F^\mu(x)-g^{\mu\nu}H^{\alpha\rho}(x-y)F_\alpha(x)\right]\,.
\end{aligned}
\end{equation}
Here the functions $F$ and $H$ are given by:
\begin{equation}
\begin{aligned}
F^\mu(x)\equiv \frac{x^\mu}{s_{d-1}|x|^{d}}\,,\quad 
H^{\mu\nu}(x) \equiv \frac{1}{s_{d-1}|x|^{d}} \left(\eta^{\mu\nu}-\frac{dx^\mu x^\nu}{x^2} \right) \,.
\end{aligned}
\end{equation}

We can further evaluate the contribution to the correlator coming from $T_{\text{imp}}$
\begin{equation}
\begin{aligned}
&\langle \mathcal{O}_Q(0) T_{\text{imp}}^{\mu\nu}(x)J^\rho(y)\mathcal{O}_{-Q}(\infty)\rangle \\
& = i\chi Q \left(g^{\mu\nu}\partial^2-\partial^\mu\partial^\nu \right) \left[F^\rho(x-y)\left(G(y)-G(x) \right)+F^\rho(y)G(x-y)+(Q-1)F^\rho(y)G(x) \right].
\end{aligned}
\end{equation}
Here $G$ is given by \eqref{eq:PropScalarOsborn}. This correlator has a finite macroscopic limit, but at the end of the day this does not contribute to the sum rules.  The reason is that the improvement terms drop from the expressions for the charges and the sum rules stem from looking at the correlators of the charge generator and an operator that plays the role of an order parameter. In what follows, we will thus work with the non-improved $T_{\mu\nu}$ (unless otherwise stated) in order to show the saturation of sum rules. 
The same conclusion holds for the case of the free scalar field even when considering the sum rule associated with the broken dilatations \eqref{SumRuleD2}.

Finally, the right hand side of the sum rule requires us to find
\begin{equation}
\label{eq:RHSOfWardIdentity}
\langle  \mathcal{O}_Q(0) J^{d}(z=1,\bar z=1) \mathcal{O}_{-Q}(\infty)\rangle=-\frac{iQ}{s_{d-1}}\,.
\end{equation}  
Once we are equipped with the expression for the free space correlator, we take the macroscopic limit, which is the subject of the next subsection. 

\subsubsection{Macroscopic Limit}
\label{subsubsec:MacroLimitScalar}
By the state/operator correspondence, the operator $\mathcal{O}_Q$ in equation \eqref{eq:DefOQForBosons} describes a state with charge density $\rho$
on the cylinder $ S^{d-1}\times \R $. This can be seen by noting that 
\begin{equation}
\langle \Omega\vert J^0\vert \Omega\rangle_{\text{cyl}}=R^{1-d}\frac{\langle \mathcal{O}_{Q}(0)J^0(1)\mathcal{O}_{-Q}(\infty)\rangle}{\langle\mathcal{O}_{Q}(0)\mathcal{O}_{-Q}(\infty)\rangle}=\rho\,.
\end{equation}
In the macroscopic limit, the charge density is kept finite as we take $R\to\infty$ limit. The scaling dimension associated with the operator $\mathcal{O}_Q$ satisfy $\Delta_Q \sim Q \sim R^{d-1}$. The energy density reads:
\begin{equation}
\epsilon = \frac{\Delta_H}{s_{d-1}R^d}\to 0.
\end{equation}
In the macroscopic limit, we take:
\begin{equation}
\label{eq:XmuYmuMacroLim}
x^\mu= y^\mu + \frac{u^\mu}{R}\,,\qquad \text{where}\ R\to\infty,\, \text{and} \ u^\mu\ \text{is fixed}.
\end{equation}
We also set $y^\mu=\delta^{\mu}_0$.  In terms of $z,\bar{z}$ we have:\footnote{In this section we use the somewhat unfortunate notation $u=\tau+i u^1$, and when we refer to a spacetime point, we always write $u^\mu$. (In previous sections we instead used $u=\tau+i x^1$, but here $x^\mu$ is already reserved for coordinates on the plane before mapping to the cylinder and taking the macro limit. }
\begin{equation}
z = 1+\frac{u}{R}, \qquad \bar{z}=1+\frac{\bar{u}}{R}.
\end{equation}
We note that the convention of taking the macroscopic limit is different compared to \cite{Jafferis:2017zna}. We made this choice in order to ensure that there is no parity transformation implemented while taking the macroscopic limit. Altogether, the macroscopic limit of the correlation functions which are of interest for the saturation of the sum rules is given by: 
\begin{equation}
\langle \Omega \vert J^i(0)T^{\tau\tau}(u^\mu)\vert\Omega\rangle_{\text{mac}} \equiv \lim_{R\to\infty} \left(R^{1-2d}\frac{\langle \Omega(0) T^{dd}(x)J^i(1)\Omega(\infty)\rangle}{\langle\Omega(0)\Omega(\infty)\rangle}\bigg|_{x^\mu=\delta^{\mu}_d+u^\mu/R} \right),
\end{equation}
and
\begin{equation}
\langle \Omega\vert J^\tau\vert \Omega\rangle_{\text{mac}} \equiv\lim_{R\to\infty} \left(R^{1-d}\frac{\langle \Omega(0)J^d(1)\Omega(\infty)\rangle}{\langle\Omega(0)\Omega(\infty)\rangle}\bigg|_{x^\mu=\delta^\mu_d+u^\mu/R} \right).
\end{equation}
The correlators associated with the numerator on the right-hand side in both equations above correspond to the correlators given by equations \eqref{eq:CorrelationFunction1Scalars} and \eqref{eq:RHSOfWardIdentity}.
Both limits are well-behaved. Setting $y^\mu=\delta^\mu_d$, the resulting expressions (in Euclidean signature) in the macroscopic limit after covariantizing are:
\begin{equation}
\begin{aligned}
\label{eq:MacroCorrFunctT00Ji}
\langle \Omega \vert J^i(0) T^{\tau\tau}(u^{\mu})\vert\Omega\rangle_{\text{mac}}  
=-i\frac{\rho d}{
s_{d-1}}\frac{\tau {u}^i}{{u}^{d+2}}
\end{aligned}
\end{equation}
and
\begin{equation}
\begin{aligned}
\label{eq:ChargeDensityExpr}
&\langle \Omega\vert J^\tau\vert \Omega\rangle_{\text{mac}}  =-i\rho\,.
\end{aligned}
\end{equation}
One can see from the last equation that the macroscopic limit has been taken in a way so that the charge density $\rho$ is constant.

\subsubsection{Saturation of the Sum Rules}
\label{subsubsec:SaturationOfSumRulesScalar}
The sum rules are inherently statements in Minkowski signature.  For the purpose of evaluating the sum rules, we evaluate the following Wightman correlation function by doing proper analytic continuation of \eqref{eq:MacroCorrFunctT00Ji} and recalling $T^{00}=-T^{\tau\tau}$:
\begin{equation}
\label{eq:WightmanDef}
G_{T^{00},J^i}(u^0,\vec{u}) \equiv \langle \Omega \vert T^{00}(u^0,\vec{u})J^i(0)\vert\Omega\rangle\,.
\end{equation}
Here we have defined $(u^0,\vec{u})$ to denote the Minkowski coordinates. The proper analytic continuation is achieved by letting $\tau=iu^0+\epsilon$ and then take $\epsilon\to 0^{+}$. 
We find that
\begin{equation}
\label{eq:GT00Ji}
G_{T^{00},J^i}(u^0,\vec{u}) = -\frac{\rho d}{
s_{d-1}}\frac{u^0{u}^i}{{u}^{d+2}}.
\end{equation}

We have already computed the spectral density from (a very close analog of) this Green's function in section~\ref{sec:superfluid}, see \eqref{eq:FourierWightman2} and \eqref{rhoFromWightman}. Plugging into those formulas we get
%
%
%
%
%
\begin{equation}
\begin{aligned}
\label{eq:ScalarSpectralDensity}
\rho_{T^{00},J^i}(\omega,\vec{p})&=  \frac{1}{2}\rho\, p^{i} \left[\Theta(\omega)\delta(\omega-p)+\Theta(-\omega)\delta(\omega+p)\right]\\
& \underset{p \to 0}{\simeq} \rho p^i \delta(\omega) ,
\end{aligned}
\end{equation}
Thus, we find that the sum rule \eqref{SumRuleJ} associated with the broken boosts symmetry is satisfied. 
As for the sum rule associated with the broken scale invariance, we note that in the macroscopic limit, without including the improvement term in the stress-tensor \eqref{eq:BosonicStressTensor}, the following happens to be true for the free scalar 
$$\langle\Omega| T^{0i} J^{0} |\Omega\rangle=\langle\Omega| T^{00} J^{i} |\Omega\rangle, $$
where $T_{\mu\nu}$ above refers to the stress-tensor \eqref{eq:BosonicStressTensor} without the improvement part $T_{\text{imp}}$. Together with the analysis shown at the beginning of subsection \ref{subsubsec:FlatSpaceCorr}, this immediately tells us that the sum rule associated with the broken scale symmetry \eqref{SumRuleD2} is satisfied. The extra factor of $(d-1)$ in the sum rule comes from the contraction of spatial indices in the expression for $x_iT^{0i}$.

\subsection{Sum Rules from the Cylinder Vantage Point}
\label{subsec:SumRulesFromTheCylinderVantage}
In this subsection, we identify the states that are responsible for saturation of the sum rules discussed in the previous subsection. To this effect, we would like to study a sum rule in analogy to the $T^{00}T^{00}$ sum rule that was studied in the superfluid case in section \ref{sec:superfluid}. In the free scalar case, however, the energy density vanishes in the macroscopic limit and the $T^{00}T^{00}$ spectral density hence vanishes in the $p\to 0$ limit. 
Instead, we study the sum rule associated with the $T^{00}J^{0}$ correlator,  which reads:
\begin{equation}
\rho_{T^{00}J^{0}}(\om,p)\underset{p\to 0}{\simeq} - \rho p^2 \delta'(\omega) \, .
\end{equation}
The above equation follows from combining the conservation of $J$ with the $T^{00}J^{i}$ sum rule just as in section~\ref{sec:energySumrule}. 
In the large $Q$ limit, we can write  for $\tau>0$
\begin{equation}\label{eqs:T00J0free}
\begin{aligned}
\langle \Omega|T_{00} (\tau) J_{0}(0) |\Omega \rangle& =\frac{Q^2}{s_{d-1}^2R^{2d-1}}\\
&\quad + \underbrace{\frac{Q}{s_{d-1}^2R^{2d-1}}\left[\sum_{\ell=1}\frac{\ell^2}{(d-2)} e^{-\ell |\tau|/R}C^{(d/2-1)}_{\ell}(\cos\theta)\right]}_{\text{terms for the sum rule}}+\cdots\,.
\end{aligned}
\end{equation}
The RHS of the above expression gives $\langle \Omega|J_{0}(0) T_{00} (\tau) |\Omega \rangle$ for $\tau<0$.
This equation should be thought of as an analogue of \eqref{eqs:T00T00s} valid in the superfluid for  the free scalar case. Here, the dots indicate terms that are not important for reproducing the sum rule 
 in the macroscopic limit, i.e to reproduce the $p\to 0$ behavior of spectral density. Now it is easy to realize that the sum rule is saturated by states living on the Regge trajectory $\omega_\ell=\ell$. The calculation proceeds exactly the same way as in the superfluid case, with the only difference being that here $\omega_\ell=\ell$ for all $\ell$. Once again, the states $\ell\sim p R$ become important in the infinite volume limit.

Now let us understand in detail how this single Regge trajectory on the cylinder comes about from the previous calculation of the four-point correlator on the plane via Wick contraction. Schematically, we have the following type of contractions in the correlator 
\begin{equation}
\langle \wick{ \c6{\phi}^{Q-2} \c3{\phi}\c2 {\phi} \vert \c2{\partial \bar\phi(x)} \c5{\partial \phi(x)} \c3 {\bar\phi(y)}\c4{\partial \phi(y)}  \vert\c4{\bar\phi} \c5{\bar\phi} \,\c6{\bar\phi}^{Q-2}}  \rangle\sim Q^2\,, \nonumber 
\end{equation}
which gives the first line of \eqref{eqs:T00J0free}, 
 and
\begin{equation}
\langle  \wick{ \c3{\phi}^{Q-1}   \c2{\phi}\vert \c2{\partial\bar\phi(x)}\c2{\partial\phi(x)} \c2{\partial \bar \phi(y)} \c2{\phi(y)} \vert \c2{\bar\phi}\,\c3{\bar \phi}^{Q-1} }\rangle\sim Q\,, \nonumber
\end{equation} 
which gives the second line of \eqref{eqs:T00J0free}.
The contractions that yield result proportional to $Q$ survive in the macroscopic limit and eventually a part of it
\es{importantPart}{
& \frac{Q}{R^{2d-3}}\partial_\tau^2 G(\tau,\theta)  
}
 is responsible for saturating the sum rules. $G(\tau,\theta)$ is the free scalar propagator on the cylinder:
 \es{CylinderProp}{
 &G(\tau,\theta)=\frac{1}{(d-2)s_{d-1}} \left(1+e^{-2|\tau|/R}+2e^{-|\tau|/R}\cos\theta\right)^{-\frac{d-2}{2}} \,.
 }
  The underbraced term in \eqref{eqs:T00J0free} comes precisely from the expansion of $\partial_\tau^2 G(\tau,\theta)$ in terms of  Gegenbauer polynomials.

As a final remark, let us understand the single Regge trajectory in terms of single or multiparticle states. Of course, it is clear that the relevant states would be labelled by a single number $p\sim \ell R$. The question is whether this corresponds to single or multiparticle states. To proceed, recall that the relevant contribution arises from contracting one leg of $T$ with the bra $\langle\Omega|$ and one leg of $J$ with ket $|\Omega\rangle$, and then contracting the leftover leg of $T$ with left over leg of $J$. We can think of breaking apart the $TJ$ Wick contraction and inserting a complete set of states. To make this notion more precise, we start by noting that the state $|\Omega\rangle $ defines a Bose-Einstein condensate on the cylinder. One can view it as created by the zero angular momentum modes ($a_0^\dagger$) of the scalar field $\phi$, i.e. $|\Omega\rangle\sim(a_0^\dagger)^Q|0\rangle$, where $|0\rangle$ is the true vacuum. 
On the cylinder, we can annihilate states with zero angular momentum and charge $1$ from this condensate, and create a particle with angular momentum $\ell$ and charge $1$ on top of it. This resembles a bit the case of a particle-hole pair in the theory of Fermi surface (studied in section \ref{sec:Fermi}), albeit one important difference: the particle-hole excitations on  top of the Fermi surface are labelled by two numbers, the angular momentum of the particle and the angular momentum of the hole, both of which can take non-zero values. In the case of the free scalar field, however, the hole carries zero angular momentum. Thus, the particle-hole pairs are labelled by a single number and form the Regge trajectory as discussed above. Note that we need a particle-hole pair as opposed to a single particle excitation on the cylinder since
$\langle  \Omega | T  a^{\dagger}_{\ell}a_0 |\Omega\rangle$ is non-zero whereas $\langle  \Omega | T  a^{\dagger}_{\ell}|\Omega\rangle =0 $ due to the charge conservation on the cylinder.  Nonetheless, in the macroscopic limit, both $|\Omega\rangle$ and $a_0|\Omega\rangle$ define the same state with equal and finite charge density. Thus, as we take $R\to\infty$ limit, the aforementioned particle-hole pair on the cylinder (that consists of a hole carrying zero angular momentum) behaves like a single particle excitation, labelled by a single momentum vector $\vec{p}$ in the infinite volume theory. This is similar to the behavior described in section 5 of \cite{Alberte:2020eil}, in the context of the free massive particle. 

\section{Free Fermions}\label{sec:Fermi}
\label{sec:FreeFermions}
We study systems of free fermionic field theories at finite charge density $\rho$ and energy density $\epsilon$. In subsection \ref{subsec:FreeFermionsLargeCharge} we address the large charge sector of these models in three and four spacetime dimensions. In subsection \ref{subsec:RelativisticFermiGas} we consider the relativistic Fermi gas in four dimensions, calculate the spectral density, and show the saturation of the sum rules as described in section \ref{sec:NGthm} for the broken symmetries.

\subsection{Free Fermions at Large Charge}
\label{subsec:FreeFermionsLargeCharge}
In this subsection we consider free fermionic field theories with a global $U(1)$ symmetry in $d=3$ and $d=4$ dimensions. We denote the lightest operator of charge $Q$ by $\mathcal{O}_Q$, its dimension being $\Delta_Q$, and focus on the large  $Q$ limit. 
In addition, we restrict the discussion to cases in which the ground state is homogeneous and isotropic. Under this assumption, and in the limit of large charge, the lightest operator of charge $Q$ can be constructed using simple counting arguments. Similar arguments can be found in \cite{Shenker:2011zf,Aharony:2015pla}, where the leading order term in the expansion of $\Delta_Q$ was calculated for the $d=3$ case.
Our main findings in this subsection are:
\begin{enumerate}
\item There is no $Q^0$ term in the expansion for $\Delta_Q$. 
\item The energy difference between the first excited state and the ground state is $O(1)$:
\begin{equation}
\Delta_{+1,Q}- \Delta_Q = O(1).
\end{equation}
\end{enumerate}
We emphasize that the analysis described in this subsection holds only under the assumption of a homogeneous and isotropic ground state. Towards the end of this subsection, we make a comment regarding cases in which this is not the situation.

We start with the $d=3$ case. The fermionic field $\psi$ is a two component complex Grassmann spinor. Under global $U(1)$ symmetry it transforms as:
\begin{equation}
\label{eq:GlobalU1TransDef}
\psi \to e^{i\alpha}\psi.
\end{equation}
As a result of the Fermi statistics, operators of the form $\psi^n$ vanish for all $n>2$.\footnote{We use the notation $\psi^2=\ep^{ab}\psi_a\psi_b$.} Therefore, in order to construct operators of charge $Q$ under the transformation \eqref{eq:GlobalU1TransDef}, one necessarily has to include fermions dressed with derivatives, in order to construct a product with more than two fermions. The resulting operator $\mathcal{O}_Q$ is therefore expected to be of the form ${\psi^2(\partial\psi)^2(\partial^2\psi)^2}_{\cdots}\,$. 

The equation of motion reduces the number of independent physical degrees of freedom. Hence, without loss of generality, we can  eliminate $\partial_2\psi$, as it is linearly dependent on the other derivatives of $\psi$.
We define by $\mathcal{D}^n$ an operator which consists of $n$ spacetime derivatives of the following form:
\begin{equation}
\mathcal{D}^n \equiv \partial_{0}^{n_0}\partial_1^{n_1}, \qquad \text{where } n_0+n_1=n.
\end{equation}
Note that $n$ therefore represents the total number of derivatives of type $\partial_0$ and $\partial_1$.
The operator $\mathcal{O}_Q$ will consist of multiplication of all the possible terms of the form $\prod_{k=0}^n (\partial_0^{n-k}\partial_1^k\psi)^2$, where $n$ takes all integer values between $0$ to a maximal value that is determined by the requirement of having a charge $Q$.
Note that for each value of $n$, there are $n+1$ different terms that contain $n$ derivatives (from either $\partial_0$ type, $\partial_1$ type, or a mixed combination). The fermions consist of two-complex components Grassmann fermions, hence each such term can be taken with a power of two at most.   
Altogether, for large $Q$ the operator $\mathcal{O}_Q$ takes the following form:
\begin{equation}
\label{eq:OperatorConstruction3d}
\mathcal{O}_Q \propto (\psi)^2(\mathcal{D}_1\psi)^4\cdots (\mathcal{D}_{n_{\text{max}}}\psi)^{2n_{\text{max}}+2},
\end{equation} 
where $n_{\text{max}}$ is determined by the condition that the total number of fermions in the operator $\mathcal{O}_Q$ is equal to $Q$: 
\begin{equation}
\label{eq:3dQuadraticEquation}
Q =  (n_{\text{max}}+1)(n_{\text{max}}+2).
\end{equation} 
It is important to note that we are not obtaining all integer values of $Q$ through this construction, since $n_{\text{max}}$ is an integer. We discuss the operators for $Q$'s that cannot be produced through \eqref{eq:3dQuadraticEquation} later in this section: they have spin and hence do not correspond to homogeneous states on the cylinder. 

The total number of derivatives $n_{\text{der}}$ that appear in the  operator \eqref{eq:OperatorConstruction3d} is given by:
\begin{equation}
\label{eq:3dNumberOfDerivatives}
n_{\text{der}} = \frac{2}{3}n_{\text{max}}(n_{\text{max}}+1)(n_{\text{max}}+2)= \frac{2}{3}Q\, n_{\text{max}}.
\end{equation} 
Solving the quadratic equation \eqref{eq:3dQuadraticEquation} for $n_{\text{max}}$ and plugging it into \eqref{eq:3dNumberOfDerivatives}, one finds the total number of derivatives associated with the operator $\mathcal{O}_Q$ to be given by: 
\begin{equation}
\label{eq:3dCalcNder}
n_{\text{der}} = \frac{2}{3}Q^{\frac{3}{2}}-Q+\frac{\sqrt{Q}}{12}+O(Q^{-\frac{1}{2}}).
\end{equation}
The dimension $\Delta_Q$ associated with the operator $\mathcal{O}_Q$ is given by $
\Delta_Q = Q\Delta_{\psi}+1\cdot n_{\text{der}}\,$, 
where $\Delta_\psi=1$ is the dimension of the fermionic field in $d=3$. Thus, we get:
\begin{equation}
\label{eq:3dDimensionOfOper}
\Delta_Q = \frac{2}{3}Q^{\frac{3}{2}}+\frac{1}{12}\sqrt{Q} +O(Q^{-\frac{1}{2}}).
\end{equation}

Next, we turn to consider the case of a Weyl fermion in $d=4$ dimensions.\footnote{The same analysis can be automatically extended to the case of a Dirac fermion.}  The operator $\mathcal{O}_Q$ will consist of multiplication of all the possible terms of the form $\prod_{k,l=0}^n (\partial_0^{n-k-l}\partial_1^k\partial_2^l\psi)^2$, where again $n$ takes all possible integer values between $0$ to a maximal value that depends on $Q$.\footnote{Similar to the $d=3$ case, without loss of generality, we can set $\partial_3\psi$ as the term which linearly depends on the others using the equations of motion.}
For each value of $n$, there are $\frac{(n+2)(n+1)}{2}$ different terms, each can be taken with a power of $2$ at most. This yields the following expression for the large charge operator $\mathcal{O}_Q$: 
\begin{equation}
\label{eq:OpConstruction4d}
\mathcal{O}_Q \propto (\psi)^2(\mathcal{D}_1\psi)^6(\mathcal{D}_2\psi)^{12}\cdots (\mathcal{D}_{n_{\text{max}}}\psi)^{(n_{\text{max}}+1)(n_{\text{max}}+2)}.
\end{equation}
From the condition that the operator carries a charge $Q$ under the global $U(1)$ symmetry, one finds the following relation for $n_{\text{max}}$:
\begin{equation}
\label{eq:Eqfornmax4d}
Q = \frac{1}{3}\left( n_{\text{max}}+3\right)\left(n_{\text{max}}+2\right)\left(n_{\text{max}}+1\right).
\end{equation}
The total number of derivatives in the operator 
\eqref{eq:OpConstruction4d} reads:
\begin{equation}
\label{eq:4dNumOfDer}
n_{\text{der}}= \frac{1}{4}\left(n_{\text{max}}+3 \right)\left( n_{\text{max}}+2\right) \left(n_{\text{max}}+1\right)n_{\text{max}}=\frac{3}{4}n_{\text{max}}Q.
\end{equation}
Using equation \eqref{eq:Eqfornmax4d}, we get:
\begin{equation}
\label{eq:ExprforNmax}
n_{\text{der}}= \frac{3^{\frac{4}{3}}}{4}Q^{\frac{4}{3}}-\frac{3}{2}Q+\frac{1}{4\cdot 3^{\frac{1}{3}}}Q^{\frac{2}{3}}+O(Q^{-\frac{2}{3}}).
\end{equation}
The dimension $\Delta_Q$ is given by $\Delta_Q = Q\Delta_{\psi}+ n_{\text{der}}= \frac{3}{2}Q+n_{\text{der}}$. Using equation \eqref{eq:ExprforNmax}, we find the following expression for the scaling dimension $\Delta_Q$:
\begin{equation}
\label{eq:DeltaQ4d}
\Delta_Q = \frac{3^{\frac{4}{3}}}{4}Q^{\frac{4}{3}}+\frac{1}{4\cdot 3^{\frac{1}{3}}}Q^{\frac{2}{3}}+O(Q^{-\frac{2}{3}}),
\end{equation}
Note that as in the $d=3$ case, there is no $Q^0$ term in the expansion for $\Delta_Q$.

Excited states correspond to particle-hole excitations.
The lowest order excitation corresponds to removing a single fermion from the Fermi surface and replacing it with an excited fermion, with an energy slightly above the Fermi energy. In the language of operators, this problem translates to removing a single fermion with $n_{\text{max}}$ derivatives from the operator  \eqref{eq:OperatorConstruction3d} (or \eqref{eq:OpConstruction4d} in the $d=4$ case), and replacing it with a fermion that carries $n_{\text{max}}+1$ derivatives. The resulting operator, which we denote by $\mathcal{O}_{+1,Q}$, corresponds to the next-to-lightest operator that carries the same charge $Q$.  
Following \eqref{eq:3dNumberOfDerivatives} (or \eqref{eq:4dNumOfDer} in the $d=4$ case), the total number of derivatives such an operator contains is shifted by $+1$ compared to the number of derivatives associated with the operator $\mathcal{O}_Q$. 
Hence, the energy difference between the lowest energy excitation to the ground state energy satisfies:
\begin{equation}
\Delta_{+1,Q}- \Delta_Q = O(1).
\end{equation}

\begin{figure}[!h]
\centering
\includegraphics[scale=0.6]{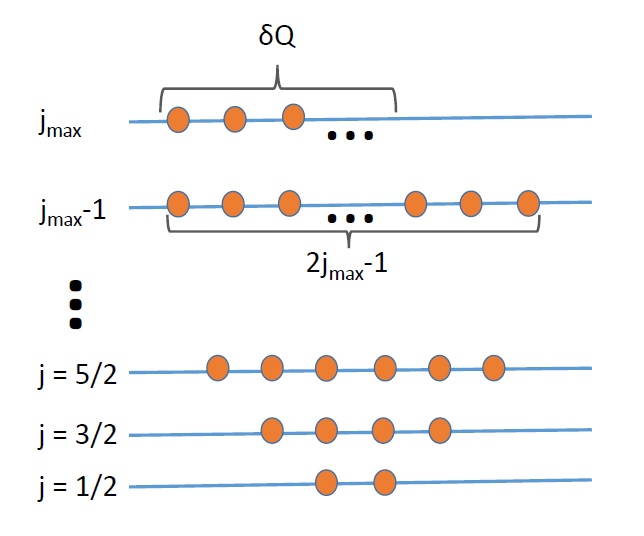}
\caption{An illustration describing the energy shells. For each $j<j_{\text{max}}$, there are $2j+1$ occupied states in each shell. At the outermost shell, which corresponds to the $j_{\text{max}}$ shell, there are $\delta Q$ occupied states, where $0\leq \delta Q \leq 2j_{\text{max}}+1$. For $\delta Q = 0$ or $\delta Q= 2j_{\text{max}}+1$, the outermost shell is filled and we recover the homogenous and isotropic ground state with an associated $\Delta_Q$ that is given by equation \eqref{eq:3dDimensionOfOper}. } \label{fig:EnergyLevels3d}
\end{figure}

Let us make a comment regarding cases in which the ground state is not homogeneous and isotropic. In terms of energy levels on the cylinder $ S^{d-1}\times \R $, this corresponds to cases in which the outermost energy shell is not fully occupied. For simplicity, we focus on $d=3$ dimensions. $\Delta_Q$ is then given by:
\begin{equation}
\label{eq:DeltaQNotFilledShell}
\Delta_Q = \sum_{j=\frac{1}{2}}^{j_{\text{max}}-1}\left(2j+1\right)\varepsilon_j+\delta Q \varepsilon_{j_{\text{max}}}, 
\end{equation}
where  $\varepsilon_j \equiv j+\frac{1}{2}$ are the energy eigenvalues on the sphere, and the $(2j+1)$ factor above represents the degeneracy.  $\delta Q$ represents the particles in the outermost, not necessarily filled energy shell (as described in figure \ref{fig:EnergyLevels3d}) and it is related to the charge $Q$ by:
\begin{equation}
\label{eq:QForNotFilledShell}
Q = \sum_{j=\frac{1}{2}}^{j_{\text{max}}-1} \left( 2j+1\right) + \delta Q.
\end{equation}

In general, $\delta Q$ can take any integer value in the range between $0$ to $2j_{\text{max}}+1$, where the latter corresponds to the case in which the outermost shell is filled and is associated with $j_{\text{max}}$, while the former corresponds to the case in which the outermost shell is the $j_{\text{max}}-1$ shell and it is also filled (see figure \ref{fig:EnergyLevels3d}).  Note that for $\delta Q =0$ or $\delta Q = 2j_{\text{max}}+1$ we simply recover the homogeneous and isotropic ground state scenario described above: $Q$ is then such that equation \eqref{eq:3dQuadraticEquation} is satisfied with an integer value of $n_{\text{max}}$, as defined above, and using the two equations \eqref{eq:DeltaQNotFilledShell}, \eqref{eq:QForNotFilledShell} one can reproduce the result \eqref{eq:3dDimensionOfOper} for $\Delta_Q$ based on the counting arguments. 
\begin{figure}[!h]
\centering
\subfloat[]{\label{fig:Fermion3dDeltaQ}{\includegraphics[width=0.35\textwidth]{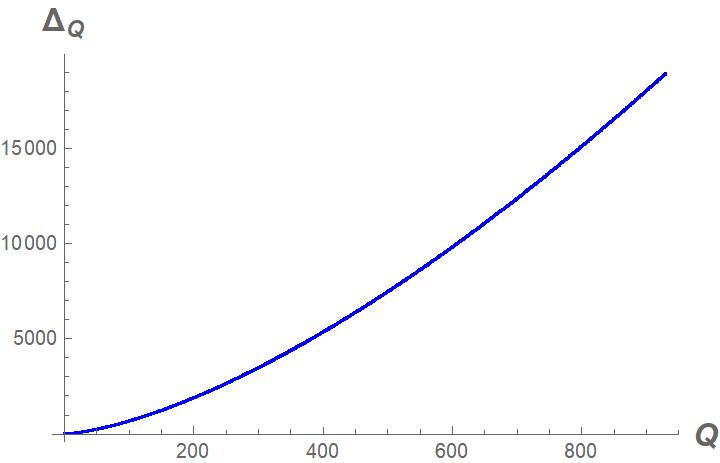}}}    \qquad \qquad
\subfloat[]{\label{fig:Fermion3dEnergyFlactuations}{\includegraphics[width=0.55\textwidth]{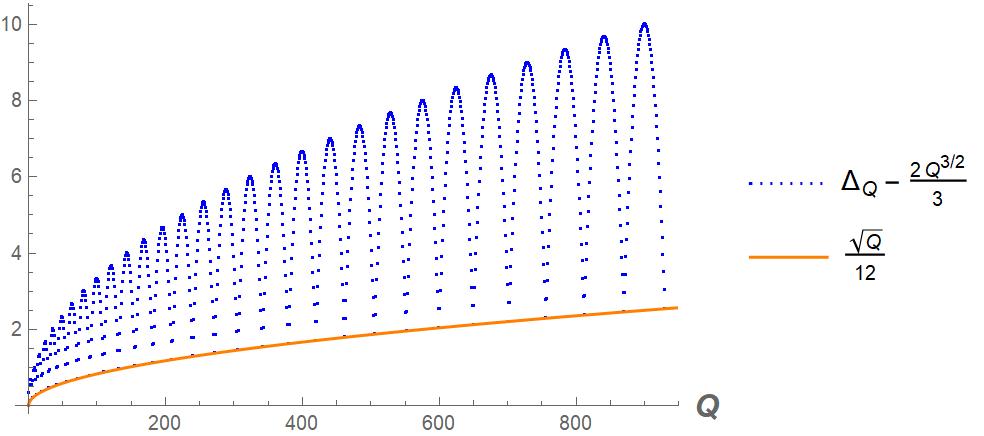}}}
\caption{An illustration describing the behavior of the scaling dimension for a general value of $Q$. Figure \ref{fig:Fermion3dDeltaQ} describes $\Delta_Q$ as a function of $Q$. In figure \ref{fig:Fermion3dEnergyFlactuations}, the blue dashed line shows $\Delta_Q-\frac{2}{3}Q^{3/2}$ as a function of $Q$, while the orange line describes $\frac{\sqrt{Q}}{12}$. The points in which the blue dashed line meets the orange line correspond to cases in which $Q$ is  such that equation \eqref{eq:3dQuadraticEquation} is satisfied with integer values of $n_{\text{max}}$. }
\label{fig:Fermion3dFig}
\end{figure}

In figure \ref{fig:Fermion3dFig}, we refer to the general $Q$ case. One can see, from figure 
\ref{fig:Fermion3dEnergyFlactuations}, that $\Delta_Q$ meets the value \eqref{eq:3dDimensionOfOper} that is associated with a homogeneous and isotropic ground state only for specific values of $Q$. These values correspond to the cases in which the outermost shell is filled, as described above. It is also interesting to notice that the fluctuations which appear in the graph of the difference between the $\Delta_Q$ to the leading order term in equation \eqref{eq:3dDimensionOfOper},  $\Delta_Q-\frac{2}{3}Q^{3/2}$ (described in figure \ref{fig:Fermion3dEnergyFlactuations}), possess an amplitude of order $O(\sqrt{Q})$. We learn that $\Delta_Q$ is not an analytic function of $Q$, if we try to define it for by using all integer $Q$'s. It is however analytic, if we only consider $Q$'s that correspond to completely filled shells, and analytically continue from these points.

The same analysis can be extended to other dimensions as well. In appendix
 \ref{app:DeltaQFreeFermion}, we discuss the $d=4$ case. We find that similar to the $d=3$ case, outside the scope of a homogeneous and isotropic ground state there are fluctuations in the difference between $\Delta_Q$ to the leading order term of \eqref{eq:DeltaQ4d}, these fluctuations are of order $O(Q^{2/3})$ (which is the same order in $Q$ as the next to leading order term in \eqref{eq:DeltaQ4d}), hence $\Delta_Q$ is not analytic in $Q$. In addition, we show that for a general number of spacetime dimensions $d$ the leading behavior of $\Delta_Q$ in the large charge limit is given by:
\begin{equation}
\label{eq:LargeChargeGeneralDLeadingOrder}
\Delta_Q\underset{Q\to\infty}{\simeq}  \frac{ 2^{\frac{1-\lceil d/2\rceil }{d-1}}}{d\Gamma(d-1)} \left[\Gamma(d)Q \right]^{d/(d-1)} \,,
\end{equation}
 where by $\lceil d/2\rceil $ we refer to the ceiling function of $d/2$.

\subsection{Relativistic Fermi Gas}
\label{subsec:RelativisticFermiGas}
In \cite{Alberte:2020eil}, non-relativistic Fermi-liquid theories were studied and were shown to satisfy the sum rule associated with the broken boost symmetry by a particle-hole continuum. We extend this analysis to the case of a relativistic Fermi gas, a state of matter that consists of many non-interacting fermions. We show that similar to the non-relativistic case, the sum rules associated with the broken symmetries are satisfied by particle-hole states.

As in \cite{Alberte:2020eil}, we are interested in cases in which the ground state of the theory itself breaks boosts, while preserving spacetime translational and spatial rotational invariance. The ground state of the theory consists of fermionic particles that occupy all momentum states with momenta $|\vec{p}| \leq p_F$, where $p_F$ is the corresponding Fermi momentum. It is therefore taken to be a tensor product of single-particle momentum eigenstates:
\es{eq:FLGroundState}{
\vert \text{GS}\rangle \equiv \mathcal{N} \prod_s \prod_{|\vec{p}\leq p_F} |\vec{p},s\rangle, \qquad |\vec{p},s\rangle\equiv c_{\vec{p}}^{s\dagger}|0\rangle,
}
where $c_{\vec{p}}^{s\dagger}$ is a fermionic creation operator creating a single particle state of momentum $\vec{p}$ and spin $s$. The constant $\mathcal{N}$ is a normalization constant and chosen such that $\langle \text{GS}\vert \text{GS}\rangle=1$. 

From the anti-commutation relations, $\lbrace c_{\vec{p}}^r\,, c_{\vec{q}}^{s\dagger} \rbrace =  (2\pi)^3\delta^{(3)}(\vec{p}-\vec{q})\,\delta^{rs} $, it is clear that the following properties of the ground state \eqref{eq:FLGroundState} hold:
\es{}{
 c_{\vec{p}}^{s\dagger}&\vert \text{GS}\rangle=0, \qquad |\vec{p}|\leq p_F, \\
 c_{\vec{p}}^s\,&\vert \text{GS}\rangle =0, \qquad |\vec{p}|> p_F.
}
The first line above simply represents Pauli's exclusion principle, while the second line states that one cannot annihilate a state which is not already contained in the Fermi ground state \eqref{eq:FLGroundState}.   

Particle-hole states are defined defined by \cite{Alberte:2020eil}:
\begin{equation}
\label{eq:ParticleHoleState}
\vert \chi\rangle = c_{\vec{p_1}}^{r\dagger}\, c_{\vec{p_2}}^{r'}\vert \text{FL}\rangle, 
\end{equation}
where  $p_2 \leq p_F$ and $ p_1 >p_F$. The annihilation operator $c_{\vec{p_2}}^{r'}$ creates a hole with momentum $\vec{p}_2$ and spin $r'$, while the creation operator $c_{\vec{p_1}}^{s\dagger}$ creates a fermionic particle with momentum $\vec{p}_1$ and spin $r$. 
The total momentum associated with the particle-hole state $\vert\chi\rangle$ is given by the difference $\vec{p}=\vec{p}_1-\vec{p}_2$. The energy associated with such a state reads $E(\vec{p},\vec{p_2})= E_{p_1}-E_{p_2}$, where $E_{p_i}=|\vec{p}_i|\equiv p_i$ is the energy of a single particle state with momentum $\vec{p}_i$. \\

In this subsection, we study the saturation of the sum rules associated with the broken boosts and dilatation for the case of relativistic free Dirac fermions in $d=4$ dimensions in flat spacetime.

\subsubsection{Matrix Elements}
The action of a free, massless Dirac fermion in $d=4$ dimensions is given by:
\begin{equation}
\label{eq:DiracAction}
S = \int d^4x \, i\bar{\psi}\slashed{\partial}\psi\, ,
\end{equation}
where $\bar{\psi}=\psi^\dagger\gamma^0$, $\slashed{\p}=\gamma^\mu\partial_\mu$.
The fermion can be written in terms of modes expansion:
\es{}{
&\psi(x) = \int \frac{d^3p}{(2\pi)^3}\,\frac{1}{\sqrt{2\omega_p}}\sum_s \left( c_{\vec{p}}^s\, u_{\vec{p}}^s\,e^{-ipx} + d_{\vec{p}}^{s\dagger}v_{\vec{p}}^s\,e^{ipx} \right),\\
&\bar{\psi}(x) = \int \frac{d^3p}{(2\pi)^3}\,\frac{1}{\sqrt{2\omega_p}}\sum_s \left( d_{\vec{p}}^s \,\bar{v}_{\vec{p}}^s\,e^{-ipx} + c_{\vec{p}}^{s\dagger}\bar{u}_{\vec{p}}^s\,e^{ipx} \right),
}
where $c_{\vec{p}}^s\,,c_{\vec{p}}^{s\dagger}$ and $d_{\vec{p}}^s\,,d_{\vec{p}}^{s\dagger}$ are the creation and annihilation operators of fermionic particles and anti-particles (respectively). They satisfy the following anti-commutation relations:
\begin{equation}
\label{eq:AntiCommRel}
\lbrace c_{\vec{p}}^r\,, c_{\vec{q}}^{s\dagger} \rbrace = \lbrace d_{\vec{p}}^r\,, d_{\vec{q}}^{s\dagger} \rbrace = (2\pi)^3\delta^{(3)}(\vec{p}-\vec{q})\,\delta^{rs}.
\end{equation}
The spinors $u_{\vec{p}}^s$ and $v_{\vec{p}}^s$ represent  the solutions of the massless Dirac equation. They satisfy:
\begin{align}
&\sum_s u_{\vec{p}}^s\, \bar{u}_{\vec{p}}^s  = \sum_s v_{\vec{p}}^s \, \bar{v}_{\vec{p}}^s = \gamma \cdot p \,, \label{eq:SpinorRelations1}\\
&\sum_s \bar{u}_{\vec{p}}^s\gamma^\mu u_{\vec{p}}^s = 4p^\mu,\label{eq:SpinorRelations2}\\
&\sum_{r,r'} \bar{u}_{\vec{p}_2}^{r'}\gamma^\mu u_{\vec{p}_1}^r\bar{u}_{\vec{p}_1}^r\gamma^\nu u_{\vec{p_2}}^{r'} = 4\left(p_2^\mu\,p_1^\nu +p_2^\nu\, p_1^\mu -g^{\mu\nu}p_1\cdot p_2 \right). \label{eq:SpinorRelations3}
\end{align}
The stress-tensor is given by:
\begin{equation}
T^{\mu\nu}(x) = \frac{i}{4}\left[ \bar{\psi}\gamma^\mu\partial^\nu\psi -\partial^\nu\bar{\psi}\gamma^\mu\psi + (\mu \leftrightarrow \nu ) \right] -\eta^{\mu\nu}\mathcal{L}.
\end{equation}
We consider the system \eqref{eq:DiracAction}  in the ground state described by  \eqref{eq:FLGroundState}. 
The energy density $\epsilon$ and pressure $P$ are defined as the vacuum expectation values of $T^{00}$ and $T^{ij}$ (respectively) with respect to the ground state \eqref{eq:FLGroundState} using:
\begin{equation}
\epsilon\equiv \langle \text{GS}\vert T^{00} \vert\text{GS}\rangle  \,,\quad\ P\delta^{ij}\equiv \langle \text{GS}\vert T^{ij} \vert\text{GS}\rangle  \,.
\end{equation}
Here we have secretly used the fact that ground state is isotropic to pull out the factor $\delta^{ij}$ in defining the pressure $P$. Using $\langle \text{GS}\vert c_{\vec{p}}^{s\dagger}c_{\vec{q}}^{s'} \vert\text{GS}\rangle = (2\pi)^3\delta^{(3)}(\vec{p}-\vec{q})$, 
we find:
\es{}{
&\epsilon = -\frac{1}{2}\int \frac{d^3q}{(2\pi)^3}\frac{q_i}{E_q}\sum_s \left( \bar{u}_{\vec{q}}^s\,\gamma^i u_{\vec{q}}^s \right) = \frac{p_F^4}{4\pi^2}\,,  \\
& P\delta^{ij} = \frac{1}{2}\int \frac{d^3q}{(2\pi)^3}\frac{1}{2E_q}\sum_s\left( q^i\bar{u}_{\vec{q}}^s\gamma^ju_{\vec{q}}^s+q^j\bar{u}_{\vec{q}}^s\gamma^iu_{\vec{q}}^s\right) \quad \implies \quad P =  \frac{1}{3}\frac{p_F^4}{4\pi^2}\, .
}
In the last step, we have used \eqref{eq:SpinorRelations2} and integrated over a sphere of radius $p_F$. 
Using the anti-commutation relations \eqref{eq:AntiCommRel}, as well as the definitions of the ground state \eqref{eq:FLGroundState} and the particle-hole state \eqref{eq:ParticleHoleState}, one can show that the only nontrivial identity involving creation-anihilation operator is given by:
\begin{align}
& \langle \text{FL} \vert c_{\vec{p}}^{s\dagger} \, c_{\vec{q}}^{s'} \vert\chi \rangle = (2\pi)^6\delta^{s'r}\delta^{r's}\delta^{(3)}(\vec{q}-\vec{p}_1)\delta^{(3)}(\vec{p}-\vec{p}_2)\,.
\end{align}
We define the following matrix elements:
\begin{equation}
\mathcal{T}^{\mu\nu}(x) \equiv  \langle \text{FL} \vert T^{\mu\nu}(x) \vert\chi \rangle \, .
\end{equation}
A straightforward calculation yields:
\es{}{
\mathcal{T}^{00}(0)& = \frac{1}{4}\frac{1}{\sqrt{E_{p_1}E_{p_2}}} \left( E_{p_1}+E_{p_2}\right)\bar{u}_{\vec{p}_2}^{r'}\gamma^0 u_{\vec{p}_1}^r,\\
\mathcal{T}^{0i}(0) &= \frac{1}{8}\frac{1}{\sqrt{E_{p_1}E_{p_2}}} \left[\left( E_{p_1}+E_{p_2}\right)\bar{u}_{\vec{p}_2}^{r'}\gamma^i u_{\vec{p}_1}^r+\left(p_1^i+p_2^i\right)\bar{u}_{\vec{p}_2}^{r'}\gamma^0 u_{\vec{p}_1}^r\right] .
}
Using the relation \eqref{eq:SpinorRelations3}, we get:
\begin{align}
\label{eq:T00T0iNotSimplified}
\mathcal{T}^{00}\mathcal{T}^{0i^*}(\vec{p}_1,\vec{p}_2) &=  \frac{1}{8}\frac{1}{E_{p_1}E_{p_2}}\left(E_{p_1}+E_{p_2} \right)^2\left(E_{p_2}p_1^i+E_{p_1}p_2^i \right)\\
& +\frac{1}{8}\frac{1}{E_{p_1}E_{p_2}}\left(E_{p_1}+E_{p_2} \right)\left(p_1^i+p_2^i \right)\left(E_{p_2}E_{p_1}+\vec{p}_1\cdot\vec{p}_2 \right), \nonumber\\
\label{eq:T00T00NotSimplified}
\mathcal{T}^{00}\mathcal{T}^{00^*}(\vec{p}_1,\vec{p}_2) &= \frac{1}{4}\frac{(E_{p_1}+E_{p_2})^2}{E_{p1}E_{p_2}}\left(E_{p_1}E_{p_2}+\vec{p}_1\cdot\vec{p}_2\right),
\end{align}
where we have defined $\mathcal{T}^{\mu\nu}\mathcal{T}^{\rho\sigma^*}(\vec{p}_1,\vec{p}_2) \equiv\sum_{r,r'}\mathcal{T}^{\mu\nu}(0)\mathcal{T}^{\rho\sigma^*}(0)$.

\subsubsection{Saturation of the Sum Rules}
Expanding \eqref{eq:T00T0iNotSimplified}, \eqref{eq:T00T00NotSimplified} in small $\vec{p}$ (where $\vec{p}\equiv \vec{p}_1-\vec{p}_2$), we get: 
\begin{equation}
\label{eq:T00T0iExpan}
\mathcal{T}^{00}\mathcal{T}^{0i^*}(\vec{p},\vec{p}_2) = p_2p^i+\frac{(\vec{p}\cdot\vec{p}_2)p_2^i}{p_2}+2p_2p_2^i+O(p^2),
\end{equation}
and:
\begin{equation}
\label{eq:T00T00Expan}
\mathcal{T}^{00}\mathcal{T}^{00^*}(\vec{p},\vec{p}_2) = 2p_2^2+2\vec{p}_2\cdot\vec{p}+O(p^2).
\end{equation}
In order to evaluate the spectral density we need to integrate over $\vec{p}_2$. Note that in the limit of small $\vec{p}$, the energy associated with the state of momentum $\vec{p}$ is given by:
\begin{equation}\label{eq:energy}
E(\vec{p},\vec{p}_2) = E_{p_1}-E_{p_2} = p\cos(\theta)+O(p^2),
\end{equation}
where $\theta$ is the angle between the vectors $\vec{p}_2$ and $\vec{p}$. 

\begin{figure}[!h]
\centering
\subfloat[]{\label{fig:FermiFig1a}{\includegraphics[width=0.25\textwidth]{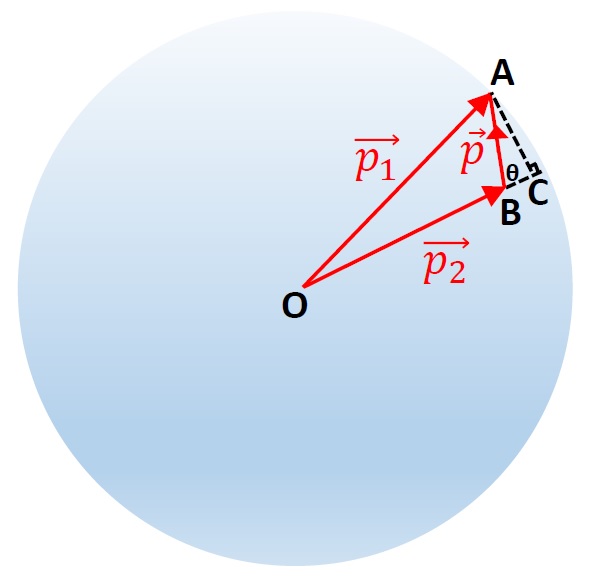}}}    \qquad \qquad
\subfloat[]{\label{fig:FermiFig1b}{\includegraphics[width=0.25\textwidth]{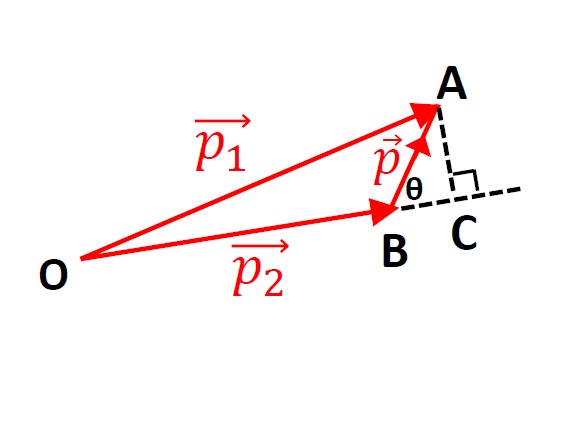}}}
\caption{An illustration describing the momentum space notation corresponding to the integral in the expression for the spectral density \eqref{eq:GeneralSpectralDensFromMatrixFer} for a specific configuration of $\vec{p}_1,\vec{p}_2$: as $p\to 0$, $p_1 \to p_F^+$, and $p_2+p\cos\theta \approx p_F$. Thus the variable $\delta p_2\equiv p_2-p_F$ ranges from $-p\cos\theta$ to $0$ in $p\to0$ limit. This will be useful in evaluation \eqref{eqs:int}.
Figure \ref{fig:FermiFig1a} describes a Fermi surface of radius $p_F$ in this specific configuration in momentum space, while figure \ref{fig:FermiFig1b} zooms in on the triangle $AOB$.  }
\label{fig:FermiFig}
\end{figure}

The spectral function $\rho_{T^{00}T^{0\mu}}$ to the leading order is given by the following : 
\begin{equation}
\label{eq:GeneralSpectralDensFromMatrixFer}
\rho_{T^{00}T^{0\mu}}(\omega,\vec{p})= \mathcal{G}_{T^{00}T^{0\mu}}(\omega,\vec{p})\mp \mathcal{G}^{*}_{T^{00}T^{0\mu}}(-\omega,-\vec{p}),
\end{equation}
where the $\mp$ sign takes the values of $-$ for $\mu=0$ and $+$ for spatial indices $\mu=i$, and $\mathcal{G}_{T^{00}T^{0\mu}}(\omega,\vec{p})$ involves the following integral (see figure \ref{fig:FermiFig}):
\begin{equation}
\begin{aligned}\label{eqs:int}
&\mathcal{G}_{T^{00}T^{0\mu}}(\omega,\vec{p}) \equiv\frac{1}{(2\pi)^3} \int d^3p_2\ \delta\left(\omega-E(\vec{p},\vec{p}_2)\right) \mathcal{T}^{00}\mathcal{T}^{0\mu*}(\vec{p},\vec{p}_2)  \\\
&=  \left[\frac{ p_F^2}{(2\pi)^2}\int_{0}^{\pi/2} d\theta\ \sin\theta\ \delta(p\cos\theta-\omega) \mathcal{T}^{00}\mathcal{T}^{0\mu*}(p,p_F,\cos\theta) \int_{-p\cos\theta}^0d\delta p_2\right]+\cdots\\
&=\frac{\chi_{[0,p]}(\om)}{4\pi^2}\frac{\omega p_F^2 }{p}  \mathcal{T}^{00}\mathcal{T}^{0\mu*}(p,p_F, \omega/p).
\end{aligned}
\end{equation}
Note that $\mathcal{T}^{00}\mathcal{T}^{0\mu*}(\vec{p},\vec{p}_2)$ is a function of $p, p_2$ and the angle between the vectors, i.e.~of $\cos\theta$. At this point we change the integration variable from $p_2$ to $\delta p_2=p_2-p_F$ and from fig.~\ref{fig:FermiFig}, we read off the limit of integral in $p\to 0$ limit. For the same reason, we have kept only the leading order terms in $\delta p_2$ of $\mathcal{T}^{00}\mathcal{T}^{0\mu*}(\vec{p},\vec{p}_2)$ in the second line, which amounts to replacing $p_2$ with $p_F$ and we made the angle dependence explicit. We also use \eqref{eq:energy} inside the delta function, which subsequently sets $\cos\theta=\omega/p$ leading to the term $\mathcal{T}^{00}\mathcal{T}^{0\mu*}(p,p_F, \omega/p) \equiv \mathcal{T}^{00}\mathcal{T}^{0\mu*}(p,p_2, \cos\theta)\vert_{p_2=p_F,\,\cos\theta=\omega/p}$. The function $\chi_{[0,p]}(\om)$ is the characteristic function of the interval $[0,p]$. 
  
We start with the calculation of $\rho_{T^{00}T^{00}}(\omega,p)$. From \eqref{eq:T00T00Expan}, we read: $
\mathcal{T}^{00}\mathcal{T}^{0\mu*}\left(p,p_F,\omega/p\right)= 2p_F^2+ 2p_F \omega $, then:
\begin{equation}
\begin{aligned}
\rho_{T^{00}T^{00}}(\omega,\vec{p})
&= \frac{\chi_{[-p,p]}(\om)}{2\pi^2}\frac{\omega p_F^4 }{p} +\cdots \underset{p\to 0}{\simeq}-(\epsilon+p)p^2\delta'(\omega)\,,
\end{aligned}
\end{equation}
as it should be, in accordance with \eqref{cons}.

Next, we turn to calculate $\rho_{T^{00}T^{0i}}$. For this purpose, it is convenient to define $\vec{p}=p\hat{z}$, and keep $\vec{p_2}$ arbitrary. From \eqref{eq:T00T0iExpan}, we read: 
$\mathcal{T}^{00}\mathcal{T}^{0z*}(p,p_F,\omega/p) = p_Fp(1+\omega^2/p^2)+2p_F^2\omega/p\,$. Plugging it into the expression for the spectral density \eqref{eq:GeneralSpectralDensFromMatrixFer}, after covariantizing the result for $\rho_{T^{00}T^{0z}}$, we find the following expression for the spectral density $\rho_{T^{00}T^{0i}}$:

\begin{equation}
\begin{aligned}
\rho_{T^{00}T^{0i}}(\omega,\vec{p})\underset{p\to 0}{\simeq} \frac{\chi_{[-p,p]}(\om)}{2\pi^2}\,\frac{\omega^2 p_F^4 }{p^3}p^{i} =\chi_{[-p,p]}(\om) \frac{3(\epsilon+P)}{2}\,\frac{\omega^2 }{p^3}p^{i}\,.
\end{aligned}
\end{equation}
Using the above result, it is straightforward to check the saturation of the sum rules. One finds:
\begin{equation}
\begin{aligned}
\label{eq:SimplExprSumRule}
\frac{\partial  \rho_{T^{00}T^{0i}}}{\partial p^j}(\omega,\vec{p})=\underset{p\to 0}{\simeq} (\epsilon+P)\delta(\omega)\delta^i_j,
\end{aligned}
\end{equation}
where we have used 
$$\chi_{[-p,p]}(\om)
\left(\frac{\omega^2}{p^3}\right) =\left[\Theta(\omega+p)-\Theta(\omega-p)\right]
\left(\frac{\omega^2}{p^3}\right) \underset{p\to 0}{\simeq} \frac{2}{3} \delta(\omega)\,.$$
Therefore, the sum rule associated with the broken boosts \eqref{SumRule3} is satisfied. 
Using \eqref{eq:SimplExprSumRule} one easily finds: 
\begin{equation}
\frac{\partial  \rho_{T^{00}T^{0i}}}{\partial p^i}(\omega,\vec{p})\underset{p\to 0}{\simeq}\,3(\epsilon+P)\delta(\omega) = 4\epsilon\delta(\omega)\,,
\end{equation}
thus, the sum rule \eqref{SumRuleD2} associated with the broken dilatations is satisfied (with $d=4$).  

While in this section we have been studying the CFT of a free fermion, the analysis is in fact applicable to any (possibly interacting) CFT state around which the effective theory is a free Fermi surface.  To our knowledge it is not presently known if a such a free Fermi surface is a natural end point under the RG evolution around heavy states. It would be interesting to investigate it along the lines of~\cite{Polchinski:1992ed,Shankar1994}.

\section{2d CFTs at Large Charge} \label{sec:boostbreak}

\subsection{Boost Breaking in 2d CFTs}

We will consider 2d CFTs on the cylinder with circle of radius $R$: 
 $ds^2=d\tau^2+R^2(d\theta)^2$, with $\theta\simeq \theta+2\pi$. 
The advantage of the 2d setup is that we can construct the correlation functions of the EM tensor explicitly and verify the existence of the large volume (macroscopic)  limit. The low-energy states responsible for the boost Nambu-Goldstone theorem can be also identified. Remarkably, many of the things we find are similar to the superfluid discussion in section~\ref{sec:superfluid}.
 
\subsection*{TJ correlator}
We take an arbitrary state $|\Omega\rangle$ which corresponds to a spinless highest weight state in the Verma module with dimension $\Delta$. Let $\Phi$ be some primary and consider first the four-point function in flat space ($h_\Phi,\bar h_\Phi, h_\Omega,\bar h_\Omega$ stand for the obvious scaling dimensions and we assume $h_\Omega=\bar h_\Omega$.)

\begin{equation}\label{eq:TPhi}
\langle\Omega(0)T(z) \Phi(1) \Omega(\infty)\rangle={Ch_\Omega\over z^2}+{Ch_\Phi\over z(z-1)^2}~,
\end{equation}
where
\begin{equation}
\langle\Omega(0) \Phi(1) \Omega(\infty)\rangle=C~.
\end{equation}
In order to transform this to the cylinder with a circle of radius $R$ we need to plug $z= e^{u/R}$ and take $T(z)\to R^2 z^{-2} T(u)$ and $\Phi(1)\to R^{ \Delta_\Phi} \Phi(0)$.\footnote{The transformation of the EM tensor $T(z)\to R^2 z^{-2} T(u)$ is missing a constant -- the famous ground state energy from the Schwartzian. This is unimportant for us because we are considering the correlation functions in heavy states and hence we drop this constant. } Therefore the following Euclidean correlation function is found on the cylinder with coordinate $u=\tau+iR\theta$:
$$\langle\Omega|T(u) \Phi(0) |\Omega\rangle={Ch_\Omega\over R^{2+\Delta_\Phi}}+{Ch_\Phi\over 2R^{2+\Delta_\Phi} (\cosh(u/R)-1)}~.$$
The constant piece is necessary to account for the propagation of the state  $|\Omega\rangle$. 

Let us now investigate the macroscopic limit. For concreteness, we take $\Delta_\Phi=1$ and $\Phi$ to be a conserved current $(1,0)$ operator.  $C/(\pi R)=\rho$ is the charge density in the state  $|\Omega\rangle$ while $h_\Omega/(\pi R^2)=\epsilon$ is the energy density. Evidently, to achieve a nontrivial macroscopic limit, we need to take $C$ to scale as a positive power of $R$ -- i.e., we change the state $\Omega$ as a function of $R$ such that $C\over R$ is finite in the $R\to\infty$ limit. This is the same as having a constant charge density. Secondly, the constant ${Ch_\Omega\over R^{3}}$
is proportional to $\rho\epsilon$ which we should also hold fixed. Then the macroscopic limit becomes 
$$\langle\Omega|T(u)J(0) |\Omega\rangle={\pi^2}\rho\epsilon+{\pi \rho\over   u^2}~.$$
Similarly, 
$$\langle\Omega|T(u)\bar J(0) |\Omega\rangle={\pi^2}\rho\epsilon~.$$

In components this becomes (assuming $C$ is real) at separated points in Euclidean signature (where $T=\pi(T_{tt}-T_{tx})$, $\bar T=\pi(T_{tt}+T_{tx})$, similary $J=\pi(J_t-J_x)$, $\bar J=\pi(J_t+J_x)$):\footnote{To conform with both the conventions of the 2d CFT literature and the usual notion of $T_{\mu\nu}$ in higher dimensions, we use the definitions $T\equiv -{4\pi\ov\sqrt{g}} \,{\de S\ov \de g^{uu}}$ and $T_{\mu\nu}\equiv {2\ov\sqrt{g}} \,{\de S\ov \de g^{\mu\nu}}$.}
\begin{equation}
\begin{aligned}
&\langle\Omega|T_{tt}(\tau,x) J_t(0)|\Omega\rangle=\rho\epsilon+\frac{\rho}{2\pi}{(\tau^2-x^2)\over (\tau^2+x^2)^2}\,,\\
&\langle\Omega|T_{xt}(\tau,x) J_x(0)|\Omega\rangle=\frac{\rho}{2\pi}{\tau^2-x^2\over (\tau^2+x^2)^2}\,,\\
&\langle\Omega|T_{tt}(\tau,x) J_x(0)|\Omega\rangle=-i\frac{\rho}{\pi}{\tau x\over (\tau^2+x^2)^2}\,,\\
&\langle\Omega|T_{xt}(\tau,x) J_t(0)|\Omega\rangle=-i\frac{\rho}{\pi}{\tau x\over (\tau^2+x^2)^2}\,.
\end{aligned}
\end{equation}
We do not write the matrix elements containing $T_{xx}$ as they are the same up to a sign as those containing $T_{tt}$. 

Analytic continuation to Minkowski signature is implemented in the equations above by setting $\tau=it \pm \epsilon$ with $\epsilon\to0^{+}$. The $\pm$ correspond to different (Minkowski) time ordering of the operators:
\begin{equation}
\begin{aligned}
&\langle\Omega|T_{tt}(t,x) J_x(0)|\Omega\rangle=- \frac{\rho}{2\pi}{tx\over x^2+t^2}\left({\partial\over \partial x}{1\over t+x-i\epsilon}-{\partial\over \partial x}{1\over t-x-i\epsilon}\right)\\
&\langle\Omega| J_x(0)T_{tt}(t,x)|\Omega\rangle=- \frac{\rho}{2\pi}{tx\over x^2+t^2}\left({\partial\over \partial x}{1\over t+x+i\epsilon}-{\partial\over \partial x}{1\over t-x+i\epsilon}\right)
\end{aligned}
\end{equation}

Now we use the familiar identity (with $\epsilon\to 0^+$ assumed): ${1\over u+i\epsilon} - {1\over u-i\epsilon} =-2\pi i \delta (u)$ to find an expression for the commutator in position space: 
\begin{equation}
\langle\Omega|[T_{tt}(t,x), J_x(0)]|\Omega\rangle= -i\rho{tx\over x^2+t^2}\left(\delta'(t+x)+\delta'(t-x)\right)~.
\end{equation}
The support of the commutators on the light-cone is of course due to the non-dissipative nature of the excitations pertinent to this problem. In frequency and momentum space we find 
\begin{equation}
\begin{aligned}
\langle\Omega|[T_{tt}(\om,k), J_x(0)]|\Omega\rangle&= \int_{-\infty}^{\infty} dt\,dx\ e^{i\om t-i kx} \langle\Omega|[T_{tt}(t,x), J_x(0)]|\Omega\rangle \\
&= -\pi \rho k (\delta(\om+k)+\delta(\om-k))~\,.
\end{aligned}
\end{equation}
 The spectral density is given by
\es{TttJxFunny}{
\rho_{T_{tt}J_x}(\om, k) =-\frac{1}{2} \rho k (\delta(\om+k)+\delta(\om-k)) \underset{k\to0}{\simeq} -\rho k \delta(\om)\,,
}
as required by the sum rule \eqref{SumRuleJ} for the boost. (We remind the reader that in our conventions $g_{xx}=-1$, hence $k_x=-k^x=-k$, which explains the sign in \eqref{TttJxFunny}.)

Similarly, one can show that 
\begin{equation}
\rho_{T_{xt}J_t}(\om, k)=- \frac{1}{2} \rho k (\delta(\om+k)+\delta(\om-k)) \underset{k\to0}{\simeq} -\rho k \delta(\om)\,,
\end{equation}
and verify the sum rule \eqref{SumRuleD2} for the dilatation.


\subsection*{TT correlator}
The correlator involving the energy momentum tensor is given by
\begin{equation}
\langle\Omega(0)T(z) T(1) \Omega(\infty)\rangle={h^2_\Omega\over z^2}+{2h_\Omega\over z(z-1)^2}+{c/2\over (z-1)^4}~.
\end{equation}
This leads to an amplitude on the circle of radius $R$:
\es{ToRep}{
\langle\Omega|T(u) T(0) |\Omega\rangle={h^2_\Omega\over R^{4}}+{h_\Omega\over R^{4} (\cosh(u/R)-1)}+{c\over 8R^4(\cosh(u/R)-1)^2}~.
}
The macroscopic limit requires to keep $h_\Omega/R^2\equiv \pi\epsilon$  fixed and $c$ fixed. Then we obtain in the macroscopic limit: 
\es{Aim}{
\langle\Omega|T(u) T(0) |\Omega\rangle={\pi^2\epsilon^2}+{2\pi\epsilon\over  u^2}+{c\over 2u^4}~.
}
Similarly 
$$\langle\Omega|T(u) \bar T(0) |\Omega\rangle={\pi^2\epsilon^2}~.$$
We can now extract the energy-density correlator with itself as before by inserting $T=-2\pi T_{uu}=-\pi(T_{\tau \tau}-iT_{\tau x})=\pi (T_{tt}-T_{tx})$ and expanding to find:
\begin{equation}
\langle\Omega|T_{tt}(\tau,x)  T_{tt}(0) |\Omega\rangle=\epsilon^2+\frac{2\epsilon (\tau^2-x^2)}{ \pi(x^2+\tau^2)^2}+\frac{c}{4\pi^2}\frac{(x^4+\tau^4-6x^2\tau^2)}{4\pi^2(x^2+\tau^2)^4}
\end{equation}

The commutator is then given by (similar to the calculation of $TJ$ commutator)
\begin{equation}\langle\Omega|[T_{tt}(t,x),  T_{tt}(0)] |\Omega\rangle= i\epsilon\left(\delta'(x+t)-\delta'(x-t)\right) -i \frac{c}{24\pi}\left(\delta'''(x+t)-\delta'''(x-t)\right)\,.
\end{equation}
This can be now transformed to frequency and momentum space to find  
\begin{equation}
\begin{aligned}
\langle\Omega|[T_{tt}(\om,k),  T_{tt}(0)] |\Omega\rangle&= \int_{-\infty}^{\infty} dt\, dx\ e^{i\om t-i kx} \langle\Omega|[T_{tt}(t,x), T_{tt}(0)]|\Omega\rangle \\
&=2\pi\epsilon k\left(\delta(\om-k)-\delta(\om+k)\right) +c\pi k^3 /6\left(\delta(\om-k)-\delta(\om+k)\right)~.
\end{aligned}
\end{equation}

The spectral density is given by
\es{spectralsmallk}{
\rho_{T_{tt}T_{tt}} \underset{k\to 0}{\simeq} -2\epsilon k^2\delta'(\omega) - \frac{ k^4}{6}\le[c\,\delta'(\omega)+2\ep\, \delta'''(\omega)\ri]+\cdots\,.
}
The sum rule \eqref{t00t00} is saturated by the first term, as $\epsilon+P=2\epsilon$ in $2$ dimension.

To understand which terms contribute to the sum rule we must study in detail the intermediate states by inserting a complete set of states. It is obvious that the only states $\langle\Omega'|$ for which $\langle\Omega'|T_{tt}(0) |\Omega\rangle\neq0$ are in the Verma module of $|\Omega\rangle$ (left or right descendants but not both). The usual basis of states $L_{-N}^{n_N}\cdots L_{-1}^{n_1}|\Omega\rangle$ is inconvenient to use since it is not orthonormal and the matrix elements are difficult to compute. Instead we use the oscillator basis of~\cite{zamolodchikov1986two}, nicely reviewed in~\cite{Besken:2019bsu} (we are using their notations) and we start by computing the wave function 
\es{Ubasis}{
\langle U| T(u) |\Omega\rangle&={1\over R^2}\sum_{n\leq 0}e^{-nu/R} \langle U| L_n |\Omega\rangle  ={1\over R^2}\sum_{n\leq 0}e^{-nu/R}{\cal L}_{-n}\cdot 1 \\
&={1\over R^2}\left[h-\sum_{k=2}^\infty e^{-ku/R}\sum_{p=1}^{k-1}p(k-p)u_{p}u_{k-p}+2\sum_{k=1}^\infty e^{-ku/R} k(\mu k -i\lambda) u_k\right]~.}
where $u$ is our usual coordinate on the cylinder of radius $R$ and $u_k$ is an infinite set of variables collectively denoted $U$. $\mu,\lambda$ are related to $h_\Omega$ and $c$  via $c=1+24\mu^2$, $h_\Omega=\lambda^2+\mu^2$. In this basis the monomials are orthogonal with norm
$$(1,1)=1~,\quad (u_k,u_l)=\delta_{k,l}S_{k,1}~,\quad (u_k^2,u_l^2)=\delta_{k,l}S_{k,2}~,\quad (u_ku_p,u_qu_{l})=
\delta_{k,q}\delta_{p,l} S_{k,1}S_{l,1}+\delta_{k,l}\delta_{p,q} S_{k,1}S_{l,1}~. $$
In the last term we assumed that the four  indices are not all the same. We used $S_{k,j}=(2k)^{-j}\Gamma(j+1)$. We can unify the third and fourth formulas and write:
$$(1,1)=1~,\quad (u_k,u_l)=\delta_{k,l}{1\over 2k}~,\quad (u_ku_p,u_qu_{l})=
\left(\delta_{k,q}\delta_{p,l}++\delta_{k,l}\delta_{p,q}\right) {1\over 4kl}~. $$
Inserting a complete set of states, we find that only need to insert ``single particle'' and ``two particle'' states:  
$$\langle\Omega|T(u) T(0) |\Omega\rangle=\langle\Omega|T(u)|\Omega\rangle\langle\Omega|T(0) |\Omega\rangle+\sum_{k=1} 2k\langle\Omega|T(u)|u_k\rangle\langle u_k|T(0) |\Omega\rangle$$ $$+4\sum_{k< l}kl \langle\Omega|T(u)|u_ku_l\rangle\langle u_ku_l|T(0) |\Omega\rangle+2\sum_{k}k^2 \langle\Omega|T(u)|u_k^2\rangle\langle u_k^2 |T(0) |\Omega\rangle~.$$
The first term on the right hand side gives $h/R^4$. The second term gives $${4\over R^4} \sum_{k=1}^\infty 2k e^{-ku/R} k(\mu k -i\lambda) {1\over 2k} k(\mu k +i\lambda) {1\over 2k}={2\over R^4} \sum_{k=1}^\infty  e^{-ku/R} k(\mu^2 k^2 +\lambda^2)~.$$
This is easily summed to give 
$${1\over R^4}(\lambda^2+\mu^2){1\over (\cosh(u/R)-1)}+{3\over R^4} \mu^2 {1\over  (\cosh(u/R)-1)^2 }~.$$
This accounts for almost the whole answer in \eqref{ToRep}, except that the coefficient of the second term above is off. (In the full answer it is $c/(8 R^4)=(3\mu^2+1/8)/R^4$ instead of $3\mu^2/R^4$ that we obtained from the one particle exchange.) 
 One can check that the difference is made up by the two-particle states. As we have seen above in~\eqref{Aim}--\eqref{spectralsmallk}, the first term, 
${1\over R^4}(\lambda^2+\mu^2){1\over (\cosh(u/R)-1)}$ is enough to saturate the boost Nambu-Goldstone theorem in the macroscopic limit. Therefore, the boost Nambu-Goldstone theorem is saturated by one-particle states $|\Omega\rangle$ and $|u_k\rangle$. 
Note that of these states, only $|u_1\rangle$ is an ordinary conformal descendant of $|\Omega\rangle$ (indeed, it is proportional to the action of ${\cal L}_{-1}$ on $|\Omega\rangle$).  

In frequency space the commutator directly on the cylinder as a result of these one-particle exchanges is (we denote by $p=n/R$, with $n$ an integer, the momentum on the circle of radius $R$) 
\begin{equation}\label{sd}\langle\Omega|[T(\om,p),T(0)] |\Omega\rangle= {2n\over R^4} (\mu^2n^2+\lambda^2)\, \delta\le(\om-{n\ov R}\ri)+\cdots
\end{equation}
All the states that appear in the intermediate channel have energy and momentum that are related by $\om_n=p_n=n/R$. This is because they are excitations of the ground state given by the action of a holomorphic EM tensor.

As always, the spectral density, being a sum of delta functions, needs some smearing before it can be written in the infinite volume limit. By contrast, correlation functions for $u\ll R$ land themselves to a nice macroscopic limit more directly. 
From the spectral density~\eqref{sd} we see that any one individual state, even if it has $n\sim R$ and $\lambda^2\sim R^2$ as required in the macroscopic limit, leads to a vanishing spectral density in the macroscopic limit (the contribution is suppressed as $1/R$). As was discussed in detail in section~\ref{subsec:gap}, we have to smear over a band of states centered around $n\sim R$ and $\lambda^2\sim R^2$ to recover the correct result.

Note the nice analogy to the superfluid effective theory and mean field theory: in \eqref{eqs:T00T00s}
and \eqref{eqs:T00J0free}  we have contributions from what is analogous to one-particle states in the Verma module 
$|u_n\rangle$, of which $|u_1\rangle$ is a conformal descendant of the ground state. These are sufficient to reproduce the boost sum rule. The contribution comes from the states with energy $n\sim R$ while the ground state has energy $\sim R^2$. 

\subsection{A Compact Boson Effective Theory}

It is tempting to try and reproduce the above results with an effective theory. One candidate is the superfluid EFT of section~\ref{sec:superfluid} specialized to $d=2$, which could apply to the lowest energy state at fixed (large) $U(1)$ charge. In CFTs with a discrete operator spectrum it is a fundamental result \cite{Affleck:1985jc} that the $U(1)$ symmetry is enhanced to a $u(1)\times u(1)$ Ka\v{c}-Moody algebra, and the Energy-Momentum tensor of the full theory decomposes to two separately conserved Energy-Momentum tensors. 

Below we present an effective theory in $d=2$ with the following properties: 
\begin{itemize} 
\item The $U(1)$ symmetry is of course not spontaneously broken -- it is in the usual ``log-ordered'' phase which is common in $d=2$. 
\item It has one compact boson but it allows for an energy-momentum tensor with {\it arbitrary} central charge.
\item The $U(1)$ symmetry can be promoted to a Ka\v{c}-Moody symmetry only if the central charge is $c=1$. 
\end{itemize} 

Besides describing the compact boson large charge limit, which is somewhat trivial, the theory we present is potentially interesting for situations 
where the $U(1)$ symmetry of a CFT does not enhance to Ka\v{c}-Moody. Such CFTs must have a continuous spectrum\footnote{The simplest such example is a noncompact complex scalar $\Phi$, with the $U(1)$ symmetry rotating around the origin of the target space $\mathbb{C}$ and the associated current
\es{J}{
J_\mu=i\le(\Phi^\dagger\p_\mu\Phi-\Phi\p_\mu\Phi^\dagger\ri)
} which does not get enhanced to a Ka\v{c}-Moody symmetry.} and one might worry that the large charge limit would be necessarily more complicated.

%

A  more realistic and interesting application for our EFT is to describe superfluids with a boundary, or equivalently, the large charge limit of 3d boundary CFT (BCFT). Indeed, the aforementioned shift in the central charge of a single compact boson will be crucial for matching the boundary trace anomaly of the superfluid. For some literature on boundary trace anomalies see~\cite{Henningson:1999xi, Schwimmer:2008yh,Solodukhin:2015eca,Jensen:2015swa,Herzog:2015ioa,Herzog:2017kkj,Wang:2021mdq}.
We leave the development of this direction to future work.

\subsubsection{First Look at the Large Charge Effective Theory }

Let us start with the EFT \eqref{eqs:schi} specialized to $d=2$:\footnote{By $d^2x$ we mean $d\tau dx=\frac12 du d\bar{u}$.}
\begin{equation}\label{leadingterm}
S_0={\kappa\ov \pi} \int d^2x\ \partial \varphi\bar \partial \varphi~,
\end{equation}
where $\varphi$ is a compact scalar $\varphi\sim \varphi+2\pi$.
 We expand the theory around
$\varphi= -i\mu \tau+\pi$ such that $\kappa\mu/(2\pi)=\rho$ and $\kappa \mu^2/(4\pi)=\epsilon$ and try to match the correlators we found from the theory for the fluctuations: $\kappa/\pi \int d^2x\ \partial\pi\bar\partial\pi$. The stress tensor of the theory $S_0$ in  \eqref{leadingterm} is $T=-\ka (\p\varphi)^2$, which in terms of the fluctuations takes the form
\es{leadingT}{
T={\kappa\mu^2\ov 4}+i\kappa\mu \partial\pi-\kappa (\partial\pi)^2\,.
}
The stress tensor two point function can be computed using the propagator
\es{PropCorr}{ 
\langle\partial \pi(u)\partial\pi(0)\rangle = -{1\over 2\kappa u^2}\,,
}
and we get
\es{TTleading}{
\langle T(u)T(0)\rangle ={\pi^2\epsilon^2}+{2\pi\epsilon\over u^2}+{1\over 2 u^4} \,.
}
The first two pieces are exactly right but the third one is not. This is because our effective theory has central charge 1 instead of $c$. We have so far merely reproduced the known result that the compact boson has central charge 1.

If we want to reproduce \eqref{TTleading} exactly, we face the question of how to make a single compact boson $\varphi$ have $c\neq 1$, which seems at first sight impossible. A similar in spirit problem arises in the quantization of the effective string~\cite{Polchinski:1991ax} and the solution here is similar. We are allowed to add singular terms to the effective action since we are anyhow expanding around a nontrivial background:
\es{HigherDer}{
S_1={\beta\ov \pi} \int d^2x\ {\partial^2\varphi\bar\partial^2\varphi\over \partial\varphi\bar\partial\varphi}~.
}
$\beta$ will turn out to be proportional to the shift in central charge. Expanded about the superfluid solution we find that this leads to a contribution to the effective action of the fluctuations 
\es{HigherDer2}{
S_1=-{4\beta\over  \pi\mu^2} \int d^2x\ \partial^2\pi\bar\partial^2\pi+O(1/\mu^{4})\,.
}

On the one hand, as will be discussed in the next section, $S_1$ is somewhat trivial; one manifestation of this is that it does not lead to a modification of the propagator \eqref{PropCorr} up to $O(1/\mu^{4})$ (this is true since $S_1$ vanishes on shell).
On the other hand, the conformal symmetry of the action $S=S_0+S_1$ is modified:  with precision $O(1/\mu^{3})$ the deformed action $S$ is invariant under the corrected conformal transformations
\es{CorrectedTF}{
\de \varphi(u,\bar{u})=\lam(u)\p \varphi-{\beta\ov2\ka}\,{\p^2\lam(u)\ov {\p}\varphi}\,,
}
(there is an independent antiholomorphic copy of the symmetry) and the stress tensor that generates this symmetry is
\es{CorrectedEMT}{
T&=-{\ka}(\p\varphi)^2+\beta\,{\p\varphi\,\p^3\varphi-(\p^2\varphi)^2\ov(\p\varphi)^2}+O(1/\mu^{3})\\
&={\ka \mu^2\ov 4}+i\ka\mu\p\pi-\ka(\p\pi)^2+{2i\beta\ov \mu}\,\p^3\pi+O(1/\mu^{2})\,.
}
The stress tensor two-point function can now be straightforwardly computed using Wick contractions using the propagator \eqref{PropCorr}. We get
\es{TTsubleading}{
\langle T(u)T(0)\rangle =\eqref{TTleading}+{12\beta\over u^4} +O(1/\mu)\,.
}
If we set $\beta=(c-1)/24$, we recover \eqref{Aim} to $O(1/\mu)$.
To work out the predictions of the EFT to higher orders, we need a more systematic approach, which we turn to next. 

We will also see below that there is no holomorphic current, i.e.~a weight $(1,0)$ primary operator of the Virasoro symmetry generated by the deformed $T$ of \eqref{CorrectedEMT} unless $\beta=0$.

\subsubsection{Systematic Development of the Effective Theory}

There is a systematic procedure to construct all terms allowed by symmetry in the effective action. We define the Weyl invariant metric $\hat{g}_{\mu\nu}\equiv g_{\mu\nu} \abs{\p \varphi}^2$, where $\abs{\p \varphi}^2\equiv-g^{\mu\nu}\p_\mu\varphi\p_\nu\varphi$. Then in the derivative expansion we can write the following terms:
\es{derivexp}{
S_\text{deriv}={\kappa\ov \pi} \int d^2 x\ \sqrt{\hat{g}}\le[1+ \al_{4,1} \hat{R}^2+\al_{4,2} \hat\nabla^\mu\hat\nabla^\nu\hat{R} \,\p_\mu \varphi\p_\nu \varphi+\dots\ri]\,,
}
where $\alpha_{k,i}$ is the coefficient of the $i$th term at $k$th derivative order. We used the leading order equation of motion $\nabla^2 \varphi=0$, that in $2d$ the Riemann tensor has only one independent component, $\hat{R}$, and that $\int d^2 x\sqrt{\hat{g}}\hat{R}$ is a topological invariant, the Euler characteristic of the manifold to reduce the number of terms in \eqref{derivexp}. There is one famous term that is missing from $S_\text{deriv}$, since it is not a local Weyl invariant in itself, but transforms with a shift that is a total derivative. The Wess-Zumino term takes the form~\cite{Polyakov:1981rd,Komargodski:2011xv}:
\es{derivexp2}{
S_\text{WZ}&=\al_2 \int d^2 x\ \sqrt{{g}}\le[(\p_\mu\tau)^2+\tau R\ri]\,,\\
\tau&\equiv -\log \abs{\p \varphi}\,.
}
Note that $\tau$ here is a composite dynamical field (as opposed to a background field, which is the more common case in the literature). 
 
 Let us now take $g_{\mu\nu}$ to be flat. In complex coordinates the leading order equation of motion is $\p\bar\p\varphi=0$, and by dropping terms proportional to it, we realize that $S_\text{WZ}=S_1$ from \eqref{HigherDer} with $\beta=\pi\al_2/4$. So we make contact with the considerations in the previous section.  The equation of motion
also implies that {\it on shell} $\varphi(u,\bar{u})=\chi(u)+\bar\chi(\bar u)$ and 
 \es{FlatHat}{
 d\hat{s}^2&=\hat{g}_{\mu\nu}dx^\mu dx^\nu\\
 &=-\p\chi\bar\p\bar\chi \, g_{\mu\nu}dx^\mu dx^\nu\\
 &=-\p\chi\bar\p\bar\chi \, dz d\bar{z}\\
 &=-d\chi d\bar\chi\,,
} 
i.e.~$\hat{g}_{\mu\nu}$ is flat. This then implies that on-shell all curvature invariants that we used to build $S_\text{deriv}$ in \eqref{derivexp} vanish. In addition, as we have seen before, $S_\text{WZ}$ is also a total derivative modulo the equations of motion. The EFT is nontrivial despite the action having no terms which are nonzero on-shell. This is because the EM tensor could (and should) receive various corrections. 

Since all the higher terms in the effective action vanish on-shell,  we can use a powerful general result in EFTs that there exists a field redefinition $\varphi\to\tilde \varphi$, that makes the action quadratic.
\es{derivexp3}{
S={\kappa\ov \pi} \int d^2 x\ \p\tilde \varphi\bar\p\tilde\varphi\,.
}
We can then expand around the superfluid background by taking $\tilde \varphi=-i\mu\tau+\tilde \pi$. 
The relation between $\pi$ and $\tilde \pi$ is (see also \cite{Drummond:2004yp}):
\es{FieldRedef}{
\pi=\tilde \pi-{2\beta\ov\ka \mu^2}\, \p\bar\p \tilde \pi+O(1/\mu^{3})\,.
}
The symmetry transformation of \eqref{CorrectedTF} and the corresponding stress tensor \eqref{CorrectedEMT} become
\es{CorrectedTF2}{
\de \tilde \pi(u,\bar{u})=&-{i\mu\ov2}\,\lam(u)-{i\beta\ov2\ka\mu}\,{\p^2\lam(u)}\\
&+\lam(u)\p \tilde \pi+O(1/\mu^{2})\,,\\
T(u)=&{\ka \mu^2\ov 4}+i\ka\mu\p\tilde\pi+{2i\beta\ov \mu}\,\p^3\tilde\pi\\
&-\ka(\p\tilde\pi)^2+O(1/\mu^{2})\,,
}
where in the first lines we collected terms that shift  $\tilde \pi$ and hence their generators are linear in $\tilde \pi$, while  the second lines correspond to conformal transformations. (In the expression of $T$ we dropped terms proportional to the equation of motion.) To $O(\mu^0)$ the symmetry  is just a combination of the shift and conformal symmetry of the free compact boson, and correspondingly $T$ is just a sum of the conventional current and stress tensor. At higher orders the symmetry and its generator becomes more exotic. To the order we wrote down formulas the computation of the stress tensor correlator is identical in the $\tilde\pi$ and $\pi$ variables, but the introduction of $\tilde\pi$ streamlines the computation at higher orders.

 In summary, we are faced with the problem of constructing a tensor in a derivative expansion from a free scalar governed by the action $S=\kappa/\pi \int d^2 x\ \p\tilde \pi\bar\p\tilde\pi$. The stress tensor is supposed to obey the OPE
 \es{TTOPE}{
 T(u)T(0)={c\ov 2 u^4}+{2\ov u^2}\,T(0)+{1\ov u}\,\p T(0)+\text{regular}\,,
 }
which can be achieved order by order in $1/\mu$. The remarkable fact is that $c$ is tunable.\footnote{There is another stress tensor $T=\ka(\p\phi)^2+V\p^2\tilde\phi$ (that of the linear dilaton CFT) that produces a tunable central charge $c=1+6V^2/\ka$ from a free scalar action for the noncompact scalar $\phi$; our setup with a compact $\varphi$ is different as a linear dilaton term is forbidden. } We could have started from this formulation of the problem, but for physical intuition and to make contact with the literature, we took a detour.

Using the {\it Mathematica} package OPEdefs \cite{Thielemans:1991uw}, by imposing the stress tensor OPE, we found that the first few orders of the stress tensor are:
\es{CorrectedEMT2}{
T(u)=&{\ka \mu^2\ov 4}+i\ka\mu\p\tilde\pi-\ka(\p\tilde\pi)^2+{1 \ov \mu}\,\le[{i(c-1)\ov 12}\,\p^3\tilde\pi+\ga_1\, \p\tilde\pi \p^2\tilde\pi \ri]\\
&+{1 \ov \mu^2}\,\le[{(c-1)\ov 6}\,\le((\p^2\tilde\pi)^2+\p\tilde\pi\p^3\tilde\pi\ri)+\ga_2 \,\p^4\tilde\pi+\ga_3\, (\p\tilde\pi)^2 \p^2\tilde\pi \ri]+O(1/\mu^3)\,,
}
where $\ga_i$ are arbitrary coefficients (they are not quite Wilson coefficients, since, as we remarked, the action does not admit terms beyond the free kinetic term). We now attempt to construct a $(1,0)$ holomorphic primary in the $1/\mu$ expansion by imposing the OPEs:
\es{JOPE}{
T(u)j(0)&={1\ov u^2}\, j(0)+{1\ov u}\, j'(0)+\text{regular}\,,\\
j(u)j(0)&={\ka\ov 2u^2}+\text{regular}\,.
}
We succeed to $O(\mu^0)$, but at $O(1/\mu)$ the most general Ansatz 
\es{JAnsatz}{
j={\ka\mu\ov 2}+i\ka \p\tilde\pi +{1 \ov \mu}\,\le[\lam\,\p^3\tilde\pi+\ga_1\, \p\tilde\pi \p^2\tilde\pi \ri]+O(1/\mu^2)
}
leads to a contradiction with  \eqref{JOPE}: the OPE with $T$ wants to set $\lam=-i(c-1)/12$, while the OPE with $j$ to $\lam=0$. (Of course these are consistent for $c=1$.)
This is how the general theorem of \cite{Affleck:1985jc} manifests itself in our concrete computation. The absence of a holomorphic current in effective string theory is due to similar reasons~\cite{Polchinski:1991ax}.

\subsubsection{Ground State Energy}

While there are undetermined coefficients in the stress tensor~\eqref{CorrectedEMT2}, it turns out that the vacuum energy is universal in this 2d EFT to all orders in the large charge expansion. 

The argument consists of two simple steps. First we note that while the conformal transformations implemented by $T$ in \eqref{CorrectedEMT2} are exotic (as displayed in \eqref{CorrectedTF2}), $L_0+\bar{L}_0$ generates ordinary time translations on the cylinder. This can be seen either from the transformation law it generates for constant $\lam(u)$ in \eqref{CorrectedTF2}, or by noticing that all higher order terms in $1/\mu$ are total derivatives,
\es{Ttotder}{
T(u)=&{\ka \mu^2\ov 4}+i\ka\mu\p\tilde\pi-\ka(\p\tilde\pi)^2+\p \psi\,,\\
\psi\equiv&{1 \ov \mu}\,\le[{i(c-1)\ov 12}\,\p^2\tilde\pi+{\ga_1\ov 2}\, (\p\tilde\pi)^2 \ri]+{1 \ov \mu^2}\,\le[{(c-1)\ov 6}\,\p\tilde\pi \p^2\tilde\pi+\ga_2 \p^3\tilde\pi+{\ga_3\ov 3}\, (\p\tilde\pi)^3  \ri]+O(1/\mu^3)\,.
}
The cylinder partition function in the fixed $Q$ sector can be written as
\es{cylZ}{
Z_Q[\beta]&=\Tr_\text{$Q$ fixed}\le[e^{-{\beta\ov R}\le(L_0+\bar L_0-{c\ov 12}\ri)}\ri]\\
&=\int_\text{$Q$ fixed} D\tilde\varphi\ e^{-S_\text{free}[\tilde\varphi]}\,.
}
While our manipulations above were in classical field theory, we can argue that keeping track of the Jacobian of the field redefinition $\varphi \to \tilde \varphi$ would not change the conclusion that there exists a field redefinition that makes the theory free. The Jacobian is a local Weyl invariant functional of $\varphi$, hence can be exponentiated and written in terms of $\hat{g}_{\mu\nu}$. Since we have written all these terms in the action \eqref{derivexp}, keeping track of the Jacobian only changes coefficients in the action. Then there must exist a field redefinition that makes the action free. This argument is reminiscent of the classic argument of \cite{David:1988hj,Distler:1988jt}.

In the $\beta\to\infty$ limit from the representation as a trace, we see that we pick up the ground state energy $\exp\le[-{\beta\ov R}\le(\De_Q-{c\ov 12}\ri) \ri]$. Evaluating the path integral at fixed charge gives
\es{Zpath}{
Z_Q[\beta]&\to \exp\le[-\beta \le({\ka \mu^2 R\ov 2}-{1\ov 12R}\ri) \ri]\,,
}
where the second term is the Casimir energy of the free real scalar. Using the relation \eqref{eqs:densitymu} with $c_1=\ka/(4\pi)$, we conclude that 
\es{DimPred}{
\De_Q={Q^2\ov 2 \ka}+{c-1\ov 12}+O(e^{-Q})\,.
}
We hope that this prediction can be tested in a situation where our effective theory would apply (e.g. in the context of BCFT).

\pagebreak

We note that there exists another method for computing the dimension of the lowest dimension large charge scalar, analogous to the approach of~\cite{Drummond:2004yp}. This method gives the same result as~\eqref{DimPred}.

\section*{Acknowledgements}

We thank Gabriel Cuomo, Bruno Le Floch, Petr Kravchuk and Cumrun Vafa for very useful discussions.  ZK, MM, and ARM are supported in part by the Simons Foundation grant 488657 (Simons Collaboration on the Non-Perturbative Bootstrap) and the BSF grant no. 2018204. The work of ARM was also supported in part by the Zuckerman-CHE STEM Leadership Program. SP acknowledges the support from DOE grant DE-SC0009988. 


\appendix

\section{The Zoo of Correlation Functions}
\label{app:GreensFunctions}
Consider the Euclidean time-ordered two point function of two (bosonic)\footnote{For $A,\,B$ fermionic operators, one needs to replace the commutator in \eqref{Gs} with anti-commutators.} operators $A,\, B$:
\es{GAB1}{
G_{AB}(\tau)=-\expval{T_{\tau}\,A(\tau)B(0)}\,,
}
where the symbol $T_{\tau}$ denotes the time-ordering operator, defined such that $T_{\tau}\,A(\tau)B(0) = A(\tau)B(0)$ if $\tau>0$ and $B(0)A(\tau)$ if $\tau<0$. 
We define the Fourier transform of this function as $G_{AB}(i\nu)=\int_{-\infty}^\infty d\tau \ G_{AB}(\tau) e^{i\nu \tau}$ for real $\nu$.
It is a fundamental result, derived from inserting a complete set of states, that this can be analytically continued to give a function $G_{AB}(\omega)$ on the complex $\om$ plane (first sheet) except for the real $\om$ axis, where it has a branch cut. The function has the following spectral representation:
\es{GABSpect}{
G_{AB}(\om)=\int_{-\infty}^\infty d\om'\ {\rho_{AB}(\om')\ov \om -\om'}\,.
}
$\rho_{AB}(\omega)$ is called the spectral density associated with the operators $A$ and $B$.

All other Green's functions can be computed from $G_{AB}(\om)$ using an analytic continuation. We are interested in the following Green's functions:
\es{Gs}{
G^{(R)}_{AB}(t)&\equiv -i\Theta(t)\expval{[A(t),B(0)]},\\
G^{(A)}_{AB}(t)&\equiv i\Theta(-t)\expval{[A(t),B(0)]},\\
G^{(comm)}_{AB}(t)&\equiv \expval{[A(t),B(0)]}=i\le(G^{(R)}_{AB}(t)-G^{(A)}_{AB}(t)\ri)\,,
}
where $G^{(R)}_{AB}(t)$ is the retarded function, $G^{(A)}_{AB}(t)$ is the advanced function and $G^{(comm)}_{AB}(t)$ is defined by the third line above as the correlator of the commutator $[A(t),B(0)]$. Its Fourier transform is  proportional (up to normalization) to the spectral function $\rho_{AB}(\omega)$. 
From the above expressions, it is clear that the Fourier transformed function $G^{(R)}_{AB}(\om)$ is analytic in the upper half plane of complex $\omega$, while $G^{(A)}_{AB}(\om)$ is analytic in the lower half plane of complex $\omega$. 
The following relations hold in momentum space:
\es{Grels}{
G^{(R)}_{AB}(\om)&=G_{AB}(\om+i\ep)\\
G^{(A)}_{AB}(\om)&=G_{AB}(\om-i\ep)\\
G^{(comm)}_{AB}(\om)&=i\le[G^{(R)}_{AB}(\om)-G^{(A)}_{AB}(\om)\ri]\\
&=i\le[G_{AB}(\om+i\ep)-G_{AB}(\om-i\ep)\ri]\\
&=i\int_{-\infty}^\infty d\om'\ \rho_{AB}(\om') \le[{1\ov \om -\om'+i \ep}-{1\ov \om -\om'-i \ep}\ri]\\
&=i\int_{-\infty}^\infty d\om'\ \rho_{AB}(\om') \le[-2\pi i \de(\om-\om')\ri]\\
&=2\pi  \rho_{AB}(\om)\,.
}

As an example, let us see how these identities work out for the case of the complex free scalar field. For the free scalar, the Euclidean propagator is given by 
\begin{equation}
G_{\bar{\phi}\phi}(i\nu, \vec{p})=\frac{-1}{\nu^2+|\vec{p}|^2}
\end{equation}
One analytically continues to to obtain a function defined on the complex $\omega$ plane except of the real axis: 
\begin{equation}
G_{\bar{\phi}\phi}(\omega, \vec{p})=\frac{1}{\omega^2-|\vec{p}|^2}\,, \quad\omega\notin\mathbb{R}
\end{equation}

Now we can use \eqref{Grels} by choosing $A=\bar\phi$ and $B=\phi$:
\begin{equation}
\begin{aligned}
G_{\bar\phi\phi}^{(c o m m)}(\omega) &=G_{\bar\phi\phi}^{(R)}(\omega)-G_{\bar\phi\phi}^{(A)}(\omega)\\
&= i\left[G_{\bar{\phi}\phi}(\omega+i\epsilon, \vec{p})-G_{\bar{\phi}\phi}(\omega-i\epsilon, \vec{p})\right]\\
&=i\left[\frac{1}{(\omega+i\epsilon)^2-|\vec{p}|^2}-\frac{1}{(\omega-i\epsilon)^2-|\vec{p}|^2}\right]\\
&=\frac{i}{2|\vec{p}|}\left[\frac{1}{(\omega+i\epsilon)-|\vec{p}|}-\frac{1}{(\omega+i\epsilon)+|\vec{p}|}-\frac{1}{(\omega-i\epsilon)-|\vec{p}|}+\frac{1}{(\omega-i\epsilon)+|\vec{p}|}\right]\\
&=\frac{2\pi }{2 |\vec{p}|} \left[\delta(\omega-|\vec{p}|)-\delta(\omega+|\vec{p}|)\right]\\
&= (2\pi) \text{sgn}(\omega) \delta(p_\mu p^\mu),
\end{aligned}
\end{equation}
and the spectral function reads:
\begin{equation}\label{eqs:freespec}
\rho_{\bar{\phi}\phi} (\omega) =  \text{sgn}(\omega) \delta(p_\mu p^\mu).
\end{equation}

 \eqref{eqs:freespec} can also be obtained using an equivalent description of analytically continuing the space-time coordinates. While the above method is used in sec.~\ref{sec:NGthm}, the following method is used in the rest of the paper.

We start with the Wightman correlation function for free scalar in Minkowski spacetime:
\begin{align}
\langle 0 \vert \bar{\phi}(t, \vec{x})\phi(0) \vert 0\rangle & = \frac{1}{(d-2)s_{d-1}}\left[(\epsilon+it)^2+\vec{x}\cdot\vec{x} \right]^{-\Delta_{\phi}}, \qquad \epsilon\to 0^+, \\
\langle 0 \vert {\phi}(0)\bar{\phi}(t, \vec{x}) \vert 0\rangle & =\frac{1}{(d-2)s_{d-1}}\left[(-\epsilon+it)^2+\vec{x}\cdot\vec{x} \right]^{-\Delta_{\phi}}, \qquad \epsilon\to 0^+.
\end{align}
The Fourier transformed Wightman function $G^W_{\bar{\phi}\phi}(\omega,\vec{p})$ reads: 
\begin{equation}
\label{eq:Gphiphi}
\begin{aligned}
G^W_{\bar{\phi}\phi}(\omega,\vec{p})& = \int d^dx \  e^{i(\omega t -\vec{p}\cdot\vec{x})}\langle 0 \vert \bar{\phi}(t, \vec{x})\phi(0) \vert 0\rangle \\
& =2\pi\Theta(\omega)\Theta(p_\mu^2)\frac{4^{\Delta_*-\Delta_{\phi}}\Gamma(\Delta_*)}{\Gamma(\Delta_{\phi})}\frac{(p_\mu^2)^{\Delta_{\phi}-\Delta_*-1}}{\Gamma(\Delta_{\phi}-\Delta_*)},
\end{aligned}
\end{equation}
where we used the notation $p_\mu^2 \equiv p_{\mu}p^{\mu}=\omega^2-\vec{p}\cdot\vec{p}$ and $\Delta_*=\frac{d-2}{2}$ is the unitarity bound, i.e. $\Delta_{\phi}\geq\Delta_*$ is saturated by the free scalar field. In the limit where $\Delta_{\phi}\to \Delta_*^+$, one gets: 
$$
\Theta(p_\mu^2)\frac{(p_\mu^2)^{\Delta_{\phi}-\Delta_*-1}}{\Gamma(\Delta_{\phi}-\Delta_*)}\to \delta(p_\mu^2).
$$
Thus, for the free scalar field we find:
\begin{equation}
G^W_{\bar{\phi}\phi}(\omega,\vec{p}) = 2\pi\Theta(\omega)\delta(p_\mu^2)\,.
\end{equation}
The commutator correlator is given by: 
\es{commWightman}{
G^{(comm)}_{\bar\phi\phi}(\omega,\vec{p})&=
\int d^dx e^{i(\omega t -\vec{p}\cdot\vec{x})}\langle 0 \vert \left[ \bar{\phi}(t,\vec{x}),\phi(0) \right] \vert 0 \rangle\\
& = G^W_{\bar{\phi}\phi}(\omega, \vec{p})-[G^W_{\bar{\phi}\phi}(-\omega,-\vec{p})]^* = 2\pi \text{sgn}(\omega)\delta(p_\mu^2)\,.
}
From above, one can rederive the spectral function \eqref{eqs:freespec}.

 \section{$\Delta_Q$ in the Free Fermionic Phase }\label{app:Fermiondims}
\label{app:DeltaQFreeFermion}

In this appendix, we extend the analysis described at the end of subsection \ref{subsec:FreeFermionsLargeCharge} for the case of $d=4$ dimensions.
Considering the $d=4$ theory of free fermions on the cylinder $ S^3\times \R$, we have: 
\begin{equation}
\Delta_Q = \sum_{j=1/2}^{j_{\text{max}}-1} g_j^{(d=4)} \varepsilon_j^{(d=4)}+\delta Q \varepsilon_{j_{\text{max}}}^{(d=4)},
\end{equation}
where $g_j^{(d=4)}$ represents the degeneracy and it is given by:
\begin{equation}
g_j^{(d=4)} = \left(j+\frac{3}{2}\right)\left(j+\frac{1}{2}\right),
\end{equation}
and $\varepsilon_j^{(d=4)} = j+1$ are the energy eigenvalues. The charge $Q$ is given by:
\begin{equation}
Q = \sum_{j=1/2}^{j_{\text{max}}-1} g_j^{(d=4)}+\delta Q,
\end{equation}
where $\delta Q$ represents the amount of occupied states in the outermost energy shell and it can take values in the range $0\leq\delta Q\leq \left(j_{\text{max}}+\frac{3}{2}\right)\left(j_{\text{max}}+\frac{1}{2}\right)$.
\begin{figure}[!h]
\centering
\subfloat[]{\label{fig:Fermion4dDeltaQ}{\includegraphics[width=0.35\textwidth]{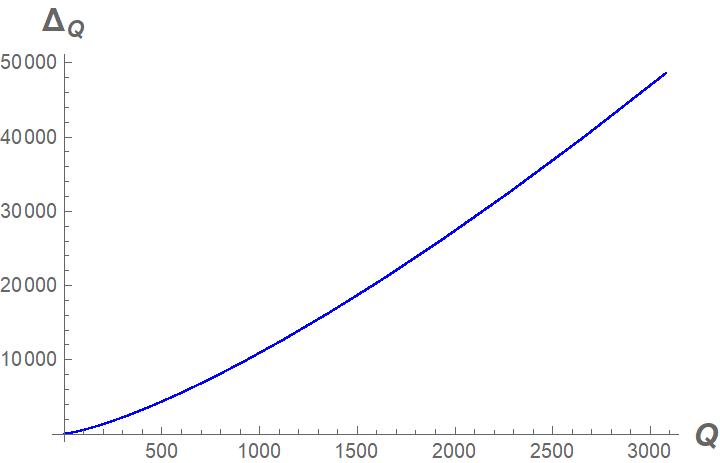}}}    \qquad \qquad
\subfloat[]{\label{fig:Fermion4dEnergyFlactuations}{\includegraphics[width=0.55\textwidth]{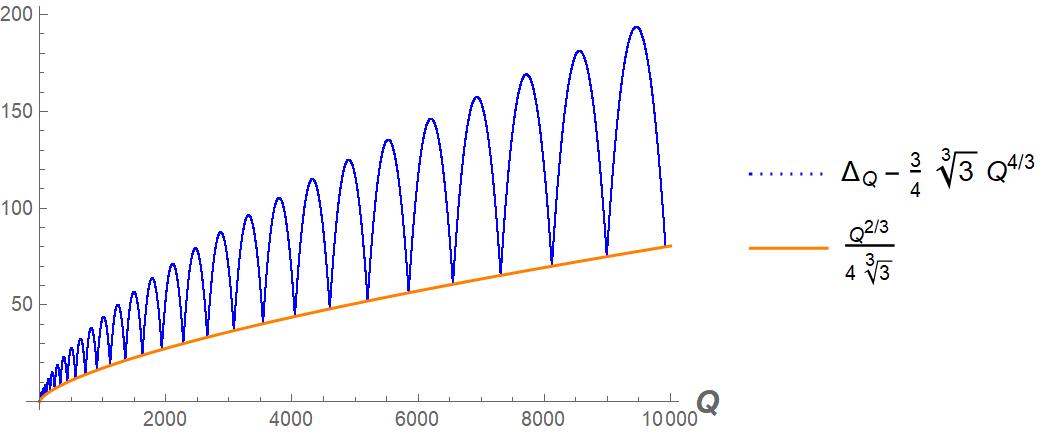}}}
\caption{An illustration describing the behavior of the scaling dimension for a general value of $Q$ in the $d=4$ case. Figure \ref{fig:Fermion4dDeltaQ} describes $\Delta_Q$ as a function of $Q$. In figure \ref{fig:Fermion4dEnergyFlactuations}, the blue dashed line corresponds to $\Delta_Q-\frac{3^{4/3}}{4}Q^{4/3}$ as a function of $Q$, while the orange line describes $\frac{Q^{2/3}}{4\cdot 3^{1/3}}$. The points in which the blue dashed line meets the orange line correspond to cases in which $Q$ is  such that equation \eqref{eq:Eqfornmax4d} is satisfied with integer values of $n_{\text{max}}$. }
\label{fig:Fermion4dFig}
\end{figure}

In figure \ref{fig:Fermion4dFig}, we describe the general $Q$ case. One can see, from figure 
\ref{fig:Fermion4dEnergyFlactuations}, that $\Delta_Q$ meets the value \eqref{eq:DeltaQ4d} that is associated with a homogeneous and isotropic ground state only for specific values of $Q$. These values again correspond to  cases in which the outermost shell is filled. 
Note that similar to the $d=3$ case, there are fluctuations in the difference between $\Delta_Q$ to the leading order term of \eqref{eq:DeltaQ4d}, these fluctuations are of order $O(Q^{2/3})$ (which is the same order in $Q$ as the next to leading order term in \eqref{eq:DeltaQ4d}), hence $\Delta_Q$ is not analytic in $Q$.

Let us make a comment regarding the leading order term in the expansion for $\Delta_Q$ for a general number of spacetime dimensions $d$. The energy eigenvalues for a general $d$ are given by:
\begin{equation}
\varepsilon_j^{(d)} = j + \frac{d-2}{2}.
\end{equation}
In the $Q\to \infty$ limit, the behavior of $g_j^{(d)}$ is governed by large $j$, and can be obtained for any dimension (one can systematically study this using Hilbert series \cite{Henning:2017fpj,Melia:2020pzd}):
\begin{equation}\label{largek}
g_j^{(d)} \underset{j\to\infty}{\simeq} \frac{2}{\Gamma(d-1)} j^{d-2} \text{dim} \left[\underbrace{1/2,1/2,\cdots}_{r-1\ \text{times}}\right]=\frac{j^{d-2}}{\Gamma(d-1)}  2^{\lceil \frac{d}{2}\rceil -1}\,,
\end{equation}
where $r$ is the rank of $SO(d)$.
Since we know that $j_{\text{max}}(Q)$ is an increasing function of $Q$ and $g_j^{(d)}$ is increasing with $j$, the large $Q$ behavior is controlled by large $j$ behavior of  $g_j^{(d)}$. At this point, we can use \eqref{largek} and find:
\begin{equation}
Q \simeq \sum_{j=1/2}^{j_{\text{max}}} \frac{j^{d-2}}{\Gamma(d-1)}  2^{\lceil \frac{d}{2}\rceil -1} \underset{j_{\text{max}}}{\simeq} \frac{j_{\text{max}}^{d-1}}{\Gamma(d)}  2^{\lceil \frac{d}{2}\rceil -1}.
\end{equation}
This can be solved in the leading order to obtain: 
\begin{equation}
\label{eq:JmaxGeneralD}
j_{\text{max}}(Q)= \left[\Gamma(d) 2^{1-\lceil \frac{d}{2}\rceil}Q\right]^{1/(d-1)}.
\end{equation}
This leads to: 
\begin{equation}
\Delta_Q=\sum_{j}^{j_{\text{max}}}\epsilon^{(d)}_jg^{(d)}_j \underset{Q\to\infty}{\simeq} \left[\sum_{j}^{j_{\text{max}}(Q)}\frac{j^{d-1}}{\Gamma(d-1)}  2^{\lceil \frac{d}{2}\rceil-1 }\right]=\frac{j_{\text{max}}(Q)^{d}}{d\Gamma(d-1)}  2^{\lceil \frac{d}{2}\rceil-1}\,.
\end{equation}
Substituting back the value of $j_{\text{max}}(Q)$ using \eqref{eq:JmaxGeneralD} in the above equation, we find the result \eqref{eq:LargeChargeGeneralDLeadingOrder} for the leading order term in $\Delta_Q$ in a general number of spacetime dimension. 

\section{Contact Terms in Energy-Momentum Correlators}\label{app:contact}

In the paper we have studied some general constraints on the spectrum of the theory in the situation that the boost symmetry is spontaneously broken. 
This essentially boiled down to studying the commutators of the conformal charges with the energy-momentum operator. For a conformal Killing vector $\xi^\mu$, \begin{equation}\label{CKV}\partial^\mu\xi^\nu+\partial^\nu\xi^\mu={2\over d}\eta^{\mu\nu}\partial\cdot \xi~,\end{equation}
we can define a corresponding conserved conformal charge $Q^{\xi} = \int d^{d-1}x\ \xi^\mu T_{\mu 0}$, where the integral is over a space-like slice.

The general transformation rule of the Energy-Momentum tensor (assuming $d\geq 3$) is 
\begin{equation} \label{commu} [Q^\xi , T_{\rho\sigma}] = \xi^\lambda \partial_\lambda T_{\rho\sigma} +{\partial\xi^\mu\over \partial x^\rho} T_{\mu\sigma}+{\partial\xi^\nu\over \partial x^\sigma} T_{\nu\rho}+{d-2\over d}\partial\cdot \xi T_{\rho\sigma}~.  \end{equation} 
The above equation is equivalent to assigning some operator contact terms in the products 
$T_\mu^\mu(x) T_{\rho\sigma}(x')$ and $\partial^\mu T_{\mu\nu}(x)T_{\rho\sigma}(x')$. Indeed, since 
$\partial^\nu( \xi^\mu T_{\mu \nu})=\xi^\mu \partial^\nu T_{\mu\nu}+{1\over d}\partial\cdot \xi T_\mu^\mu $ (where we used the equation satisfied by conformal Killing vectors~\eqref{CKV}), using the Stokes theorem and the fact that $Q^{\xi}$ is invariant under small deformations we can  compute 
$[Q^\xi , T_{\rho\sigma}]$ from the contact terms in the products 
$T_\mu^\mu(x) T_{\rho\sigma}(x')$ and $\partial^\mu T_{\mu\nu}(x)T_{\rho\sigma}(x')$.

A very important fact is that the contact terms in the products 
$T_\mu^\mu(x) T_{\rho\sigma}(x')$ and $\partial^\mu T_{\mu\nu}(x)T_{\rho\sigma}(x')$ are not fixed uniquely by the commutators $[Q^\xi , T_{\rho\sigma}]$. To understand why this is so, note that we can add contact terms 
\begin{equation}\label{ambigs}
\begin{aligned}
&T_{\mu\nu}(x)T_{\rho\sigma}(x')\\
&\sim A\delta^{d}(x-x')\left(\eta_{\mu\nu}T_{\rho\sigma}+\eta_{\rho\sigma}T_{\mu\nu}\right)+B\delta^{d}(x-x')\left(\eta_{\mu\rho}T_{\nu\sigma}+\eta_{\mu\sigma}T_{\nu\rho}+\eta_{\nu\rho}T_{\mu\sigma}+\eta_{\nu\sigma}T_{\mu\rho}\right)~
\end{aligned}
\end{equation}
with two arbitrary coefficients $A,B$. While these contact terms clearly modify the contact terms in the products 
$T_\mu^\mu(x) T_{\rho\sigma}(x')$ and $\partial^\mu T_{\mu\nu}(x)T_{\rho\sigma}(x')$, they do not modify the commutators $[Q^\xi , T_{\rho\sigma}]$. The latter statement is true as can be verified by a direct computation or by observing that $[Q^\xi , T_{\rho\sigma}]$ is a separated-points observable and hence it can only be sensitive to the contact terms that truly originate from separated points physics.

Nevertheless the  contact terms in the products 
$T_\mu^\mu(x) T_{\rho\sigma}(x')$ and $\partial^\mu T_{\mu\nu}(x)T_{\rho\sigma}(x')$ can be completely fixed once the Energy-Momentum tensor correlators are precisely defined. 

A local QFT can be coupled to a space-time metric $g$ and hence we can define the generating functional of connected correlators $W[g]$. The functional $W[g]$ is not entirely determined by the underlying theory. It suffers from an ambiguity by local terms, for instance, in 3+1 dimensions we may add $\int d^4x \sqrt g R^2$ to $W[g]$. ($R$ is the Ricci scalar.) We can also add higher powers of $R$ suppressed by a cutoff if we are dealing with an effective theory. For reasons that will become clear shortly, such ambiguities should be called ultra-local. These ambiguities will not affect our discussion below or the coefficients $A,B$ in~\eqref{ambigs}. 
The Energy-Momentum correlators can be extracted, by definition, via functional derivatives of $W[g]$:  
\begin{equation}\label{funder}\langle T_{\mu_1\nu_1}(x_1) \cdots T_{\mu_n\nu_n}(x_n) \rangle= {(-2)^n\over \sqrt {g(x_1)}\cdots \sqrt {g(x_n)}} {\delta \over \delta g^{\mu_1\nu_1}(x_1)}   \cdots {\delta \over \delta g^{\mu_n\nu_n}(x_n)}   W~.\end{equation}
Equivalently, the Energy-Momentum correlators can be worked out from the  expansion of the generating functional around some given background metric $g$ as follows. We define $$g'^{\rho\sigma}=g^{\rho\sigma}+\delta g^{\rho\sigma}~.$$
The first few terms in the expansion are
\begin{equation}
W[g']=W[g]-{1\over 2}\int d^dx \sqrt {g(x)} \delta g^{\mu\nu} \langle T_{\mu\nu}\rangle+\frac{1}{8} \int d^dx\int d^dy \sqrt {g(x)}\sqrt{ g(y)} \delta g^{\mu\nu}\delta g^{\rho\sigma} \langle T_{\mu\nu}T_{\rho\sigma}\rangle$$ $$-{1\over {8\cdot 3!} }\int d^dx\int d^dy\int d^dz \sqrt {g(x)} \sqrt {g(y)}\sqrt {g(z)}\delta g^{\mu\nu}\delta g^{\rho\sigma}\delta g^{\phi\chi} \langle T_{\mu\nu}T_{\rho\sigma}T_{\phi\chi}\rangle+...~.
\end{equation}
The functional derivative is defined in the obvious way 
\begin{equation}
{\delta \over\delta g^{\mu\nu}(x) } g^{\alpha\beta}(y)=\frac12 \left(\delta^\alpha_\mu\delta^\beta_\nu+\delta^\alpha_\nu\delta^\beta_\mu\right)\delta^d(x-y)~.
\end{equation}

The definition~\eqref{funder} is not unique, however, it has the advantage that Bose symmetry is manifestly obeyed both at separated and coincident points. The definition~\eqref{funder} is standard in the literature. See for instance~\cite{Osborn:1999az,Baume:2014rla} and references therein.

We now see why scheme ambiguities such as $\int d^4x \sqrt g R^2$ do not matter for the coefficients $A,B$ in~\eqref{ambigs}. Indeed, scheme ambiguities lead to $c$-number contact terms in $TT$ and not operator ones. Sometimes operator contact terms are referred to as semi-local while $c$-number contact terms are referred to as ultra-local. (Operator contact terms are referred to as semi-local since in the presence of additional operators there would be delta functions over some positions but not all positions.)

To understand the consequences of diffeomorphism invariance we start with some metric and perform a 
change of variables  
$$ds^2=g_{\mu\nu}(x)dx^\mu dx^\nu = {\partial x^\mu\over \partial x'^\rho }{\partial x^\nu\over \partial x'^\sigma }g_{\mu\nu}(x) dx'^\rho dx'^\sigma$$
As usual now we can view the new metric as a function of $x'$, 
$$ g'_{\rho\sigma}(x')={\partial x^\mu\over \partial x'^\rho }{\partial x^\nu\over \partial x'^\sigma }g_{\mu\nu}(x(x'))~. $$
In the absence of gravitational anomalies (which we will assume throughout for simplicity) the metrics $g,g'$ give rise to the same $W[g]=W[g']$. 
 The invariance of the effective action under diffeomorphisms implies a differential equation 
 \begin{equation}\label{diffeoWI}\int d^dx \left(\nabla^\rho \eta^\sigma+\nabla^\sigma\eta^\rho\right) {\delta\over \delta g^{\rho\sigma}(x) }W  =0\end{equation}
that is valid for any vector field $\eta$. (Note that there is no $\sqrt g$ in this integral because the functional derivative of $W$ behaves as a density.) Equation~\eqref{diffeoWI} is of course equivalent to the conservation equation $\nabla^{\rho}\langle T_{\rho\sigma}\rangle=0$. Next we must vary this equation 
 $$\delta \int d^dx \left(\nabla^\rho \eta^\sigma+\nabla^\sigma\eta^\rho\right) {\delta\over \delta g^{\rho\sigma}(x) }W  =0~.$$
 
A somewhat tedious computation of the variation above at the end of which we set the metric to be flat again, reduces to the precise prescription of contact terms in the product~$\partial^\mu T_{\mu\nu}(x)T_{\rho\sigma}(x')$:
\begin{equation}\label{prodone} 
\begin{aligned}
&\partial^\mu T_{\mu\nu}(x) T_{\rho\sigma}(x') \\
&=-\delta^d(x-x')\partial_\nu T_{\rho\sigma}(x')+\partial^{x}_\nu\delta^d(x-x') T_{\rho\sigma}(x')+\partial^x_\rho\delta^d(x-x')T_{\nu \sigma}(x')+\partial^x_\sigma\delta^d(x-x')T_{\nu \rho}(x')~.
\end{aligned}
 \end{equation} 
While the expression above does not appear manifestly Bose invariant, it actually is. Namely, if we act with another derivative on $(\partial^{x'})^\sigma$  there is a symmetry under interchanging $x\leftrightarrow x'$ and $\rho\leftrightarrow \nu$. 

With similar logic, in conformal field theories (which obey Weyl invariance upon coupling to a background metric), one can find the contact term in the product 
$T_\mu^\mu(x) T_{\rho\sigma}(x')$:
\begin{equation}\label{prodtwo} T^\mu_\mu(x)T_{\rho\sigma}(x') =2\delta^d(x-x')T_{\rho\sigma}(x')~.\end{equation}

The statements~\eqref{prodone}, \eqref{prodtwo} contain more information than the commutators and hence can be used as additional restrictions on correlation functions, including in the situation that boost invariance is spontaneously broken. These additional restrictions are however less useful since they are tied to the definition~\eqref{funder} more than to actual separated-points physics.

\bibliographystyle{JHEP}
\bibliography{refs.bib}

\end{document}